\documentclass[reqno]{siamonline190516}

\usepackage[left=1in,right=1in,top=1in,bottom=1in]{geometry}%

\usepackage{amsmath, amssymb, amsfonts, mathtools}
\usepackage{enumerate}
\usepackage{bbm}
\usepackage{cool}
\usepackage{graphicx,epsfig}
\usepackage{makecell}
\usepackage{array}
\usepackage{blkarray}
\usepackage{multirow}
\usepackage{booktabs}
\usepackage{paralist}

\usepackage[capitalize,nameinlink]{cleveref}

\crefname{section}{section}{sections}
\crefname{subsection}{section}{sections}
\Crefname{section}{Section}{Sections}
\Crefname{subsection}{section}{sections}

\Crefname{figure}{Figure}{Figures}

\crefformat{equation}{\textup{#2(#1)#3}}
\crefrangeformat{equation}{\textup{#3(#1)#4--#5(#2)#6}}
\crefmultiformat{equation}{\textup{#2(#1)#3}}{ and \textup{#2(#1)#3}}
{, \textup{#2(#1)#3}}{, and \textup{#2(#1)#3}}
\crefrangemultiformat{equation}{\textup{#3(#1)#4--#5(#2)#6}}%
{ and \textup{#3(#1)#4--#5(#2)#6}}{, \textup{#3(#1)#4--#5(#2)#6}}{, and \textup{#3(#1)#4--#5(#2)#6}}

\Crefformat{equation}{#2Equation~\textup{(#1)}#3}
\Crefrangeformat{equation}{Equations~\textup{#3(#1)#4--#5(#2)#6}}
\Crefmultiformat{equation}{Equations~\textup{#2(#1)#3}}{ and \textup{#2(#1)#3}}
{, \textup{#2(#1)#3}}{, and \textup{#2(#1)#3}}
\Crefrangemultiformat{equation}{Equations~\textup{#3(#1)#4--#5(#2)#6}}%
{ and \textup{#3(#1)#4--#5(#2)#6}}{, \textup{#3(#1)#4--#5(#2)#6}}{, and \textup{#3(#1)#4--#5(#2)#6}}

\crefdefaultlabelformat{#2\textup{#1}#3}

\newcommand{\vvec}{\mathbf{v}}

\newcommand{\Ivec}{\mathbf{I}}

\newcommand{\Kvec}{\mathbf{K}}

\newcommand{\xvec}{\mathbf{x}}

\newcommand{\Zerovec}{\mathbf{0}}
\newcommand{\Onevec}{\mathbf{1}}

\newcommand{\Fix}{{\rm Fix}}

\newcommand{\R}{\mathbb{R}}

\graphicspath{ {images/} }

\title{Bifurcations of a neural network model with symmetry}

\author{Ross Parker\thanks{Department of Mathematics, Southern Methodist University, Dallas, TX (\email{rhparker@smu.edu}).}
\and Andrea K. Barreiro\thanks{Department of Mathematics, Southern Methodist University, Dallas, TX (\email{abarreiro@smu.edu}).}
}

\begin{document}


\maketitle

\begin{abstract}
We analyze a family of clustered excitatory-inhibitory neural networks and the underlying bifurcation structures that arise because of permutation symmetries in the network as the global coupling strength $g$ is varied. We primarily consider two network topologies: an all-to-all connected network which excludes self-connections, and a network in which the excitatory cells are broken into clusters of equal size. Although in both cases the bifurcation structure is determined by symmetries in the system, the behavior of the two systems is qualitatively different. In the all-to-all connected network, the system undergoes Hopf bifurcations leading to periodic orbit solutions; notably, for large $g$, there is a single, stable periodic orbit solution and no stable fixed points. By contrast, in the clustered network, there are no Hopf bifurcations, and there is a family of stable fixed points for large $g$.
\end{abstract}



\section{Introduction}

Today experimental techniques allow an increasingly detailed view of the physical architecture of biological neural networks. However, drawing a clear line from physical connectivity to dynamic neural activity is still a challenge. The networks in question are massive in scale and high dimensional (with billions of neurons and possibly trillions of synapse).
Neural networks also show great diversity in structure at every level, from the morphology and excitability properties of a single cell to large scale connections between brain regions. 

    One common experimental finding is that neural dynamics are surprisingly low-dimensional when compared to the overall dimensionality of the neural system \cite{ecker14,Chai_etal_Neuron_2017,Gallego_NatCom_2018,Wang_etal_NatNeuro_2018,huang2019circuit} (see Fig. 1 of \cite{Gao_CON_2015} for a summary of earlier studies). The low-dimensional manifold may even shift slowly over time, as the underlying components of the network (cells and synapses) die and are replaced \cite{gallego2020long}. Thus, a major challenge for modern mathematical neuroscience is to understand how low-dimensional dynamics emerges from the observed connectivity of the brain.
    
    Real neural networks are partially structured but also partially random. Intuitively, it’s clear that not every connection in the brain must be tuned precisely (after all, every person reading this sentence will respond to these black markings in the same way, despite significant differences between our individual brains). This has motivated the use of analytical tools of random network theory, in which one seeks to draw conclusions about an ensemble of networks. An early example is the work by Sompolinsky et al. \cite{SompCris88} which applies dynamic mean field theory to single-population firing-rate networks in which connections are chosen from a mean-zero Gaussian distribution: in the limit of large network size, the authors find that the network transitions from quiescence to chaos as a global coupling parameter passes a bifurcation value. This value coincides with the point at which the spectrum of the connectivity matrix exits the unit circle \cite{girko85,bai97} thus making the connection to random matrix theory very concrete. Later authors have sought to extend these results to correlated or block-structured matrices \cite{kadmon_HS_2015,Aljadeff_2015}, and many others have studied the spectral characteristics of partially structured connection matrices \cite{Wei12, Ahmadian_etal_2015, muir_MF_2015} with neural networks as a primary motivation.
    
    However, the results of spectral theory and nonlinear dynamics have not always neatly aligned. One network setting that has caused persistent difficulty is excitatory-inhibitory networks with strong average connections \cite{RA06}. The predictions of random matrix theory suggest chaotic, asynchronous fluctuations, whereas large-scale coherent fluctuations have been observed instead. Why? The answer may be found in the nature of the deterministic perturbation. Several authors have examined how low-rank, asymmetric perturbations create an effectively feed-forward structure that allows coherent dynamics to co-exist with random fluctuations in an orthogonal subspace \cite{delMolino_etal_PRE_2013, Darshan_etal_PRX_2018, Landau_HS_PCB_2018,Landau_HS_PRR_2021}; the dimensionality of the dynamical subspace can be related to the dimension of the low-rank perturbation in the connectivity matrix \cite{Schuessler_etal_PRR_2020,beiran2021shaping}.  
    
    In an earlier work, we found an alternative possibility \cite{Barreiro2017}.  In examining balanced E-I networks without self-coupling, we persistently observed periodic solutions which could not be explained by random matrix theory. Instead, they arose as a consequence of underlying symmetries in the connection matrix and could be predicted through the machinery of equivariant bifurcation theory. However, some pieces of our analysis remained uncompleted: we were unable at that time to give a complete stability analysis. This is important because the stable solution is what one can expect to observe in a perturbed (random) network.
    
    Here, we complete this analysis for all-to-all excitatory-inhibitory networks. We then extend this analysis to a biologically significant block-structured case, in which the excitatory cells are clustered, but inhibition is global. We find that the dynamics are strikingly different: instead of limit cycles, we predict fixed points. In both cases, the structures can be understood by considering the symmetries of the deterministic connection matrix.

\section{Mathematical model}\label{sec:model}

We consider a network in which each node represents the firing rate of a single neuron. The individual neurons are connected by sigmoidal activation functions through a connectivity matrix, which specifies both the network of neuronal connections and the weight of each connection, including whether a given neuron is excitatory (E) or inhibitory (I). With noise in the connectivity matrix, this is an idealized model in neuroscience \cite{Ginzburg_HS_1994,RA06,kadmon_HS_2015}. Here, we will consider the system without noise, but where the connection weights have important symmetries. Specifically, we study:
\begin{equation}\label{eqn:sys_Basic}
    \dot{\xvec} = 
    F(\xvec, g) := -\xvec  + \frac{1}{\sqrt{N}} H\tanh (g \xvec),
\end{equation}
for $\xvec \in \R^N$, where the global coupling strength, $g$, is used a bifurcation parameter. The network comprises a total of $N$ neurons, of which $n_E$ are excitatory and $n_I$ are inhibitory. $H$ is the $N \times N$ connectivity matrix; the diagonal entries of $H$ are all 0 to exclude self-interactions of neurons (see \cite[Sec. 2.1]{Barreiro2017} for a discussion on why self-coupling of neurons is removed). We will use the parameter $f = n_E / N$ to identify the fraction of neurons that are excitatory:  for the remainder of this paper we will use $f = 0.8$ for a 4-to-1 excitatory-to-inhibitory ratio, which is typical for cortical networks \cite{Buxh_Brain_2002}. We note that $F$ is an odd function of $\xvec$, i.e. $F(-\xvec) = -F(\xvec)$. This implies that if $\xvec(t)$ is a solution to \cref{eqn:sys_Basic}, so is $-\xvec(t)$, and that $\xvec = \Zerovec$ is a fixed point of \cref{eqn:sys_Basic} for all $g$.

We consider here networks in which the excitatory neurons are grouped into $n_C$ clusters, each containing $p$ neurons, and the inhibitory neurons are grouped into $n_{C_I}$ clusters, each containing $p_I$ neurons. 
For simplicity, we only consider the case where the excitatory clusters are the same size, and the inhibitory clusters are the same size. This restriction introduces additional symmetries into the model, which are explained below.
In addition, all connections of any given type (e.g. $E \rightarrow E$ or $E \rightarrow I$) will have the same strength. The matrix $H$ then takes the general form
\begin{equation} \label{eqn:H}
H = 
\left[ 
\begin{blockarray}{cccccccc}
\begin{block}{cccc|cccc}
\mu_{EE}\Kvec_{p} & 0 & \hdots & 0 & 
\BAmulticolumn{4}{c}{\multirow{4}{*}{$\mu_{EI}\Onevec_{n_E \times n_I}$}} \\
0 & \mu_{EE} \Kvec_{p} & \hdots & 0 &&&&\\
\vdots & \vdots & \ddots & 0 &&&&\\
0 & 0 & \hdots & \mu_{EE} \Kvec_{p} &&&&\\
\end{block} 
\cline{1-8}
\begin{block}{cccc|cccc}
\BAmulticolumn{4}{c|}{\multirow{4}{*}{$\mu_{IE}\Onevec_{n_I\times n_E}$}} &
\mu_{II} \Kvec_{p_I} & 0 & \hdots & 0 \\
&&&& 0 & \mu_{II} \Kvec_{p_I} & \hdots & 0 \\
&&&& \vdots & \vdots & \ddots & 0 \\
&&&& 0 & 0 & \hdots & \mu_{II} \Kvec_{p_I} \\
\end{blockarray}
\right],
\end{equation}

where $\Onevec_{m \times n}$ is the $m\times n$ matrix of ones, and $\Kvec_n$ is the $n\times n$ matrix with all ones off the diagonal, i.e. $\Kvec_n = \Onevec_{n \times 1} \left( \Onevec_{n \times 1}\right)^T - \Ivec_n$, with $\Ivec_n$ the $n \times n$ identity matrix. The connection weights $\mu$ are defined ``matrix-style'', e.g. $\mu_{EI}$ will denote the connection from I to E, while $\mu_{IE}$ will denote the connection from E to I. The weights are also signed, so that $\mu_{EE}, \mu_{IE} > 0$ and $\mu_{EI}, \mu_{II} < 0$: this reflects the neurobiological heuristic of \emph{Dale's Law}, which states that each neuron makes excitatory \emph{or} inhibitory connections onto its postsynaptic targets.

The model \cref{eqn:sys_Basic}, \cref{eqn:H} is equivariant under the subgroup $\Gamma_H$ of $S_N$, defined by 
\[
\Gamma_H = \underbrace{S_{p} \times \cdots \times S_{p}}_{n_C} \times \, 
\underbrace{S_{p_I} \times \cdots \times S_{p_I}}_{n_{C_I}},
\]
where $S_n$ is the group of permutations on $n$ objects (see \cref{sec:EBL} for the definition of equivariance). 
Essentially, this says that labels of the neurons within each cluster can be freely permuted. Since the clusters are of equal sizes, there are two additional symmetries in the model. The labels of the excitatory clusters and the labels of the inhibitory clusters can be freely permuted, yielding symmetry groups isomorphic to $S_{n_C}$ and $S_{n_{C_I}}$, respectively.

The linearization of \cref{eqn:sys_Basic} about $\xvec = 0$ is the matrix
\begin{equation}\label{eq:DF0}
DF(0) = \frac{g}{\sqrt{N}}H - I_N,
\end{equation}
where $I_N$ is the $N \times N$ identity matrix. The eigenvalues of $DF(0)$ are then given by $\lambda^*(g) = \frac{g}{\sqrt{N}}\lambda - 1$ for all eigenvalues $\lambda$ of $H$. As a consequence, the dynamics of the system can be understood in terms of the eigenvalues of $H$. For an eigenvalue $\lambda$ of $H$ with negative real part, the corresponding eigenvalue $\lambda^*(g)$ of $DF(0)$ will always have negative real part, irrespective of $g$. On the other hand, for an eigenvalue $\lambda$ of $H$ with positive real part, the sign of the real part of the corresponding eigenvalue $\lambda^*(g)$ of $DF(0)$ will depend on the bifurcation parameter $g$. Thus, the only bifurcations of $\xvec = 0$ involve the eigenvalues of $H$ which have positive real part. Furthermore, the multiplicities of the eigenvalues of $H$ are determined by symmetries in the underlying model \cref{eqn:sys_Basic} and the matrix $H$. These lead to symmetric bifurcations as $g$ is varied; we address this in \S 3.

The dynamics near a nonzero fixed point $\xvec^* = (x_1^*, \dots, x_N^*)^T$ of \cref{eqn:sys_Basic} also depend on the matrix $H$. The linearization of \cref{eqn:sys_Basic} about $\xvec^*$ is the matrix
\begin{equation}\label{eq:DFxstar}
    DF(\xvec^*) = \frac{g}{\sqrt{N}}H(\xvec^*)  - I_N,
\end{equation}
where 
\begin{equation}\label{eq:Hxstar}
H(\xvec^*) := H \text{diag}(\sech^2(g \xvec^* ))
\end{equation}
is obtained from the matrix $H$ by multiplying column $j$ of $H$ by $\sech^2(g x_j^*)$. We note that the diagonal entries of $H(\xvec^*)$ are 0, thus $\text{Trace } H(\xvec^*) = 0$. This implies that the eigenvalues of $H(\xvec^*)$ sum to 0.

We first studied this system in \cite{Barreiro2017}, where we analyzed all-to-all connected, balanced excitatory-inhibitory networks ($n_C = 1$ and $n_{C_I} = 1$). In this paper, we first flesh out some details about that system: we derive leading order expressions for bifurcation points in the system, for the equilibria near those bifurcation points, and for the Hopf bifurcations that spawn the clustered limit cycles we observed in \cite{Barreiro2017} (\cref{sec:E1I1}). We then extend the analysis to networks in which the excitatory population is split up into clusters ($n_C > 1$ and $n_{C_I}=1$; \cref{sec:Eclusters}). We briefly compare with networks in which the inhibitory neurons are clustered instead ($n_C=1, n_{C_I}>1$; \cref{sec:inhibitoryclusters}). 

\section{The role of symmetries and the Equivariant Bifurcation Lemma} \label{sec:EBL}

In this section, we outline the tools of equivariant bifurcation theory, and explain how they apply to the model in question. Our main tool for analyzing the solutions to \cref{eqn:sys_Basic}, \cref{eqn:H} which arise at bifurcation points when symmetries are present is the \emph{Equivariant Branching Lemma} \cite{Golubitsky2002,MR631456,GSS88Vol2,HoyleRebeccaB2006Pf:a}. Before stating the result, we introduce some terminology. 

Let $\Gamma$ be a finite group acting on $\mathbb{R}^N$; then we say that a mapping $F: \mathbb{R}^N  \rightarrow \mathbb{R}^N$ is \emph{$\Gamma$-equivariant} if $F(\gamma \xvec) = \gamma F(\xvec)$, for all $\xvec \in \mathbb{R}^N$ and $\gamma \in \Gamma$.  A one-parameter family of mappings $F: \mathbb{R}^N \times \mathbb{R}  \rightarrow \mathbb{R}^N$ is \emph{$\Gamma$-equivariant}, if it is $\Gamma$-equivariant for each value of its second argument. We say that $V$, a subspace of $\mathbb{R}^N$, is \emph{$\Gamma$-invariant} if $\gamma \vvec \in V$, for any $\vvec \in V$ and $\gamma \in \Gamma$. We furthermore say that the action of $\Gamma$ on $V$ is \emph{irreducible} if $V$ has no proper invariant subspaces, i.e. the only $\Gamma$-invariant subspaces of $V$ are $\{0\}$ and $V$ itself. 

For a group $\Gamma$ and a vector space $V$, we define the \emph{fixed-point subspace} for $\Gamma$, denoted $\Fix_V (\Gamma)$, to be all points in $V$ that are unchanged under any of the members of $\Gamma$, i.e. $\Fix_V (\Gamma) = \{ \xvec \in V : \gamma \xvec = \xvec, \forall \gamma \in \Gamma \}$. 
The \textit{isotropy subgroup of $\xvec \in V$}, denoted $\Sigma_x$, is the set of all members of $\Gamma$ under which $\xvec$ is fixed, i.e. $\Sigma_x = \{ \gamma \in \Gamma : \gamma \xvec = \xvec \}$. We then say that a subgroup $\Sigma$ is \textit{an} isotropy subgroup of $\Gamma$, if it is the isotropy subgroup, $\Sigma_x$, for some $\xvec \in V$.

Suppose we have a one-parameter family of mappings, $F(\xvec, g)$, and we wish to solve $F(\xvec, g)=0$. 
For any $(\xvec, g) \in \mathbb{R}^n \times \mathbb{R}$, let $(DF)_{\xvec,g}$ denote the $N \times N$ Jacobian matrix 
\[ \left[ \frac{\partial F_j}{\partial x_k} (\xvec, g) \right]_{j, k=1...N}
\] 
A bifurcation will occur when the Jacobian ceases to be invertible, i.e. when $(DF)_{\xvec,g}$ has a nontrivial kernel. For a $\Gamma$-equivariant mapping --- i.e. $F(\xvec,g)$ is $\Gamma$-equivariant for any value of the parameter $g$ --- we may have \textit{multiple} eigenvalues go through zero at once, because of symmetries; however, some of the structural changes that occur are qualitatively the same as those that occur in a non-symmetric system in which a single eigenvalue crosses through zero. But there is a catch: we will have \emph{multiple} such solution branches, each corresponding to a subgroup of the original symmetries. This is formalized in the following theorem:

\begin{theorem}[Equivariant Branching Lemma: paraphrased from \cite{GSS88Vol2}, Theorem 3.3 on p. 82, see also pp. 67-69] Let $F: \mathbb{R}^N \times \mathbb{R} \rightarrow \mathbb{R}^N$ be a one-parameter family of $\Gamma$-equivariant mappings with $F(\xvec_0, g_0) = \Zerovec$. Suppose that $(\xvec_0, g_0)$ is a bifurcation point and that, defining $V = \ker(DF)_{\xvec_0, g_0}$, $\Gamma$ acts absolutely irreducibly on $V$. Let $\Sigma$ be an isotropy subgroup of $\Gamma$ satisfying 
\begin{eqnarray}
{\rm dim}\; \Fix_V(\Sigma) = 1,
\end{eqnarray}
where $\Fix_V(\Sigma)$ is the \emph{fixed-point subspace} of $\Sigma$ with respect to $V$: that is, $\Fix_V(\Sigma) \equiv \{ x \in V \mid \sigma x = x \textrm{ for all }\sigma \in \Sigma \}$. Then there exists a unique smooth solution branch to $F = 0$ such that the isotropy subgroup of each solution is $\Sigma$.
\end{theorem}

As we have noted, \cref{eqn:sys_Basic}, \cref{eqn:H} is $\Gamma_H$-equivariant, where
\[
\Gamma_H = \underbrace{S_{p} \times \cdots \times S_{p}}_{n_C} \times \, 
\underbrace{S_{p_I} \times \cdots \times S_{p_I}}_{n_{C_I}},
\]
and $S_n$ is the group of permutations on $n$ objects. 
Essentially, this says that labels of the neurons within each cluster can be freely permuted. In addition, the labels of the excitatory clusters and the labels of the inhibitory clusters can be freely permuted, yielding symmetry groups isomorphic to $S_{n_C}$ and $S_{n_{C_I}}$, respectively.

The origin, $\xvec = \Zerovec$ is a fixed point for all values of $g$. As we increase $g$ from 0, we will encounter a sequence of bifurcation points, i.e. points $(\xvec_0, g_0)$ for which $DF$ has a nontrivial kernel. At each such point, we will identify the kernel $V$ and the subgroups $\Sigma$ for which a solution is guaranteed by the Equivariant Branching Lemma.

\section{Model simplification}\label{sec:simplermodel}

We can simplify the model using the fact that all cells within each excitatory cluster must be synchronized at a fixed point or periodic orbit. In the case where there is a single excitatory cluster ($n_C = 1$), if $x_1$ and $x_2$ are the activities of two excitatory cells, then a straightforward calculation (see Lemma 3 from \cite{Barreiro2017}) shows that
\begin{align}\label{eq:dt_x1t2diff}
\frac{d}{dt}|x_1 - x_2|^2 \leq -2 |x_1 - x_2 |^2.
\end{align}
The only way this can be true for a fixed point (for which $\frac{d}{dt} |x_1 - x_2|^2 =0$) or for a periodic orbit (for which $x_1(t)-x_2(t) = x_1(t+T)-x_2(t+T)$ for some period $T$) is if $x_1(t) = x_2(t)$ for all $t$. If $n_C > 1$, and $x_1$ and $x_2$ are the activities of two cells in the same excitatory cluster, equation \cref{eq:dt_x1t2diff} holds by the same argument as in \cite{Barreiro2017}, since both neurons receive the same incoming connections with the same weights. 

We are primarily interested in the case where there is a single cluster of inhibitory cells, i.e. $n_{C_I} = 1$. (We will briefly consider the case of multiple inhibitory clusters in \cref{sec:inhibitoryclusters}). If there are $n_C$ excitatory clusters containing $p$ cells each, and $n_I$ inhibitory cells (for $N = p n_C + n_I$ total cells), equation \cref{eqn:sys_Basic} reduces to the system of $n_C + n_I$ equations
\begin{equation}\label{eq:reducedsystem}
\begin{aligned}
\dot{x}_{E_j} &= -x_{E_j} + \frac{(p-1)}{\sqrt{N}}\mu_{EE} \tanh(g x_{E_j}) + \frac{1}{\sqrt{N}} \mu_{EI} \sum_k \tanh(g x_{I_k}) && j = 1, \dots, n_C \\
\dot{x}_{I_j} &= -x_{I_j} + \frac{p}{\sqrt{N}}\mu_{IE} \sum_k \tanh(g x_{E_k}) + \frac{1}{\sqrt{N}} \mu_{II} \sum_{k\neq j}  \tanh(g x_{I_k}) && j = 1, \dots, n_I,
\end{aligned}
\end{equation}
where $x_{E_j}$ is the activity for the $j$th excitatory cluster, and $x_{I_j}$ is activity for the $j$th inhibitory cell. In matrix form, the equations \cref{eq:reducedsystem} can be written
\begin{equation}\label{eq:reducedmatrixform}
\dot{\xvec} = \tilde{F}(\xvec, g) := -\xvec + \frac{1}{\sqrt{N}} \tilde{H} \tanh(g \xvec),
\end{equation}
where $\xvec = (x_{E_1}, \dots, x_{E_{n_C}}, x_{I_1}, \dots, x_{I_{n_I}})^T$, and $\tilde{H}$ is the $(n_C + n_I) \times (n_C + n_I)$ reduced matrix
\begin{equation}\label{eq:tildeH}
\tilde{H} = \left[ \begin{array}{c|c}
    \\
    (p-1) \mu_{EE} I_{n_C} & \mu_{EI} \Onevec_{n_C \times n_I}\\
    \\
    \hline
    \\
    p \mu_{IE} \Onevec_{n_I \times n_C} & \mu_{II} \mathbf{K}_{n_I} \\
    \\
    \end{array}
    \right].
\end{equation}
The system of equations \cref{eq:reducedmatrixform}, \cref{eq:tildeH} is the restriction of the original system \cref{eqn:sys_Basic}, \cref{eqn:H} with $n_{C_I}=1$ to the fixed-point subspace $\Fix(\Gamma_C)$ corresponding to the subgroup
\[
\Gamma_C = \underbrace{S_{p} \times \cdots \times S_{p}}_{n_C} \times \, 
E_{n_I}
\]
of $\Gamma_H$, where $E_{n_I}$ is the trivial subgroup of $S_{n_I}$ consisting only of the identity permutation. The reduced model \cref{eq:reducedmatrixform}, \cref{eq:tildeH} is then equivariant under the subgroup 
\begin{equation}\label{eq:Gamma}
\Gamma = S_{n_C} \times S_{n_I}
\end{equation}
of $S_{n_C+n_I}$. The special case of $n_C = 1$ (a single excitatory cluster), for which $\Gamma = S_1 \times S_{n_I}$, is considered in \cref{sec:E1I1}. We note that in this case, one of the symmetries is effectively lost, since the activity of the lone excitatory cluster is represented by a single variable.
The general case ($n_C > 1$) is considered in \cref{sec:Eclusters}.
We will only consider the reduced system \cref{eq:reducedmatrixform}, \cref{eq:tildeH} in \cref{sec:E1I1} and \cref{sec:Eclusters}.

Next, we show that no stability information is lost by only studying the reduced system. Suppose $\xvec^* = (x_{E_1}^*, \dots, x_{E_{n_C}}^*, x_{I_1}^*, \dots, x_{I_{n_I}}^*)^T$ is a fixed point of \cref{eq:reducedmatrixform}. (We will discuss the existence of such fixed points in \cref{sec:E1I1} and \cref{sec:Eclusters}).
The linearization of \cref{eq:reducedmatrixform} about $\xvec^*$ is the matrix
\begin{equation}\label{eq:DtildeFxstar}
    D\tilde{F}(\xvec^*) = \frac{g}{\sqrt{N}}\tilde{H}(\xvec^*) - I_{n_C+n_I},
\end{equation}
where 
\begin{equation}\label{eq:tildeHxstar}
\tilde{H}(\xvec^*) := \tilde{H} \text{diag}(\sech^2(g \xvec^* )).
\end{equation}
The original system \cref{eqn:sys_Basic} has a corresponding fixed point $\xvec_0^*$, in which each $x_{E_j}^*$ in $\xvec^*$ is repeated $p$ times. The following proposition shows that to analyze the stability of the fixed point $\xvec_0^*$ in the full system \cref{eqn:sys_Basic}, it suffices to determine the eigenvalues of the reduced matrix $\tilde{H}(\xvec^*)$, since the additional eigenvalues of $H(\xvec_0^*)$ are negative, and thus will not affect stability.

\begin{proposition}\label{prop:tildeHeig}
Let $\xvec^*$ be a fixed point of \cref{eq:reducedmatrixform} and $\xvec_0^*$ the corresponding fixed point of \cref{eqn:sys_Basic}, and let $H(\xvec_0^*)$ and $\tilde{H}(\xvec^*)$ be defined by \cref{eq:Hxstar} and \cref{eq:tildeHxstar}. Then
\begin{compactenum}[(i)]
    \item Every eigenvalue of $\tilde{H}(\xvec^*)$ is an eigenvalue of $H(\xvec_0^*)$.
    \item $H(\xvec_0^*)$ has $n_C$ additional real, negative eigenvalues, each with multiplicity $p-1$.
\end{compactenum}
\begin{proof}
Part (i) follows immediately from the fact that \cref{eq:reducedmatrixform} is a restriction of \cref{eqn:sys_Basic}.
For part (ii), it can be verified directly that for $j=1, \dots, n_C$, $H(\xvec_0^*)$ has an eigenvalue at $\lambda = -\mu_{EE} \sech^2(x_{E_j})$ with multiplicity $p-1$. For $j=1$, for example, the $p-1$ eigenvectors are $\vvec^1, \dots, \vvec^{p-1}$, where $v^k_1 = -1$, $v^k_{k+1} = 1$, and all other components are 0. Since $\mu_{EE} > 0$, these eigenvalues are always negative. 
\end{proof}
\end{proposition}

\begin{figure}
    \centering
    \begin{tabular}{cc}
    No clusters &
    Excitatory clusters \\
    \includegraphics[width=6cm]{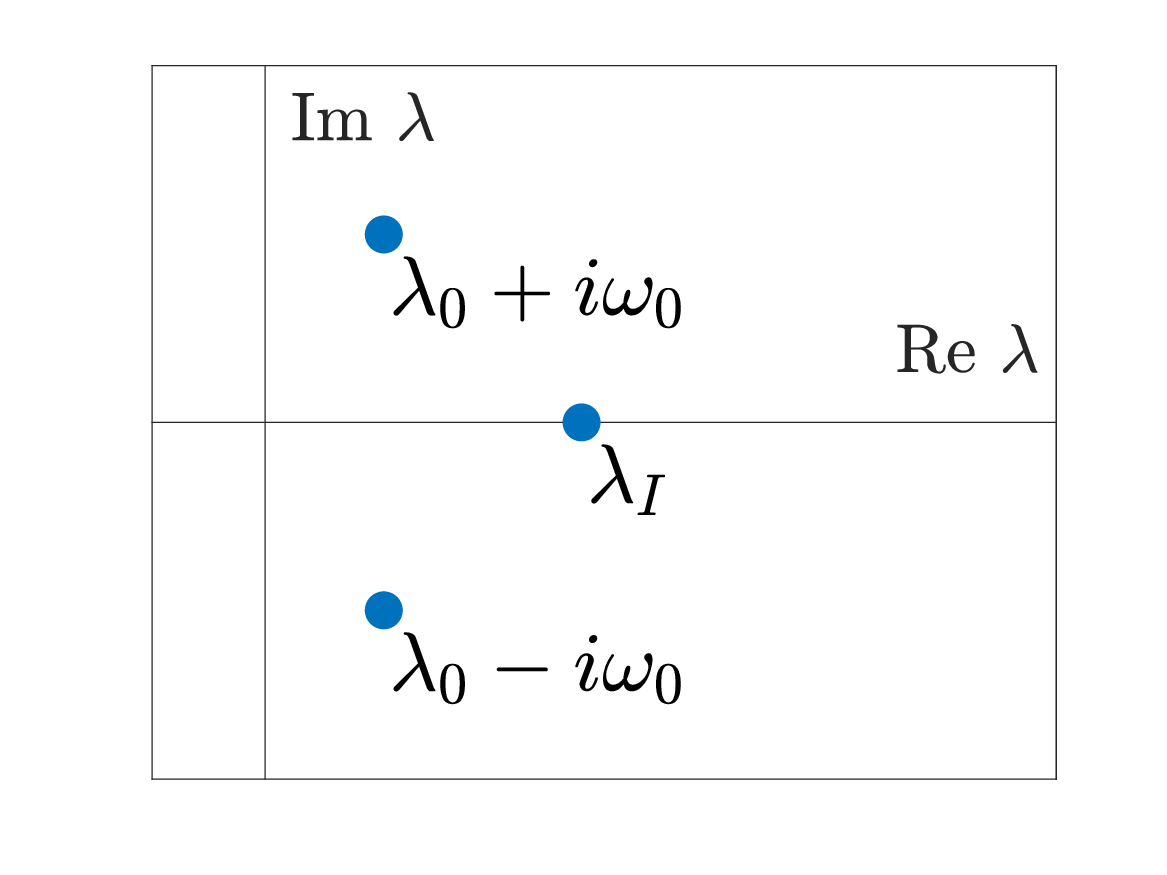} &
    \includegraphics[width=6cm]{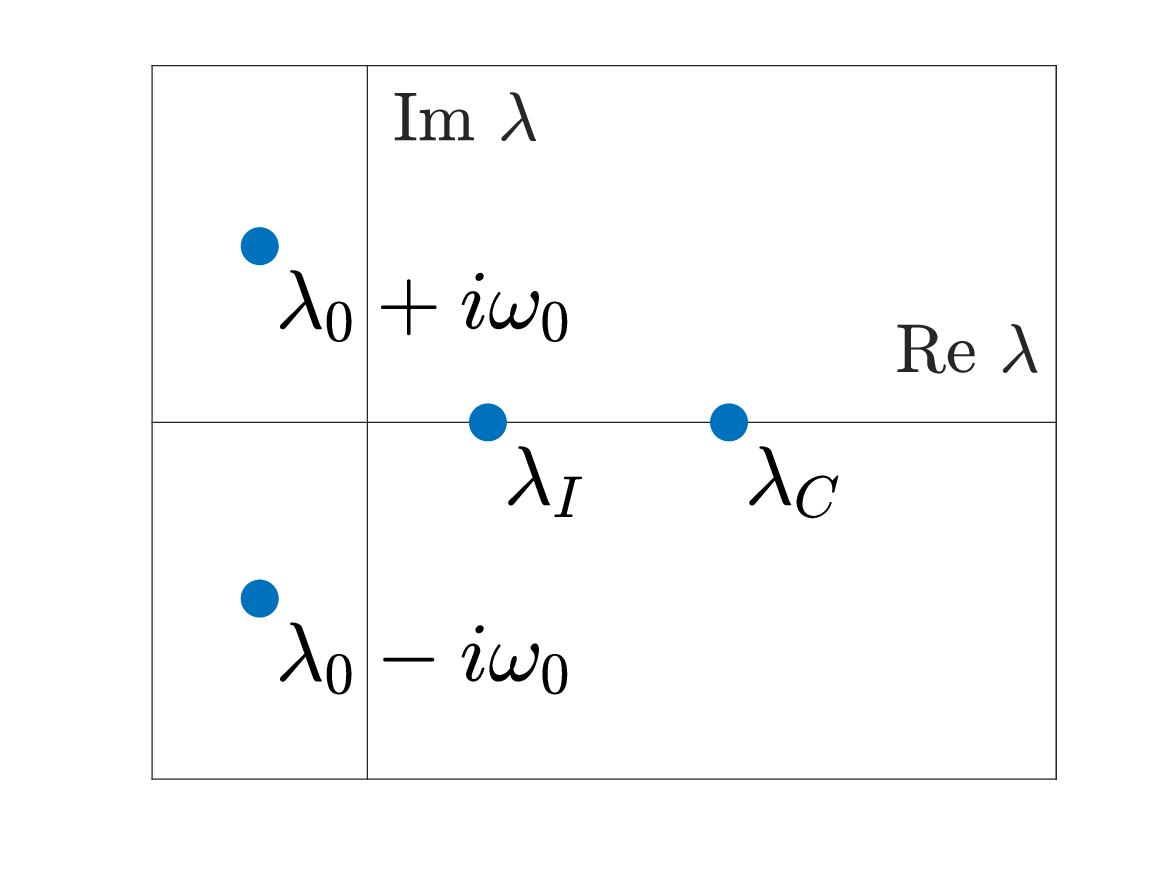}
    \end{tabular}
    \caption{Eigenvalue pattern of the matrix $\tilde{H}$ for a single excitatory and a single inhibitory cluster (left, \cref{sec:E1I1}), and multiple excitatory clusters and a single inhibitory cluster (right, \cref{sec:Eclusters}). The notation for the eigenvalues in each network model is explained in the corresponding section below.}
    \label{fig:Heigpattern}
\end{figure}

The dynamics of the full system can therefore be explained by the dynamics of the reduced system, and, in particular, in terms of the eigenvalues of the reduced matrix $\tilde{H}$ (\cref{fig:Heigpattern}). Although these patterns will be explained in detail in the corresponding sections below, we point out two crucial differences between the model with a single excitatory cluster (\cref{fig:Heigpattern}, left) and the model with multiple excitatory clusters (\cref{fig:Heigpattern}, right). For the model with multiple excitatory clusters, there is an additional positive, real eigenvalue $\lambda_C$, and the complex pair $\lambda_0 + i \omega_0$ has negative real part.

\section{Single excitatory and inhibitory cluster}\label{sec:E1I1}

The simplest case (considered in \cite{Barreiro2017}) involves a single excitatory cluster ($n_C = 1$ and $p = n_E$) and a single inhibitory cluster, in which case the matrix $\tilde{H}$ in \cref{eq:reducedmatrixform} reduces to the $(1+n_I)\times(1+n_I)$ matrix
\begin{eqnarray} \label{eq:tildeH_1d}
\tilde{H} = 
\left[ \begin{array}{c|c}
\\
(n_E - 1)\mu_{EE} & \mu_{EI} \Onevec_{1 \times n_I}\\
\\
\hline
\\
n_E \mu_{IE} \Onevec_{n_I \times 1} & \mu_{II} \mathbf{K}_{n_I} \\
\\
\end{array}
\right].
\end{eqnarray}

We choose the connection weights so that the network is \emph{balanced}; that is, the excitatory and inhibitory currents coming into each cell should approximately cancel \cite{RA06}. To achieve this balance, we set $\mu_{EI} = -\alpha \mu_{EE}$ and $\mu_{II} = -\alpha \mu_{IE}$, where $\alpha = \frac{f}{1-f}$. For simplicity, we also take $\mu_{IE} = \mu_{EE}$. The spectrum of $\tilde{H}$ is now easy to compute (see \cite{Barreiro2017}, noting that the full matrix $H$ is considered in that work). The eigenvalues of $\tilde{H}$ (left panel of \cref{fig:Heigpattern}) are
\begin{itemize}
    \item $\lambda_I := \alpha \mu_{EE} > 0$ with multiplicity $n_I - 1$.
    \item One complex pair of eigenvalues $\lambda_0 \pm i \omega_0$, with
    \[
    \lambda_0 := \mu_{EE}\frac{\alpha - 1}{2}, \quad \omega_0 := \mu_{EE}\sqrt{\alpha+1}\sqrt{n_E - \frac{\alpha+1}{4}}. 
    \]
\end{itemize}
It is straightforward to check that $0 < \lambda_0 < \lambda_I$. Since both of these are positive, there will be a bifurcation of $\xvec = 0$ involving each of these eigenvalues.

In the following sections, we will determine the bifurcations which occur as $g$ is increased, together with the structures which emerge at these bifurcation points. First, the origin loses stability in a symmetric pitchfork bifurcation, after which point there is a branch of equilibria for every possible division of the inhibitory cells into two groups. We will derive leading order formulas for these branches, as well as show which of them are initially stable. As $g$ is further increased, there is a Hopf bifurcation on each of these branches, which gives rise to a limit cycle with the same grouping pattern as the corresponding branch. Finally, at a critical value of $g$, these limit cycles coalesce into a symmetric pitchfork bifurcation of limit cycles. After this point, there is a single stable limit cycle in which there is one group of inhibitory cells and one group of excitatory cells.

\subsection{Bifurcations of the origin}\label{sec:biforigin}

As the bifurcation parameter $g$ is increased from 0, the eigenvalues $\lambda_I^*(g)$ of $DF(0)$ corresponding to $\lambda_I$ cross the imaginary axis at
\begin{equation}\label{eq:pitchlocation}
    g = g_0 := \frac{\sqrt{N}}{\alpha \mu_{EE}}.
\end{equation}
The origin $\xvec = 0$ is a stable equilibrium for $g < g_0$. At $g=g_0$, the origin loses stability in a symmetric pitchfork bifurcation, where $n_I - 1$ eigenvalues cross the imaginary axis simultaneously (see \cref{sec:symmpitch} below). As $g$ is further increased, the complex pair of eigenvalues $\lambda^*_0(g) \pm i \omega^*_0(g)$ of $DF(0)$ crosses the imaginary axis at 
\begin{equation}\label{eq:0hopflocation}
    g = g_H := \frac{ 2\sqrt{N} }{ (\alpha-1)\mu_{EE} },
\end{equation}
at which point a Hopf bifurcation occurs, giving rise to a limit cycle (see \cref{sec:periodic} below). The frequency of this limit cycle is given by the imaginary part $\omega^*_0(g)$ at $g = g_H$, which is
\begin{equation}\label{eq:omegag}
\omega^*_0(g_H) = \frac{2}{\alpha-1}\sqrt{\alpha+1}\sqrt{f N- \frac{\alpha+1}{4}}, 
\end{equation}
where we used $n_E = f N$. We note that since $\omega^*_0(g_H) = \mathcal{O}(\sqrt{N})$, $\omega^*_0(g_H) \rightarrow \infty$ as $N \rightarrow \infty$. 


\subsection{Solutions after symmetric pitchfork bifurcation}\label{sec:symmpitch}

The reader can readily check that the right-hand side of \cref{eq:reducedmatrixform}, \cref{eq:tildeH_1d} is $\Gamma$-equivariant, for $\Gamma = S_1 \times S_{n_I}$. That is, we can permute the labels on inhibitory cells without changing the equations. (The activity of the excitatory cells have been collapsed into a single variable). At $g=g_0$, $n_I-1$ eigenvalues pass through zero: the corresponding eigenspace is the set of all zero-sum vectors with support in the inhibitory cells only, i.e. 
\[ V \equiv  \ker(dF)_{\Zerovec,g^*}  = {\rm span} \, \left\{ \left[ 0 \;
\vvec_I \right] \right\}, \qquad \vvec_I \perp \Onevec_{n_{I}},
\]
which has dimension $n_I-1$.
To check that $\Gamma$ acts irreducibly on $V$, it is sufficient to show that the subspace spanned by the \textit{orbit} of a single vector $\vvec$ (defined as the set of all values  $\gamma \vvec$, for $\gamma \in \Gamma$) is full rank; this can be readily confirmed for $\vvec_I = \left[ \begin{array}{ccccc} 1 & -1 & 0 & ... & 0 \end{array} \right]$, for example. 

To determine what occurs at this pitchfork bifurcation point, we next find subgroups $\Sigma$ of $\Gamma$ which satisfy the hypothesis of the Equivariant Branching Lemma. To do this, 
we break the inhibitory cells up into precisely two clusters $I_1$ and $I_2$ of sizes $n_{I_1}$ and $n_{I_2}$, where $n_{I_1} + n_{I_2} = n_I$, and retain only permutations within each cluster. For each such decomposition, this describes a subgroup 
\begin{equation}
\Sigma_I = S_1 \times S_{n_{I_1}} \times S_{n_{I_2}}
\end{equation}
of $\Gamma$.
Assuming that (without loss of generality) the $I_1$ neurons have the indices $2,...,n_{I_1}+1$, $\Sigma_I$ has the fixed-point subspace 
\begin{eqnarray}
\Fix_V(\Sigma_I) & = & {\rm span} \, \left\{ \left[ 0 \;
\underbrace{\begin{matrix}1 & \cdots & 1\end{matrix}}_{n_{I_1}} \;
\underbrace{\begin{matrix}-\frac{n_{I_1}}{n_{I_2}} & \cdots & -\frac{n_{I_1}}{n_{I_2}} \end{matrix}}_{n_{I_2}} \right] \right\}.
\end{eqnarray}
Furthermore $\dim \Fix_V(\Sigma_I) = 1$, because it can be described as the span of a single vector. 

It follows from the Equivariant Branching Lemma that there is a branch of equilibria emerging at the symmetric pitchfork bifurcation point $g=g_0$ for all such subgroups $\Sigma_I$, i.e. for every possible division of the inhibitory cells into exactly two clusters of size $n_{I_1}$ and $n_{I_2}$, where $n_{I_1} + n_{I_2} = n_I$. We refer to these as $I_1/I_2$ branches.  Each such branch may be characterized by the number 
\begin{equation}\label{eq:beta}
\beta = \frac{n_{I_1}}{n_{I_2}},
\end{equation} 
which gives the ratio of the cluster sizes. Without loss of generality, we may take $n_{I_1} \geq n_{I_2}$, so that $\beta \geq 1$. The inhibitory cells within each of the two clusters are synchronized. The solution on each $I_1/I_2$ branch is then given as $(x_E, x_{I_1}, x_{I_2})$, where we recall from \cref{sec:simplermodel} that all excitatory cells are synchronized. Due to the odd symmetry of \cref{eqn:sys_Basic}, there is a corresponding $I_1/I_2$ branch for each $\beta$ with solution $(-x_E, -x_{I_1}, -x_{I_2})$. We will ignore this other branch for simplicity, although we note that it is this odd symmetry which permits a pitchfork bifurcation to occur.

We briefly comment on divisions of the inhibitory cells into more than two clusters. As a specific example, suppose the inhibitory cells are divided into three clusters of size $n_{I_1}$, $n_{I_2}$, and $n_{I_3}$, where $n_{I_1} + n_{I_2} + n_{I_3} = n_I$. This decomposition describes a subgroup $\Sigma_3 = S_1 \times S_{n_{I_1}} \times S_{n_{I_2}} \times S_{n_{I_3}}$ of $\Gamma$. The fixed-point subspace of $\Sigma_3$ with respect to $V$ is given by 
\begin{align*}
&{\rm span} \, \left\{ \left[ 0 \;
\underbrace{\begin{matrix}1 & \cdots & 1\end{matrix}}_{n_{I_1}} \;
\underbrace{\begin{matrix}-\frac{n_{I_1}}{n_{I_2}} & \cdots & -\frac{n_{I_1}}{n_{I_2}} \end{matrix}}_{n_{I_2}} \;
\underbrace{\begin{matrix}0 & \cdots & 0\end{matrix}}_{n_{I_3}}\right],
\left[ 0 \;
\underbrace{\begin{matrix}1 & \cdots & 1\end{matrix}}_{n_{I_1}} \;
\underbrace{\begin{matrix}0 & \cdots & 0\end{matrix}}_{n_{I_2}} \;
\underbrace{\begin{matrix}-\frac{n_{I_1}}{n_{I_3}} & \cdots & -\frac{n_{I_1}}{n_{I_3}} \end{matrix}}_{n_{I_3}} \right] \right\},
\end{align*}
which has dimension 2. Since $\dim {\rm Fix}_V(\Sigma_3) > 1$, the Equivariant Branching Lemma does not guarantee the existence of a branch of fixed points with this symmetry. In general, if the inhibitory cells are divided into $m > 2$ clusters, the fixed-point subspace for the corresponding symmetry group will have dimension $m-1>1$.
It is important to note that the Equivariant Branching Lemma does not preclude the existence of such fixed points (see the discussion in \cite[Section 4]{Barreiro2017}). Numerical experiments, however, suggest that all fixed points which are not on the primary $I_1/I_2$ branches are unstable (see \cref{sec:otherbranches}).

\subsection{Solutions along \texorpdfstring{$I_1/I_2$}{I1/I2} branches}\label{sec:solI1I2}

First, we derive leading order expressions for the equilibria along the $I_1/I_2$ branches for $g$ close to the bifurcation point $g_0$. Fix $\beta \geq 1$. To find $(x_E, x_{I_1}, x_{I_2})$ along the $I_1/I_2$ branch corresponding to $\beta$, we reduce \cref{eq:reducedsystem} to the 3-dimensional system
\begin{equation}\label{eq:reducedsystemI1I2}
 \begin{aligned}
 \begin{bmatrix} x_E\\x_{I_1}\\x_{I_2}\end{bmatrix} 
 &= \frac{\mu_{EE}}{\sqrt{N}} 
 \begin{bmatrix} (\alpha n_I - 1) & -\alpha \frac{\beta}{\beta+1} n_I & - \alpha \frac{1}{\beta+1} n_I  \\
    \alpha n_I  & -\alpha \left(\frac{\beta}{\beta+1} n_I-1\right) & - \alpha \frac{1}{\beta+1} n_I  \\
    \alpha n_I & -\alpha \frac{\beta}{\beta+1} n_I & -\alpha \left(\frac{1}{\beta+1} n_I-1\right)
 \end{bmatrix}
 \begin{bmatrix} \tanh(g x_E) \\\tanh ( g x_{I_1} ) \\\tanh(g x_{I_2})\end{bmatrix},
 \end{aligned}
\end{equation}
where $x_E$ is the activity of the synchronized excitatory cells, $x_{I_1}$ and $x_{I_2}$ are the activities of the two synchronized inhibitory clusters, and we used $n_E = \alpha n_I$. 
The system \cref{eq:reducedsystemI1I2} is the restriction of \cref{eq:reducedmatrixform} to the fixed-point subspace for the subgroup $S_1 \times \, S_{n_{I_1}} \times \, S_{n_{I_2}}$ of $\Gamma$.
For any solution $(x_E, x_{I_1}, x_{I_2})^T$ to \cref{eq:reducedsystemI1I2}, $\xvec = (x_E, x_{I_1}, \dots, x_{I_1}, x_{I_2}, \dots, x_{I_2})^T$ is an equilibrium solution to \cref{eq:reducedsystem}, where $x_{I_1}$ and $x_{I_2}$ are repeated $n_{I_1}$ and $n_{I_2}$ times, respectively. We note that any solution $(x_E, x_{I_1}, x_{I_2})$ to \cref{eq:reducedsystemI1I2} is bounded for all $g$, since the matrix in \cref{eq:reducedsystemI1I2} is constant, and $|\tanh y| \leq 1$ for all $y$. 

\begin{figure}
    \centering
    \includegraphics[width=8.25cm]{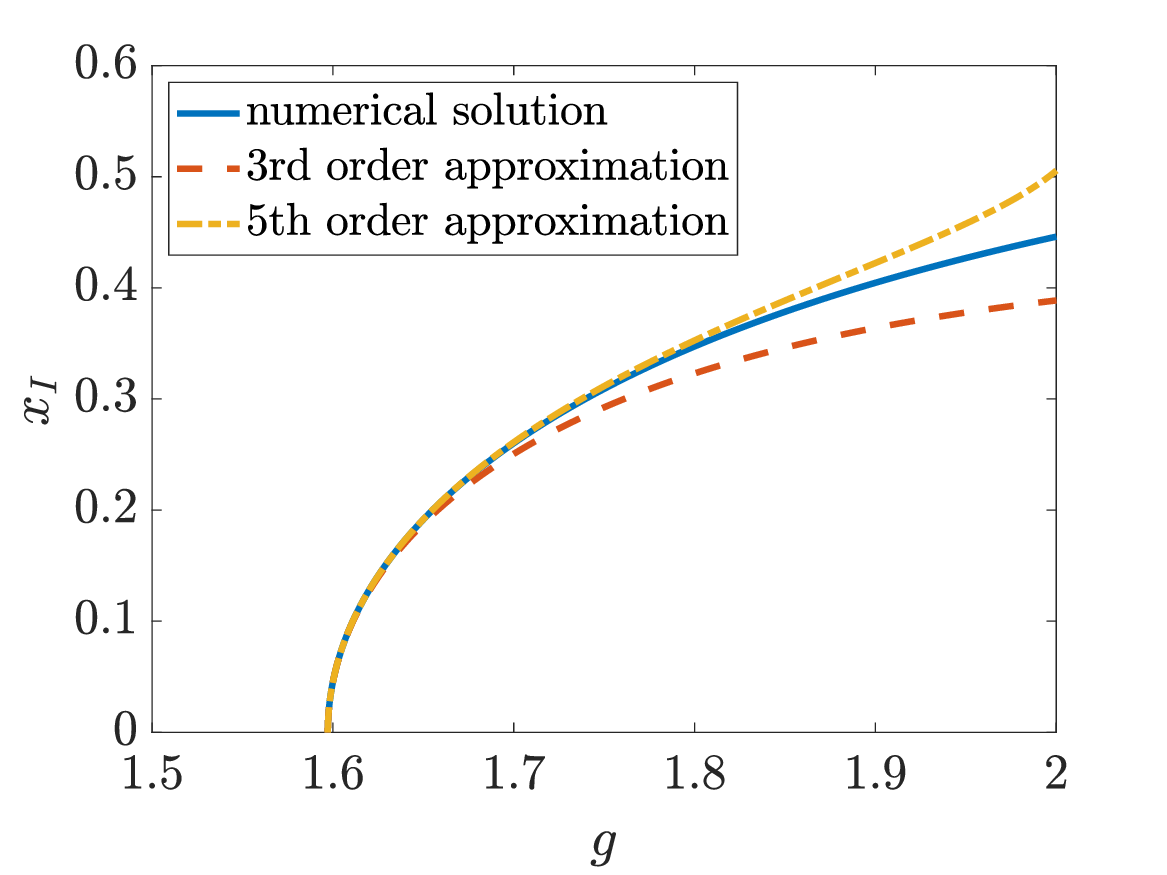}\hspace{-0.5cm}
    \includegraphics[width=8.25cm]{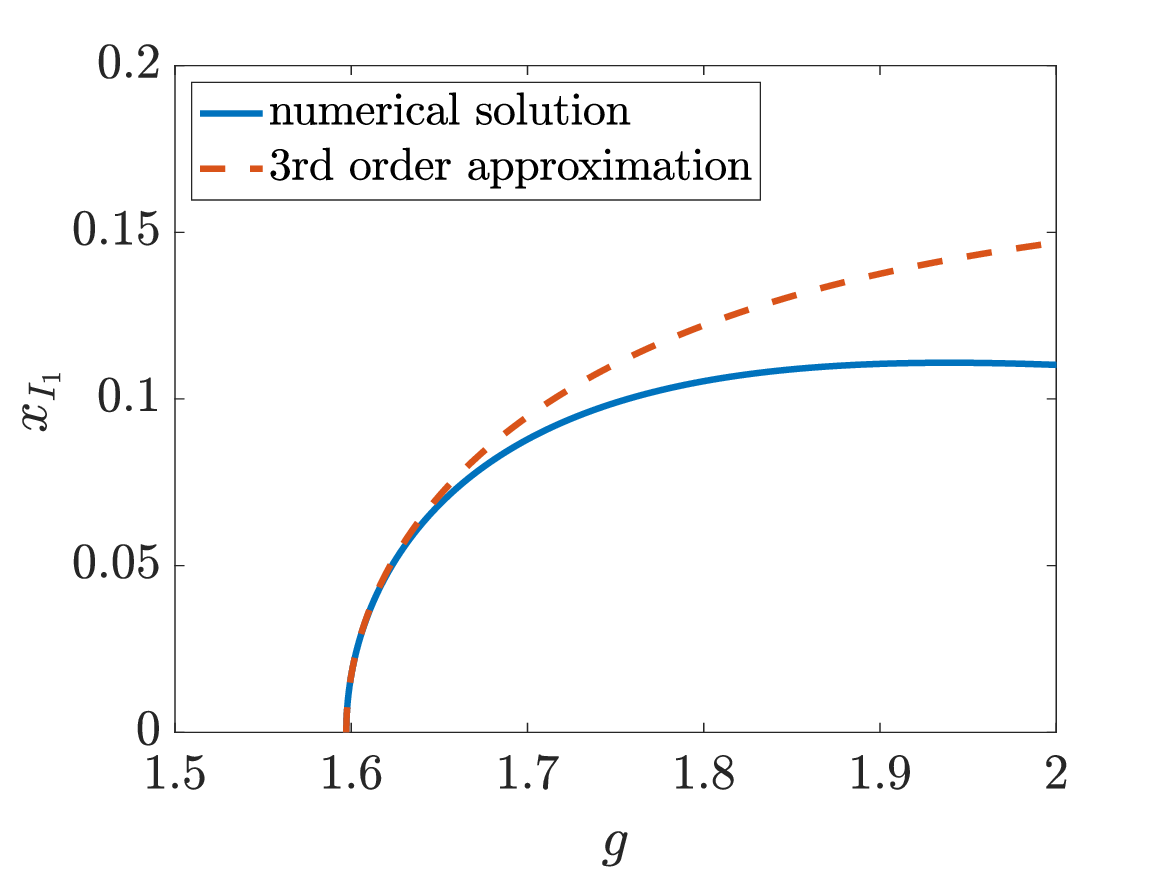} 
    \caption{Approximations to the location of $x_I$ on $I_1/I_2$ fixed point branches. Left: Third order \cref{eq:xIapprox} and fifth order  \cref{eq:xIapprox5} approximations to $x_I=x_{I_1}$ on the $\beta = 1$ (i.e. $n_{I_1}=n_{I_2}$) branch. Right: Third order approximation \cref{eq:XI1} to $x_{I_1}$ on the $\beta=3$ branch. Other parameters are: $N = 20$,  $\alpha = 4$, $\mu_{EE} = 0.7$.}
    \label{fig:xIapprox}
\end{figure}

The simplest case occurs when $n_I$ is even and $\beta = 1$, in which case $n_{I_1}=n_{I_2}$. On this branch, $x_E = 0$, and $x_{I_2} = -x_{I_1}$, i.e. there are two equally sized inhibitory populations with equal and opposite activities, and there is no excitatory cell activity. Beginning with the single remaining equation for $x_{I_1}$, and utilizing the Taylor expansion for the $\tanh$ function, we show (see detailed calculations in \cref{app:I1I2sol}) that the nonzero solution for $x_I$ is given, to leading order, by
\begin{align}\label{eq:xIapprox}
x_I &= \sqrt{ \frac{3(g - g_0) }{g^3}} && g \geq g_0.
\end{align}
By keeping up to fifth-order terms in the Taylor expansion (see detailed calculations in \cref{app:I1I2sol}), we can obtain the higher order approximation
\begin{align}\label{eq:xIapprox5}
x_I &= \frac{1}{2} \sqrt{ \frac{5}{g^2} - \frac{\sqrt{ 5 g^5( 24 g_0 - 19 g) }}{g^5}} && g \geq g_0.
\end{align}
Comparison between the third-order approximation \cref{eq:xIapprox}, the fifth-order approximation \cref{eq:xIapprox5}, and the numerical solution obtained by parameter continuation is shown in the left panel of \cref{fig:xIapprox}.

For $\beta > 1$, it is no longer true that $x_{I_2} = -x_{I_1}$. However, by making an appropriate ansatz and proceeding as described in \cref{app:I1I2sol}, we obtain the following approximations for $x_E$, $x_{I_1}$, and $x_{I_2}$ in terms of $g$, for $g$ close to $g_0$
\begin{align}\label{eq:XI1}
 x_E &= \mathcal{O}\left( \frac{1}{N^2}\right), \quad
 x_{I_1} = \sqrt{ \frac{ 3(g - g_0) }{ (1 - \beta + \beta^2 )g^3}} + \mathcal{O}\left( \frac{1}{N^2}\right), \quad
 x_{I_2} = -\beta x_{I_1} +  \mathcal{O}\left( \frac{1}{N^2}\right) && g \geq g_0.
\end{align}
Note that this reduces to \cref{eq:xIapprox} when $\beta = 1$. In addition, we note that $x_{I_1}$ and $x_{I_2}$ have opposite signs. This is, in fact, true for all $g > g_0$, as shown in \cref{app:I1I2sol}. Comparison between this approximation and the numerical solution obtained by numerical parameter continuation is shown in the right panel of \cref{fig:xIapprox}.

\subsection{Stability and bifurcations along \texorpdfstring{$I_1/I_2$}{I1/I2} branches}\label{sec:I1I2stability}

\begin{figure}
    \centering
    \includegraphics[width=6cm]{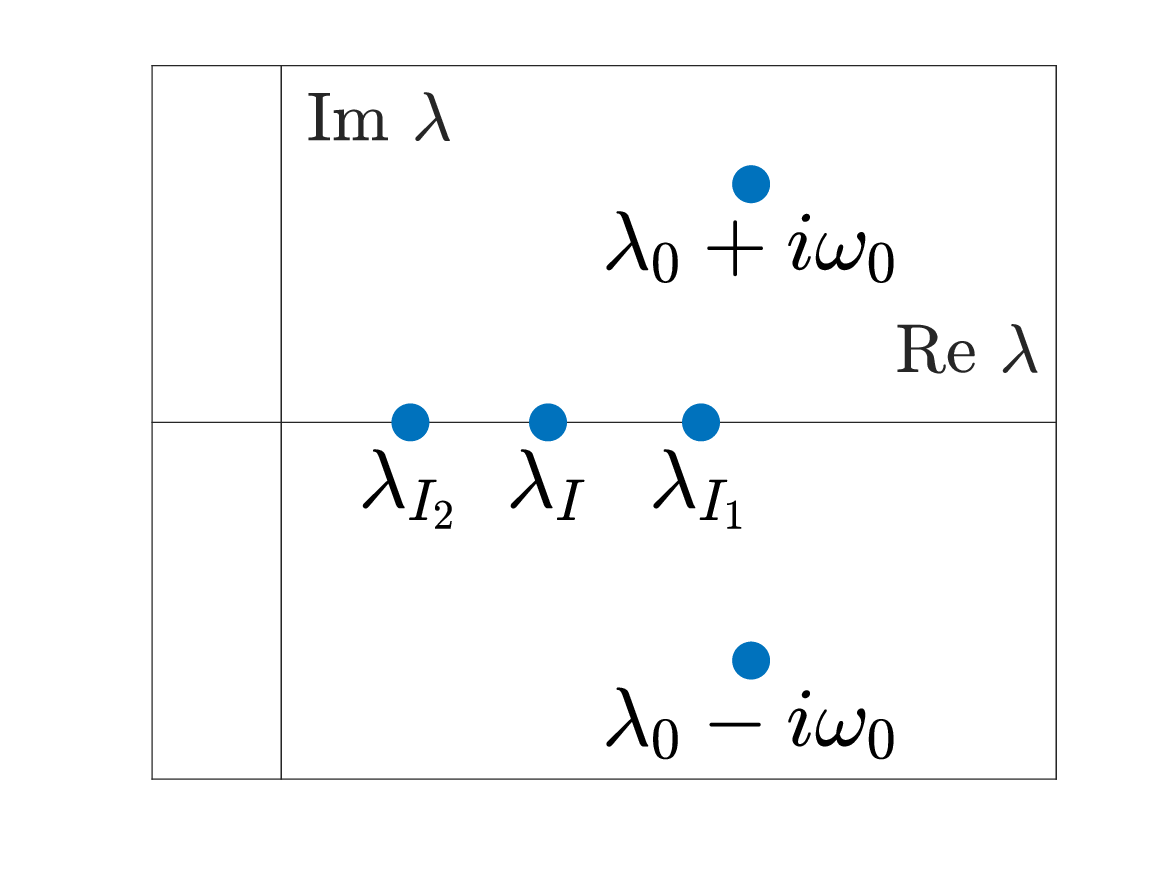}
    \caption{Eigenvalue pattern of the connectivity matrix $H(\xvec^*)$ for fixed point $\xvec^*$ on $I_1/I_2$ branch with $\beta > 1$. The notation for the eigenvalues is explained in \cref{sec:I1I2stability}.}
    \label{fig:Hstareignocluster}
\end{figure}

Now that we have obtained a leading order formula for the fixed points on the $I_1/I_2$ branches for all valid inhibitory cell ratios $\beta$, we will analyze their stability for $g$ close to the bifurcation point $g_0$. Choose any $\beta \geq 1$, so that $n_{I_1} = \frac{\beta}{\beta+1}n_I$ and $n_{I_2} = \frac{1}{\beta+1}n_I$, and let $\xvec = (x_E, x_{I_1}, x_{I_2})$ be a solution to \cref{eq:reducedsystemI1I2} for $g > g_0$. To examine the stability and bifurcations which occur along the $I_1/I_2$ branches, we look at the linearization $D\tilde{F}(\xvec^*)$, which is given by \cref{eq:DtildeFxstar}, where $\xvec^* = (x_E, x_{I_1}, \dots, x_{I_1}, x_{I_2}, \dots, x_{I_2})^T$, and $x_{I_1}$ and $x_{I_2}$ are repeated $n_{I_1}$ and $n_{I_2}$ times, respectively. As discussed above in \cref{sec:simplermodel}, stability will depend on the eigenvalues of $\tilde{H}(\xvec^*)$. A cartoon showing the location of these eigenvalues is given in \cref{fig:Hstareignocluster}. In the process of our analysis, we will show that a Hopf bifurcation occurs along each $I_1/I_2$ branch, and will find a leading order formula for its location.

To locate the eigenvalues of $\tilde{H}(\xvec^*)$, we first linearize the three-dimensional system \cref{eq:reducedsystemI1I2} about the fixed point $\xvec = (x_E, x_{I_1}, x_{I_2})$ to get the Jacobian
\begin{equation}\label{eq:J3forI1I2}
J_3(\xvec) = \frac{g}{\sqrt{N}} H_3(\xvec) - I_3
\end{equation}
where 
\begin{equation}\label{eq:H3}
H_3(\xvec) = \mu_{EE}
 \begin{bmatrix} (\alpha n_I - 1) \sech^2(g x_E) & -\alpha \frac{\beta}{\beta+1} n_I \sech^2(g x_{I_1}) & - \alpha \frac{1}{\beta+1} n_I \sech^2(g x_{I_2}) \\
    \alpha n_I \sech^2(g x_E) & -\alpha \left(\frac{\beta}{\beta+1} n_I-1\right) \sech^2(g x_{I_1}) & -\alpha \frac{1}{\beta+1} n_I \sech^2(g x_{I_2}) \\
    \alpha n_I \sech^2(g x_E) & -\alpha \frac{\beta}{\beta+1} n_I \sech^2(g x_{I_1}) & -\alpha \left(\frac{1}{\beta+1} n_I-1 \right) \sech^2(g x_{I_2})
 \end{bmatrix}
\end{equation}
and $I_3$ is the $3 \times 3$ identity matrix. We have the following proposition relating the eigenvalues of $H_3(\xvec)$ and $\tilde{H}(\xvec^*)$.

\begin{proposition}\label{prop:H3eig}
Let $\xvec = (x_E, x_{I_1}, x_{I_2})$ be a solution of \cref{eq:reducedsystemI1I2} and $\xvec^*$ the corresponding fixed point of \cref{eq:reducedmatrixform}, and let $H_3(\xvec)$ and $\tilde{H}(\xvec^*)$ be defined by \cref{eq:H3} and \cref{eq:tildeHxstar}. Then
\begin{compactenum}[(i)]
    \item Every eigenvalue of $H_3(\xvec)$ is an eigenvalue of $\tilde{H}(\xvec^*)$.
    \item $\tilde{H}(\xvec^*)$ has the following additional eigenvalues:
    \begin{itemize}
        \item $\lambda_{I_1} := \mu_{EE} \alpha \sech^2(g x_{I_1})$ with multiplicity $n_{I_1}-1$.
        \item $\lambda_{I_2} := \mu_{EE} \alpha \sech^2(g x_{I_2})$ with multiplicity $n_{I_2}-1$.
    \end{itemize}
\end{compactenum}
\begin{proof}
Part (i) follows immediately from the fact that \cref{eq:reducedsystemI1I2} is a restriction of \cref{eq:reducedmatrixform}.
For part (ii), if $n_{I_1} > 1$, then it can be verified directly that $\tilde{H}(\xvec^*)$ has an eigenvalue $\lambda_{I_1} = \mu_{EE} \alpha \sech^2(g x_{I_1})$ with multiplicity $n_{I_1}-1$. The corresponding eigenvectors are $\vvec^1, \dots, \vvec^{n_{I_1}-1}$, where $v^k_2 = -1$, $v^k_{k+2} = 1$, and all other components are 0. If $n_{I_2} > 1$, the eigenvalue $\lambda_{I_2}$ can be similarly obtained.
\end{proof}
\end{proposition}

\begin{figure}
    \centering
    \begin{tabular}{cc}
    \includegraphics[width=7.8cm]{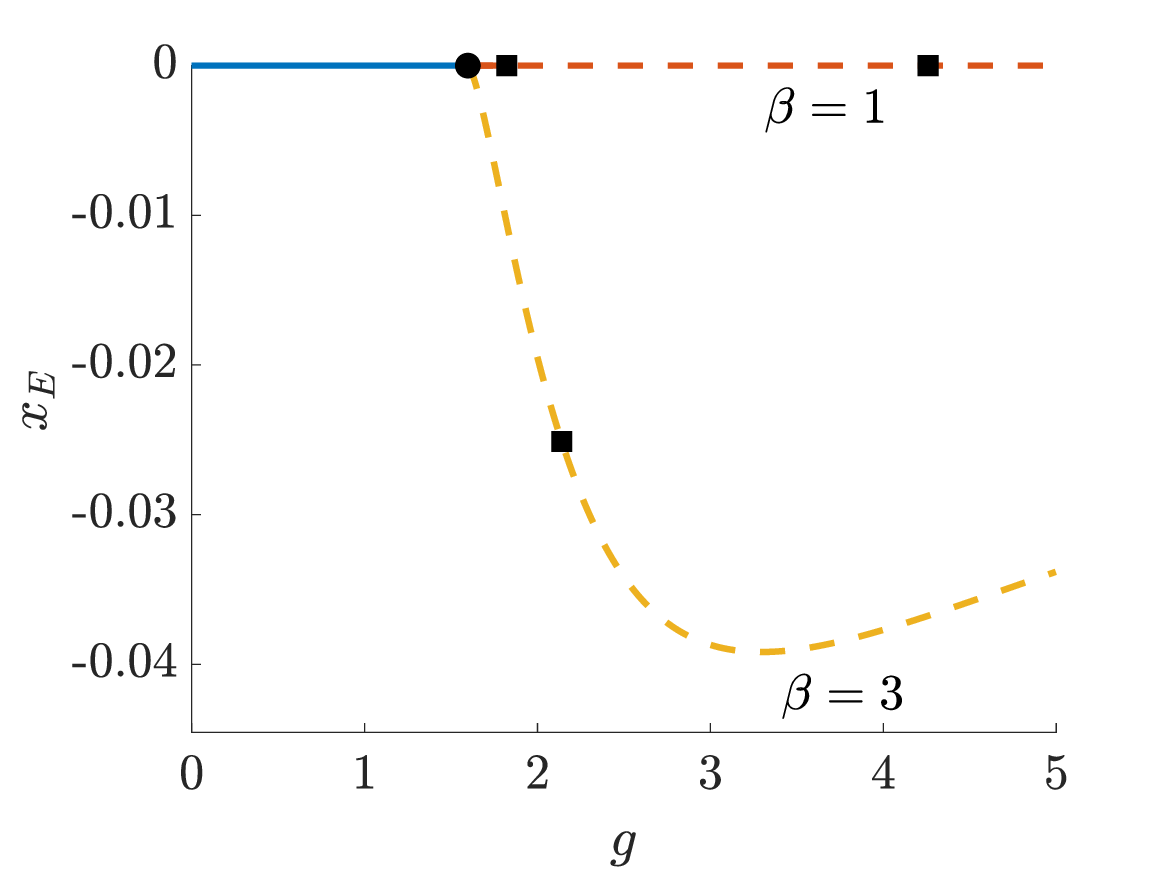} &
    \includegraphics[width=7.8cm]{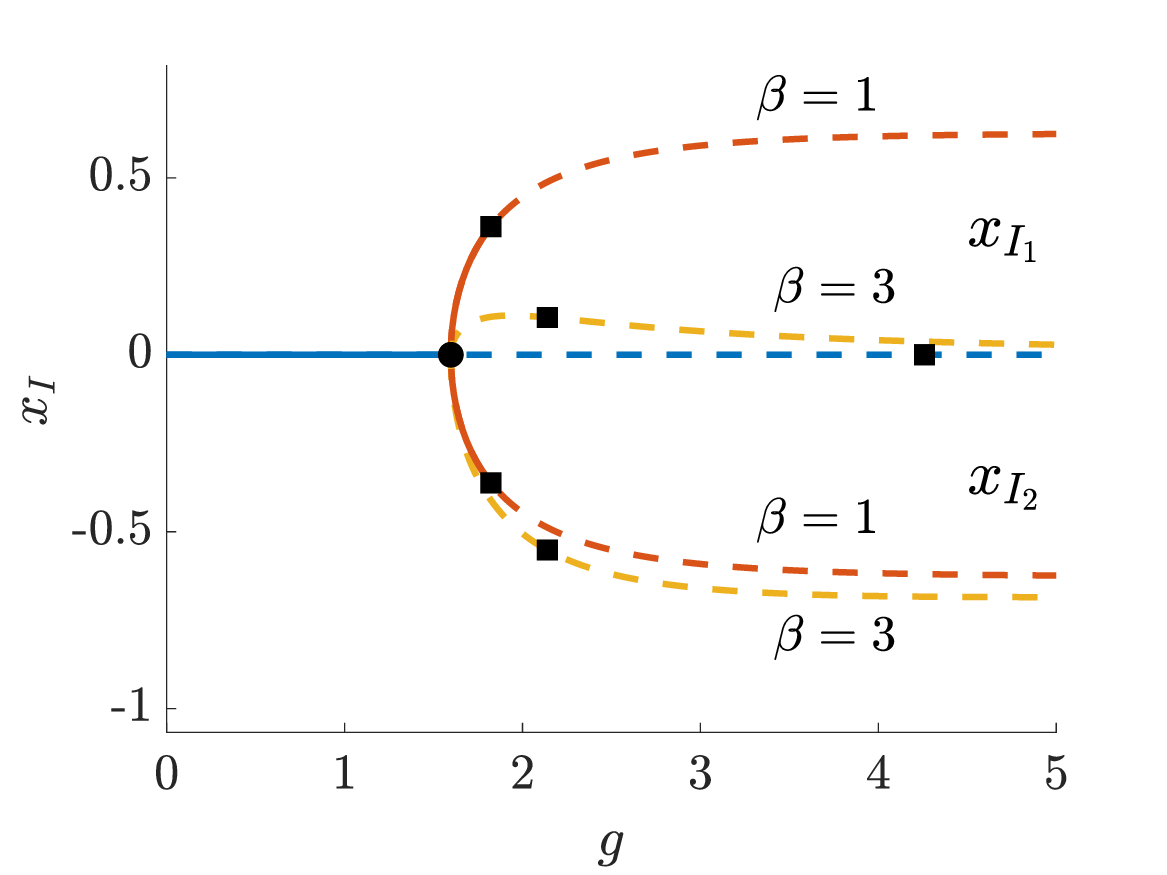} \\
    \includegraphics[width=7.8cm]{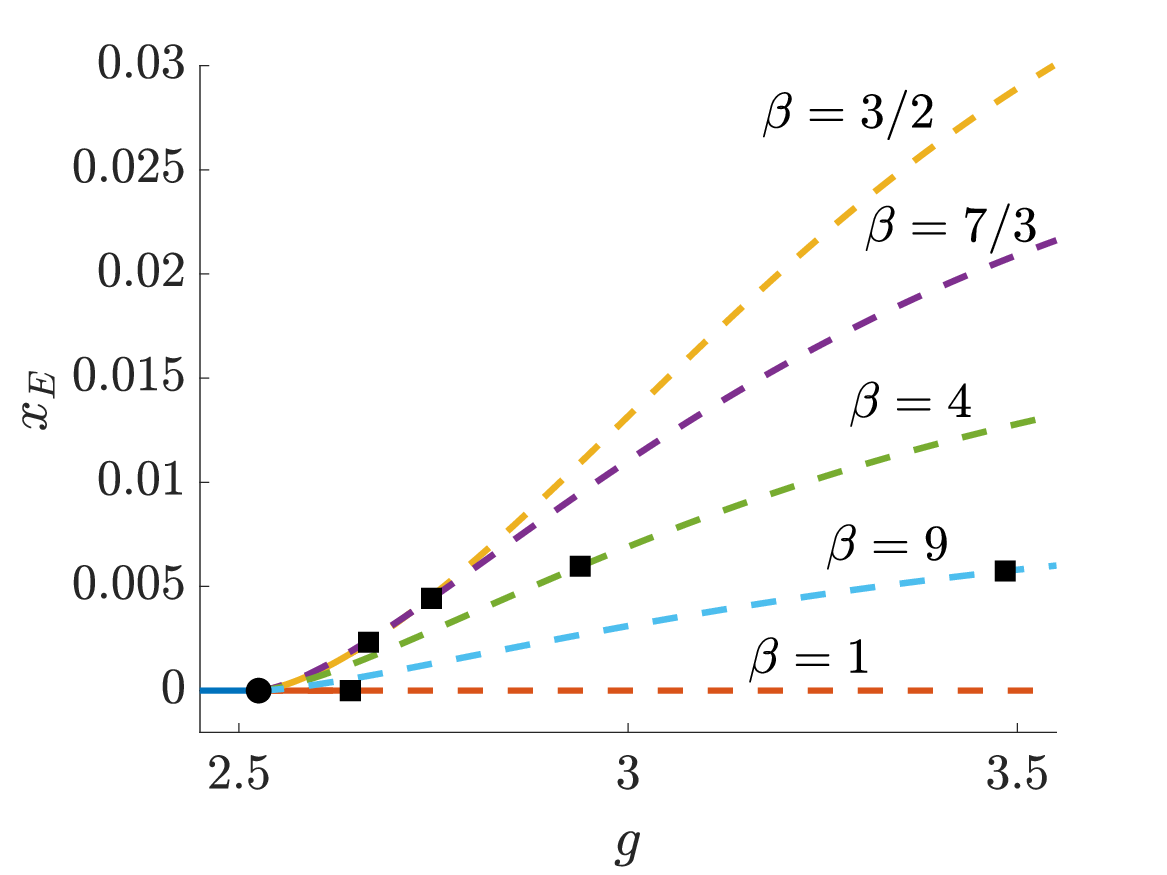} &
    \includegraphics[width=7.8cm]{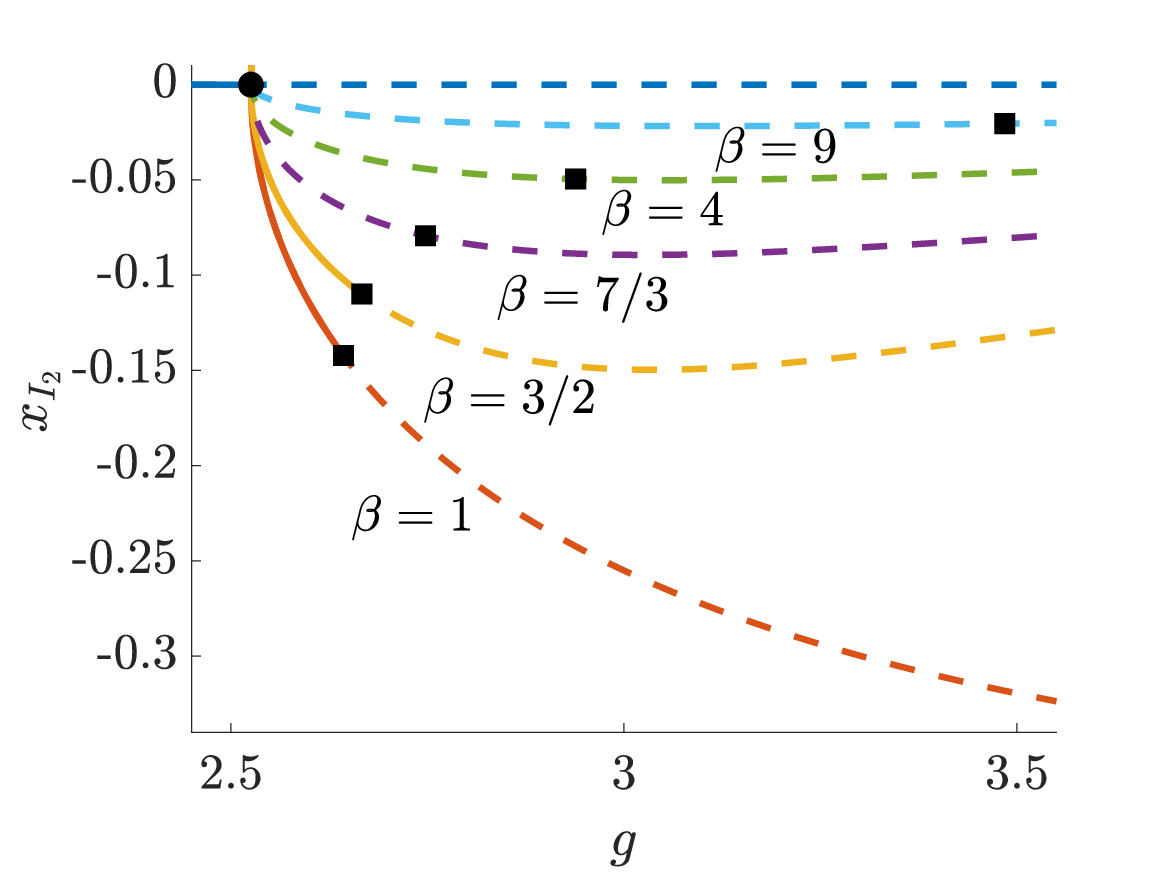}
    \end{tabular}
    \caption{Bifurcation diagram of all possible $I_1/I_2$ branches of equilibria for small $N$ networks. Top row: $N=20$. Top left: $x_E$ vs $g$. Top right: $x_{I_1}$ (above horizontal axis) and $x_{I_2}$ (below horizontal axis) vs $g$. Bottom row: $N=50$. Bottom left: $x_E$ vs $g$. Bottom right: $x_{I_2}$ \emph{only} vs $g$. Line format indicates stable (solid) vs. unstable (dashed) fixed points. The symmetric pitchfork bifurcation at $g = g_0$ is indicated with a filled circle. Hopf bifurcations are indicated with filled squares. Further bifurcations along branches are not shown to avoid clutter. Other parameters are $\alpha = 4$, $\mu_{EE} = 0.7$.}
    \label{fig:noclusterBD1}
\end{figure}

We note that the eigenvalues $\lambda_{I_1}$ and $\lambda_{I_2}$ split off from $\lambda_I$ at the pitchfork bifurcation point $g = g_0$; if $\xvec^* = 0$, then $\lambda_{I_1} = \lambda_{I_2} = \lambda_I$. To determine the stability of $\xvec^*$ for $g$ close to $g_0$, we must compute the eigenvalues $\lambda^*_{I_1}(g)$ and $\lambda^*_{I_2}(g)$ of $D\tilde{F}(\xvec^*)$ corresponding to $\lambda_{I_1}$ and $\lambda_{I_2}$. We will find (see \cref{app:I1I2stab}) that $\lambda^*_{I_2}(g)$ is always negative, while $\lambda^*_{I_1}(g)$ is negative for $\beta<2$ and positive otherwise. Therefore the fixed point is unstable for $\beta \ge 2$ (see \cref{fig:noclusterBD1}). 

\begin{figure}
    \centering
    \includegraphics[width=8.25cm]{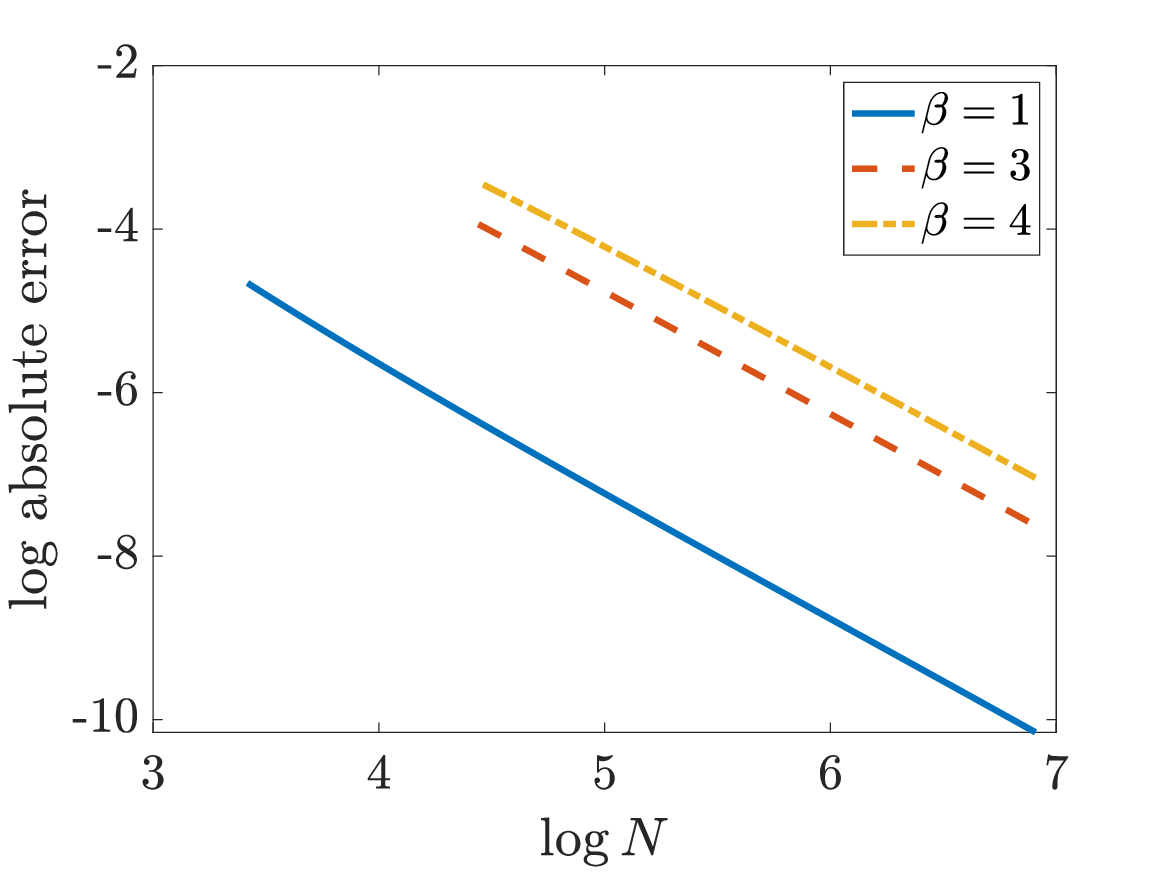} \hspace{-0.5cm}
    \includegraphics[width=8.25cm]{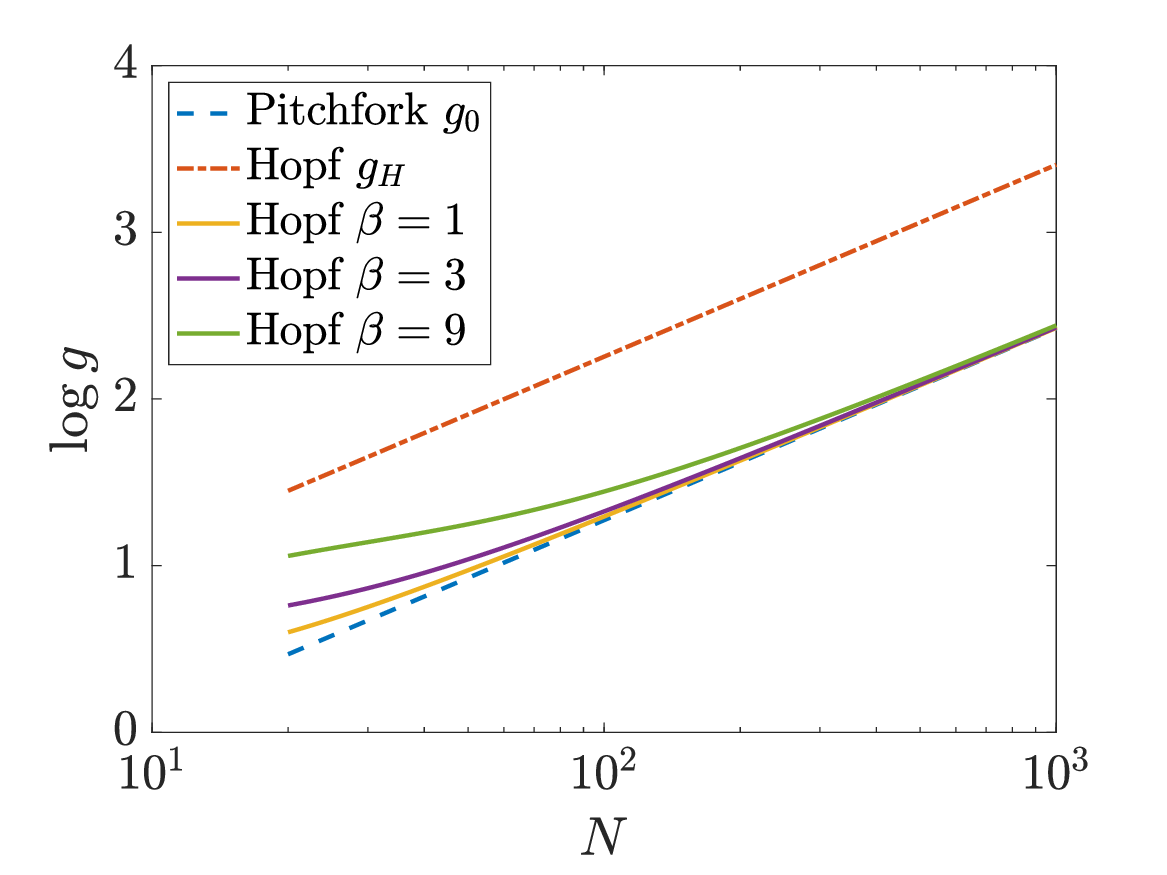}
    \caption{Locations of important bifurcations as a function of $N$. Left: Log-log plot of the absolute error of  \cref{eq:ghopfformula} vs $N$ for $\beta = 1$, 3, and 4. The slope of each line is approximately -1.5, validating the error term $\mathcal{O}\left( N^{-3/2} \right)$. Right: location of the symmetric pitchfork bifurcation $g_0$ (dashed line), Hopf bifurcation at the origin $g_H$ (dash-dotted line), and Hopf bifurcations on $I_1/I_2$ branches \cref{eq:ghopfformula} for select $\beta$ (solid lines, arranged from bottom to top in order of increasing $\beta$) as a function of $N$. Other parameters are: $\alpha = 4$, $\mu_{EE}= 0.7$. }
    \label{fig:Hopfplots}
\end{figure}

The remaining eigenvalues of $D\tilde{F}(\xvec^*)$ are the eigenvalues of $J_3(\xvec)$, given by \cref{eq:J3forI1I2}. These include one real eigenvalue and a complex pair (see \cref{app:I1I2stab} for computations). The real eigenvalue is always negative, and the complex pair crosses the real axis at a Hopf bifurcation when $g = g_{H}(\beta)$, where
\begin{equation}\label{eq:ghopfformula}
    g_{H}(\beta) = 
    \frac{\sqrt{N}}{\mu_{EE}} 
    \frac{ 2 - 5\beta + 2 \beta^2 + 3 \beta n_I}
    { \alpha(1 - 4 \beta + \beta^2) - (1 - \beta + \beta^2) + 3 \alpha \beta n_I }
    + \mathcal{O}\left( \frac{1}{N^{3/2}} \right).
\end{equation}
A plot of $g_{H}(\beta)$ versus $N$ for various $\beta$ is given in \cref{fig:Hopfplots}. We note that a Hopf bifurcation for a particular value of $\beta$ will only occur in a real network if the ratio of inhibitory cells is valid for that particular value of $N$ (e.g. for $\beta = 3$, the total number of inhibitory cells must be a multiple of 4). The leading order term of \cref{eq:ghopfformula}, as well as the order of the remainder term, agrees with results from numerical parameter continuation (\cref{fig:Hopfplots}). As $N \rightarrow \infty$, which implies $n_I = f N \rightarrow \infty$, the first terms in the numerator and denominator of \cref{eq:ghopfformula} dominate, thus $g_{H}(\beta) \rightarrow g_0$ as $N \rightarrow \infty$ for all $\beta$ (see \cref{fig:Hopfplots}). Differentiating the leading order term in \cref{eq:ghopfformula} with respect to $\beta$ and simplifying,
\begin{equation}\label{eq:gprime}
\frac{\partial}{\partial \beta} g_{H}(\beta) = \frac{ \sqrt{N} }{ \mu_{EE} }
    \frac{ 
    3(\alpha+1)(\beta^2-1)(n_I-1)
    }
    { 
        \left[ \alpha(1 - 4 \beta + \beta^2) - (1 - \beta + \beta^2) + 3 \alpha \beta n_I \right]^2
    }\:,
\end{equation}
which is 0 at $\beta = 1$ and positive for $\beta > 1$. As a consequence, $g_{H}(\beta)$ increases with $\beta$ for $\beta \geq 1$ (see \cref{fig:Hopfplots} for this ordering in $\beta$, as well as \cref{fig:noclusterBD1}). 

\subsection{Other branches of equilibria}\label{sec:otherbranches}

\begin{figure}
    \centering
    \begin{tabular}{cc}
    \includegraphics[width=7.8cm]{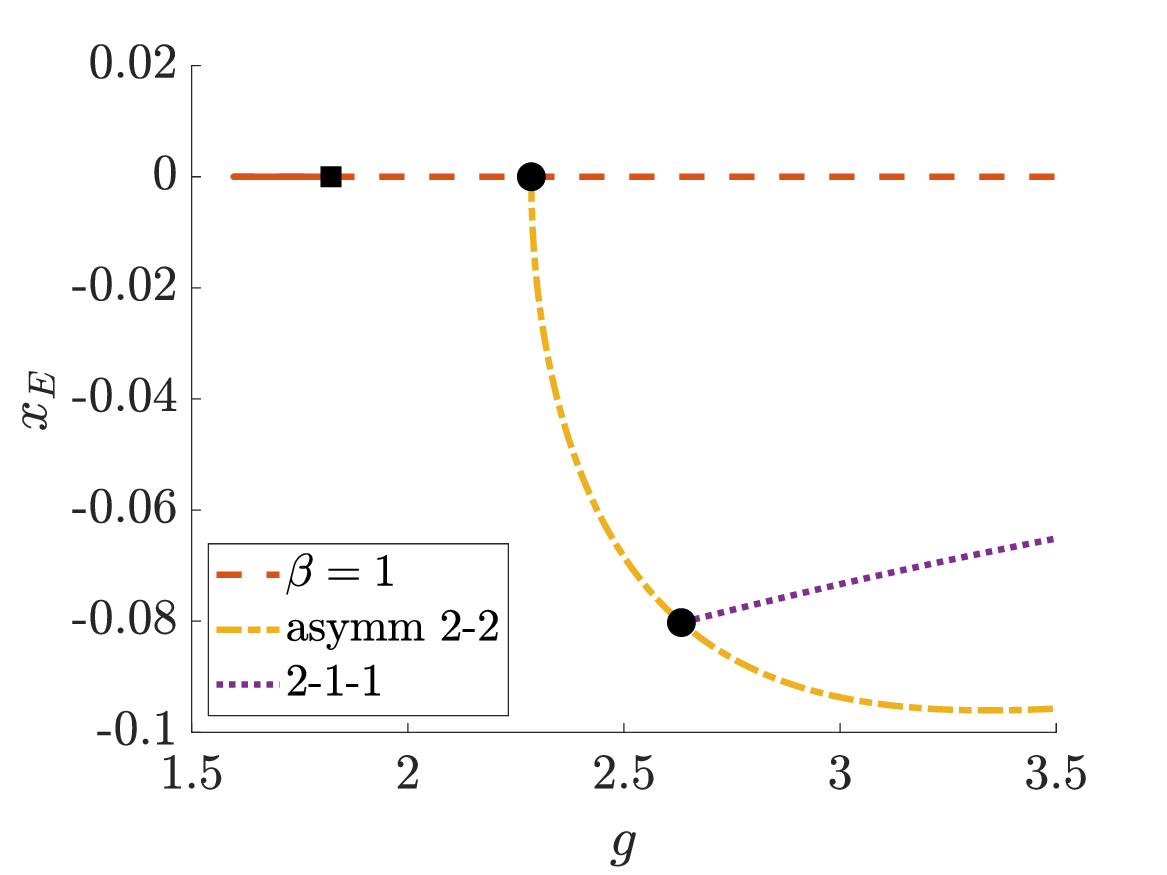} &
    \includegraphics[width=7.8cm]{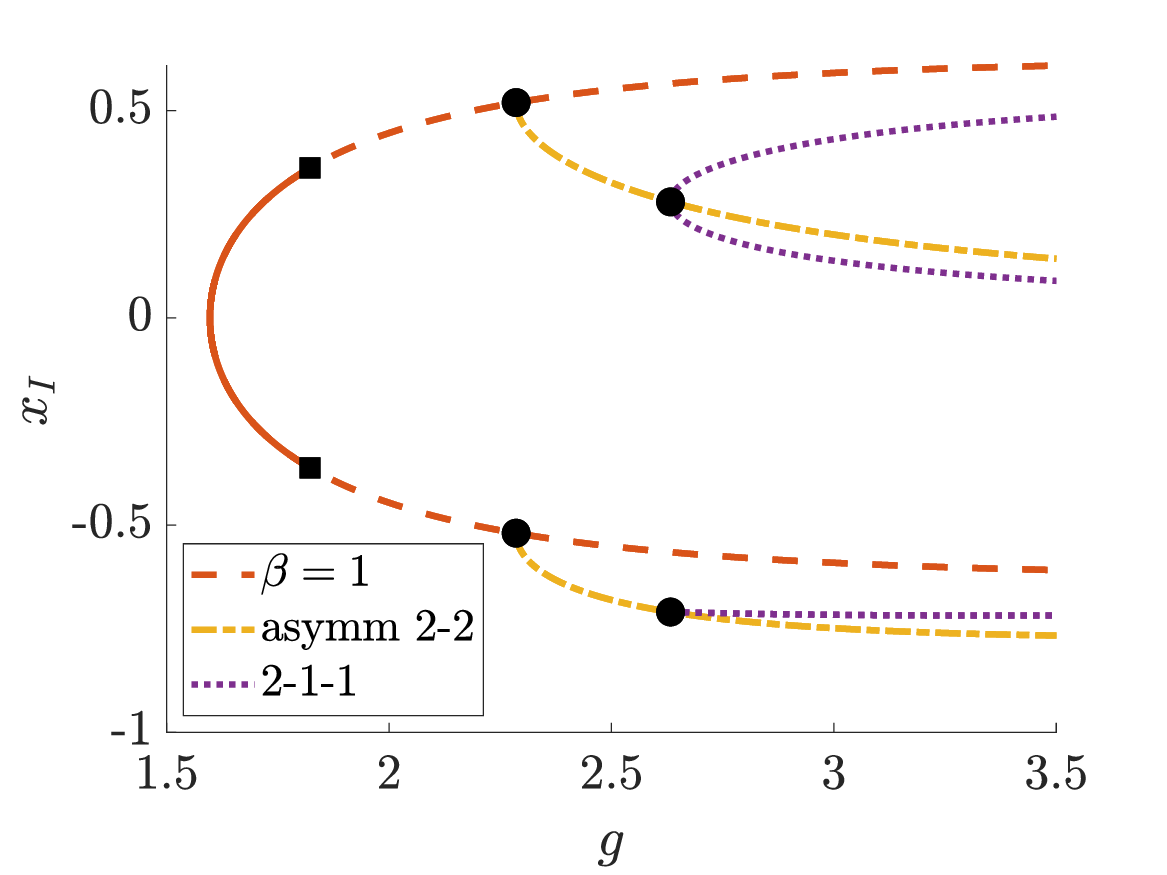}
    \end{tabular}
    \caption{Further branches of equilibria from $I_1/I_2$ branch with $\beta=1$ for $N=20$, showing excitatory cell (left) and inhibitory cell (right) activity. Line format indicates stable (solid) vs. unstable (dashed or dotted) fixed points. Branch points are indicated with a filled circle. Hopf bifurcations are indicated with filled squares. Other parameters are $\alpha = 4$, $\mu_{EE} = 0.7$.}
    \label{fig:noclusterbeta1branches}
\end{figure}

The equilibria on the $I_1/I_2$ branches, whose existence is guaranteed by the Equivariant Bifurcation Theorem and which were characterized in \cref{sec:solI1I2}, are not the only branches of equilibria. As one example, we consider what occurs on the $\beta=1$ branch for $N=20$ (see top panel of \cref{fig:noclusterBD1}). As $g$ is increased past the Hopf bifurcation, the complex pair of eigenvalues $\lambda_0 \pm i \omega_0$ collides on the positive real axis and becomes a real pair of eigenvalues $\{ \lambda_0^L, \lambda_0^R \}$, with $\lambda_0^L < \lambda_0^R$. As $g$ is further increased, $\lambda_0^L$ moves to the left, and $\lambda_0^R$ moves to the right. When $\lambda_0^L$ passes through the origin (from right to left), a symmetry-breaking bifurcation occurs (left branch point in \cref{fig:noclusterbeta1branches}). On the secondary branch, which we will call the asymmetric 2-2 branch, the excitatory activity $x_E \neq 0$, and the inhibitory pair $x_{I_1}$ and $x_{I_2}$ no longer have equal and opposite activities. As $g$ increases along this secondary branch, there is another bifurcation (right branch point in \cref{fig:noclusterbeta1branches}), which produces a branch of equilibria in which the inhibitory cells are clustered in a 2-1-1 pattern. As $N$ is increased, more complicated secondary branching patters occur, and it is unlikely that these can be systematically located and classified. That being said, numerical experiments performed on networks of varying $N$ strongly suggest that none of these secondary branches contain stable fixed points. Specifically, the only stable fixed points which have been found by numerical spectral computation are those on the primary $I_1/I_2$ branches; all other branches consist entirely of unstable equilibria. In addition, all numerical timestepping experiments starting from random initial conditions have converged to either fixed points on the primary $I_1/I_2$ branches or to periodic orbits (see \cref{sec:periodic} below).

\subsection{Periodic solutions}\label{sec:periodic}

Limit cycles arise as the bifurcation parameter $g$ passes through each Hopf bifurcation point. First, we discuss the limit cycle which bifurcates from the origin at $g = g_H$. 
At $g = g_H$, the complex pair of eigenvalues corresponding to $\lambda_0 \pm i \omega_0$ crosses the imaginary axis. The corresponding two-dimensional eigenspace is given by
\[ 
V \equiv  \ker(DF)_{\Zerovec,g_H}  = {\rm span} \, 
\left\{ \left[ 1 \; \underbrace{\begin{matrix}0 & \cdots & 0\end{matrix}}_{n_I} \right],
\left[ 0 \; \underbrace{\begin{matrix}1 & \cdots & 1\end{matrix}}_{n_I} \right]
 \right\},
\]
which is fixed by $\Gamma = S_1 \times S_{n_I}$ itself. Since $\dim \Fix_V(\Gamma) = \dim V = 2$, it follows from the Equivariant Hopf theorem \cite[Theorem 4.1]{GSS88Vol2} that there is a branch of small-amplitude, periodic solutions emanating from this Hopf bifurcation point for which the isotropy subgroup is $\Gamma$, i.e. the inhibitory neurons are all synchronized (see also \cite[Section 3.2]{Barreiro2017}). We recall that the excitatory neurons are always synchronized in the reduced model \cref{eq:reducedmatrixform} with $n_C = 1$.

\begin{figure}
    \centering
    \includegraphics[width=8.25cm]{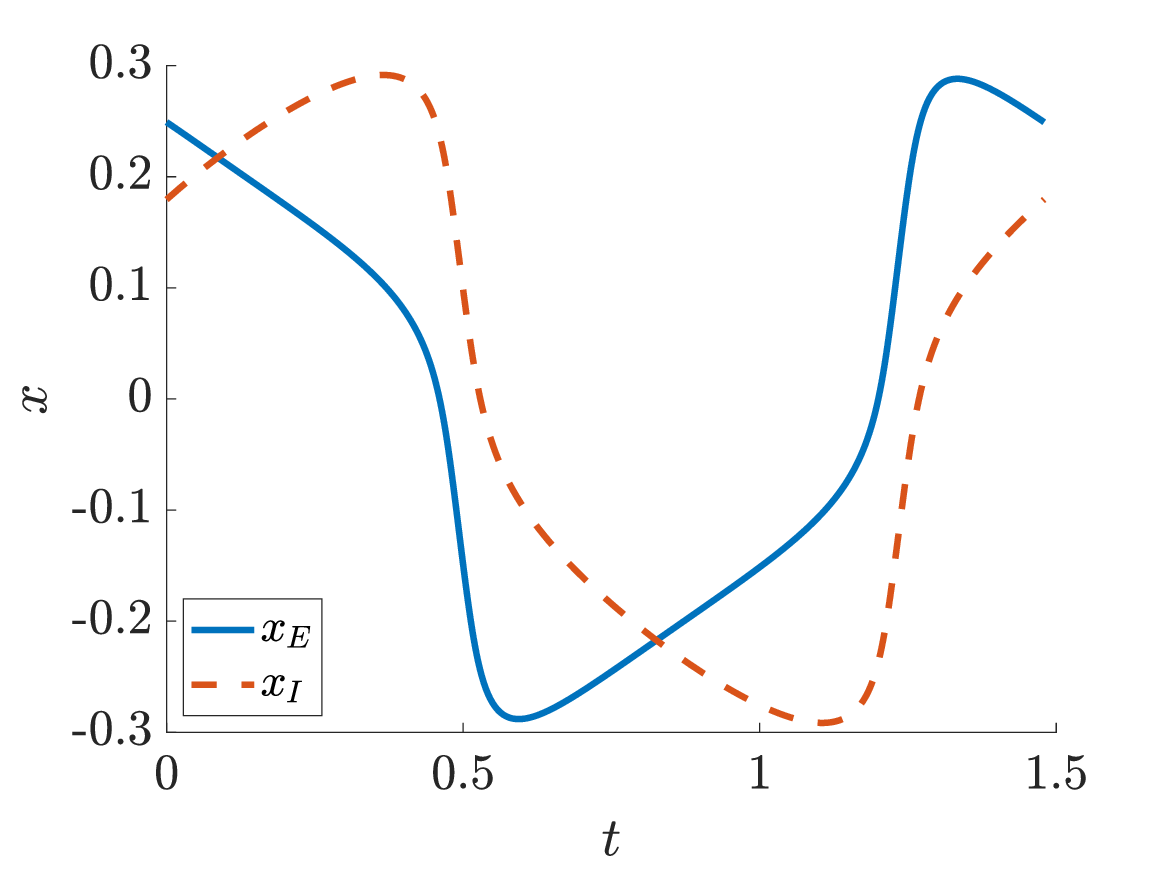}\hspace{-0.5cm}
    \includegraphics[width=8.25cm]{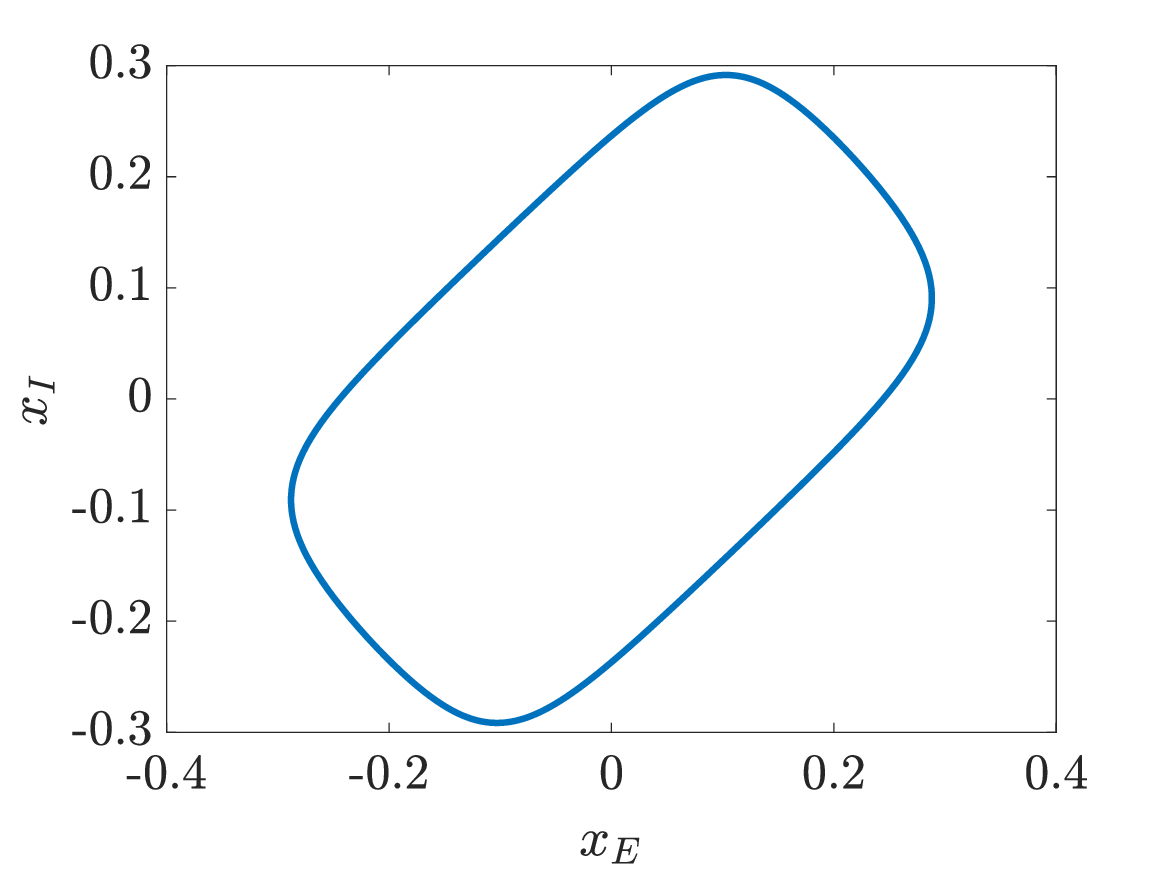}
    \caption{The $\beta = \infty$ limit cycle arising from a Hopf bifurcation at $g = g_H$. There is a single excitatory cluster with activity $x_E(t)$ and a single inhibitory cluster with activity $x_I(t)$. Parameters are: $N=20, g = 15$, $\alpha = 4$, $\mu_{EE}= 0.7$. The period of the limit cycle is 1.62.} 
    \label{fig:limitcycleorigin}
\end{figure}

Numerical computation with AUTO \cite{AUTO} validates this result, and shows that this limit cycle exists for all $g > g_H$, suggesting that the Hopf bifurcation is supercritical. Within this limit cycle, all inhibitory cells are synchronized. Since $n_{I_1} = n_I$ and $n_{I_2} = 0$, we will call this the $\beta=\infty$ limit cycle (see \cref{fig:limitcycleorigin}).
The $\beta = \infty$ limit cycle is a periodic solution to the two-dimensional system
\begin{equation}\label{eq:2dimsystem}
\begin{aligned}
\dot{x}_1 &= f_1(x_1, x_2) := -x_1 + \frac{\mu_{EE}}{\sqrt{N}}\left((n_E - 1) \tanh(g x_1) - \alpha n_I \tanh(g x_2) \right) \\
\dot{x}_2 &= f_2(x_1, x_2) := -x_2 + \frac{\mu_{EE}}{\sqrt{N}}\left( n_E \tanh(g x_1) - \alpha (n_I - 1) \tanh(g x_2) \right), 
\end{aligned}
\end{equation}
where $x_1$ represents the synchronized excitatory cell activity, and $x_2$ represents the synchronized inhibitory cell activity. In this two-dimensional system, the origin loses stability in a Hopf bifurcation at $g = g_H$ (see \cref{sec:limitcycleproof} for details). We note that equation \cref{eq:2dimsystem} is qualitatively similar to the Wilson-Cowan model for an excitatory-inhibitory pair (see section 11.3.3 of \cite{et10}) in its ``short-term memory" (STM) formulation \cite{Chow_Y_JNeurophys_2020}; both equations exhibit Hopf bifurcations and limit cycle solutions. The key difference is the use of input currents as bifurcation parameters in the Wilson-Cowan model as opposed to global coupling strength.

In the following proposition, we prove that the $\beta=\infty$ limit cycle exists for $g > g_H$, which also proves that the Hopf bifurcation is supercritical. The proof uses the Poincar\'e-Bendixson theorem, and is deferred to \cref{sec:limitcycleproof}. We note that the proposition does not address stability of the limit cycle.

\begin{proposition}\label{prop:limitcycle}
For $g > g_H$, the system \cref{eqn:sys_Basic} has a limit cycle in which all excitatory cells are synchronized, and all inhibitory cells are synchronized.
\end{proposition}

In addition to the $\beta = \infty$ limit cycle, periodic orbits arise on each $I_1/I_2$ branch as $g$ increases through the Hopf bifurcation point $g_H(\beta)$, which is given by \cref{eq:ghopfformula}. Once again, a complex pair of eigenvalues crosses the imaginary axis. The corresponding two-dimensional eigenspace is given by
\[ 
V \equiv  \ker(DF)_{\xvec,g_H}  = {\rm span} \, 
\left\{ \left[ a_1 \; \underbrace{\begin{matrix}0 & \cdots & 0\end{matrix}}_{n_{I_1}} \,
\underbrace{\begin{matrix}1 & \cdots & 1\end{matrix}}_{n_{I_2}} \right],
\left[ a_2 \; \underbrace{\begin{matrix}1 & \cdots & 1\end{matrix}}_{n_{I_1}} \,
\underbrace{\begin{matrix}0 & \cdots & 0\end{matrix}}_{n_{I_2}}\right]
 \right\},
\]
for some constants $a_1$ and $a_2$. When $\beta = 1$, this can be simplified to
\[ 
V \equiv  \ker(DF)_{\xvec,g_H}  = {\rm span} \, 
\left\{ \left[ 1 \; \underbrace{\begin{matrix}0 & \cdots & 0\end{matrix}}_{n_I} \right],
\left[ 0 \; \underbrace{\begin{matrix}1 & \cdots & 1\end{matrix}}_{n_I} \right]
 \right\},
\]

\begin{figure}
    \centering
    \includegraphics[width=8cm]{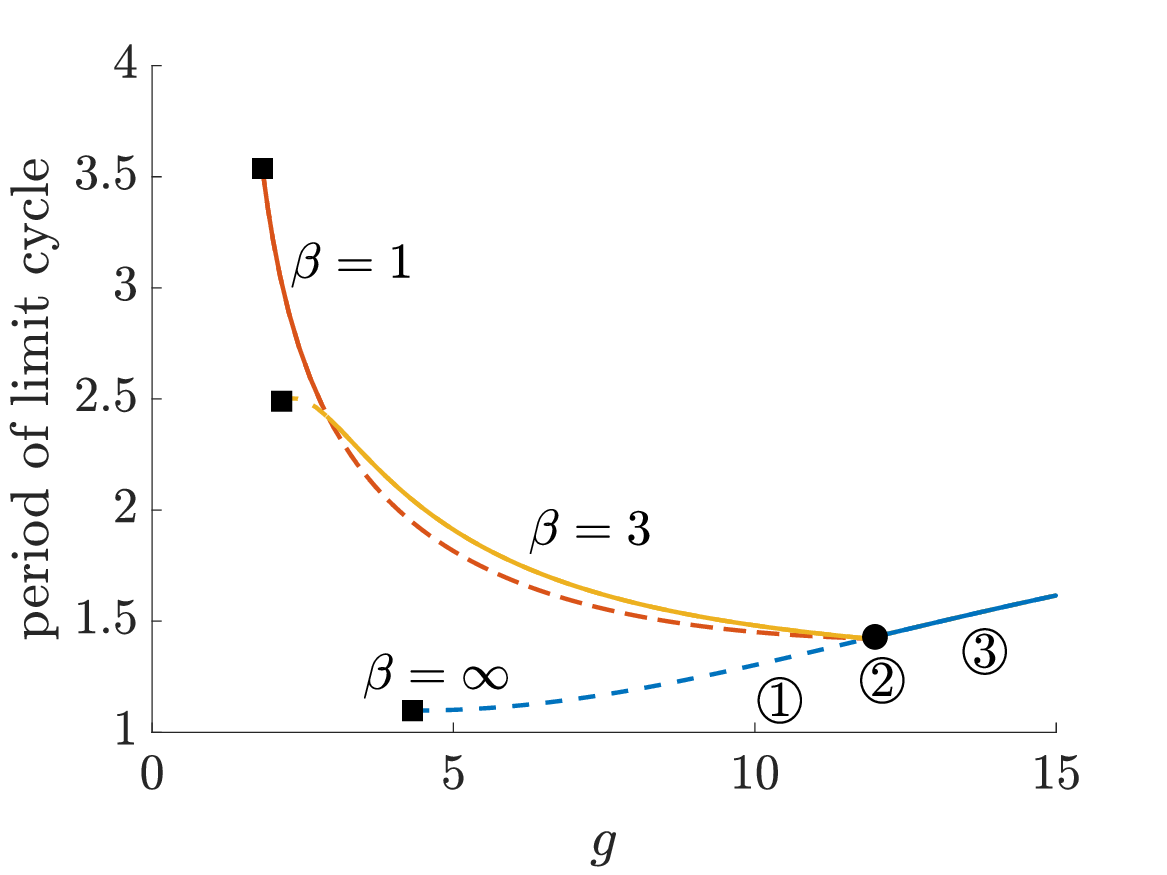}
    \includegraphics[width=8.2cm]{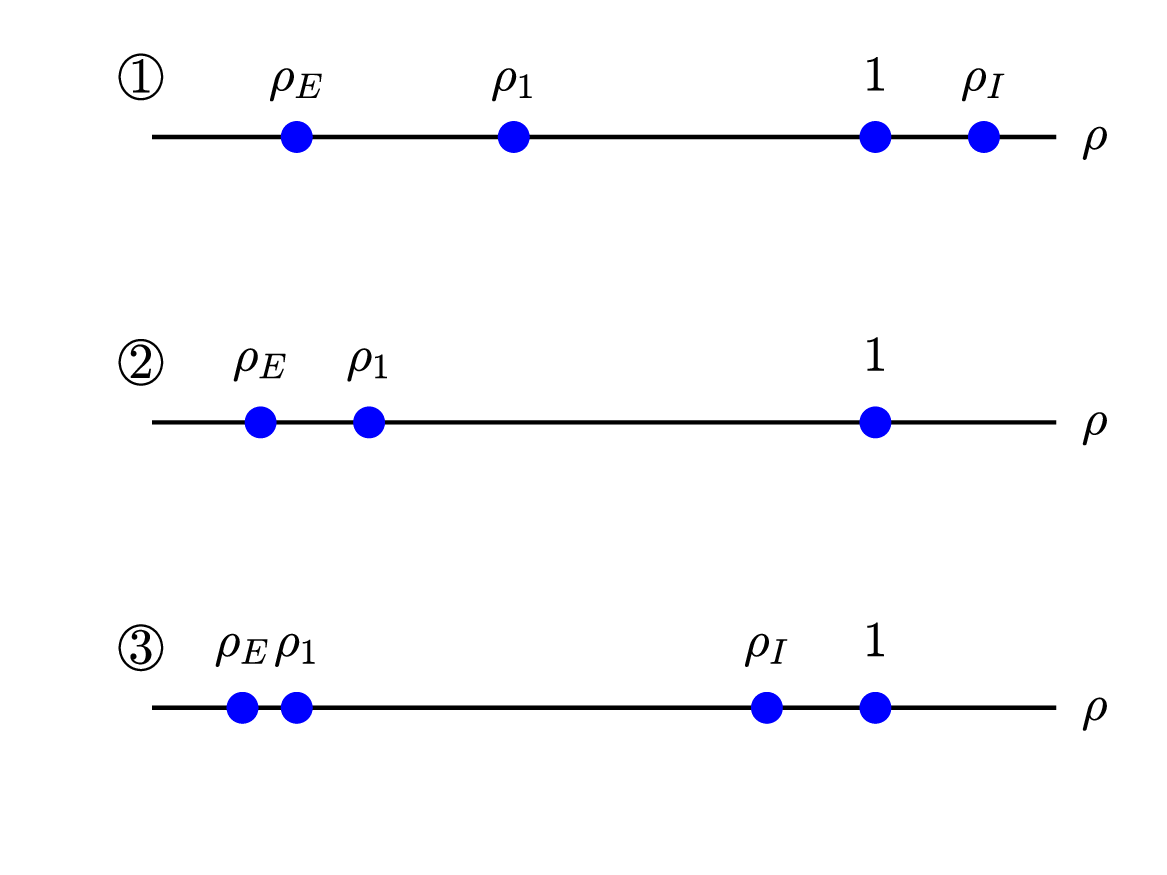}\\
    \hspace{-1cm}
    \includegraphics[width=17.25cm]{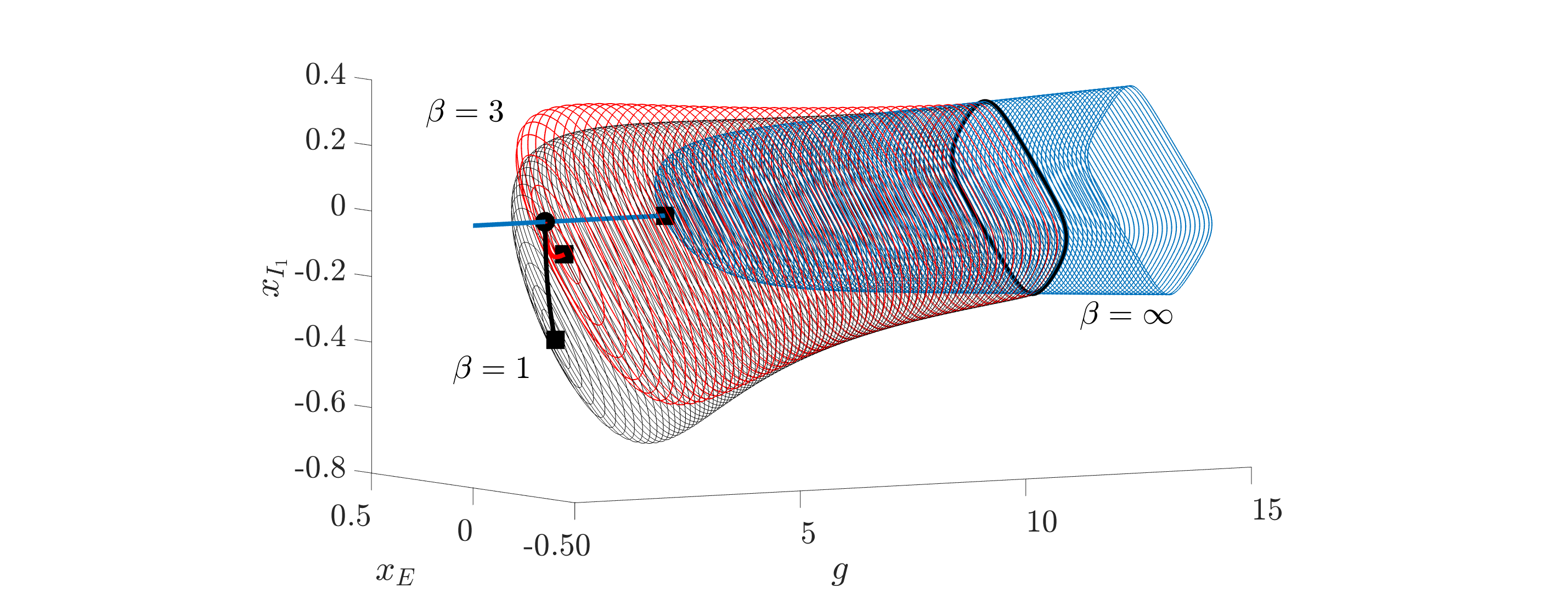}
    \caption{Each $I_1/I_2$ Hopf bifurcation spawns a branch of limit cycles which connects to the $\beta=\infty$ cycle at $g=g^*$. Top left: period of the limit cycle versus $g$. Stable limit cycles are indicated with solid lines. The symmetric pitchfork of limit cycles is indicated with a filled circle. Hopf bifurcations are indicated with filled squares, which correspond to the Hopf bifurcation points in \cref{fig:noclusterBD1}. Top right: Schematic of the Floquet eigenvalue pattern along the $\beta = \infty$ branch. The numbers 1,2, and 3 identify three representative points along the $\beta=\infty$ curve in the top left panel. Bottom: ($x_E$, $x_{I_1}$) vs. $g$ for three branches of fixed points (thick lines) and limit cycles (thin lines): $\beta = 1$ (gray), $\beta=3$ (red), and $\beta=\infty$ (blue). Other symbols are: pitchfork bifurcation at $g_0$ (filled circle), Hopf bifurcations for $\beta = 1$, $\beta = 3$, and $\beta = \infty$ (filled squares), and pitchfork bifurcation of limit cycles (dark band) at $g = g^*$. Parameters are: $N = 20$,  $\alpha = 4$, $\mu_{EE} = 0.7$.}
    \label{fig:periodvsg}
\end{figure}

\begin{figure}
    \centering
    \hspace{-1cm}
    \includegraphics[width=17.25cm]{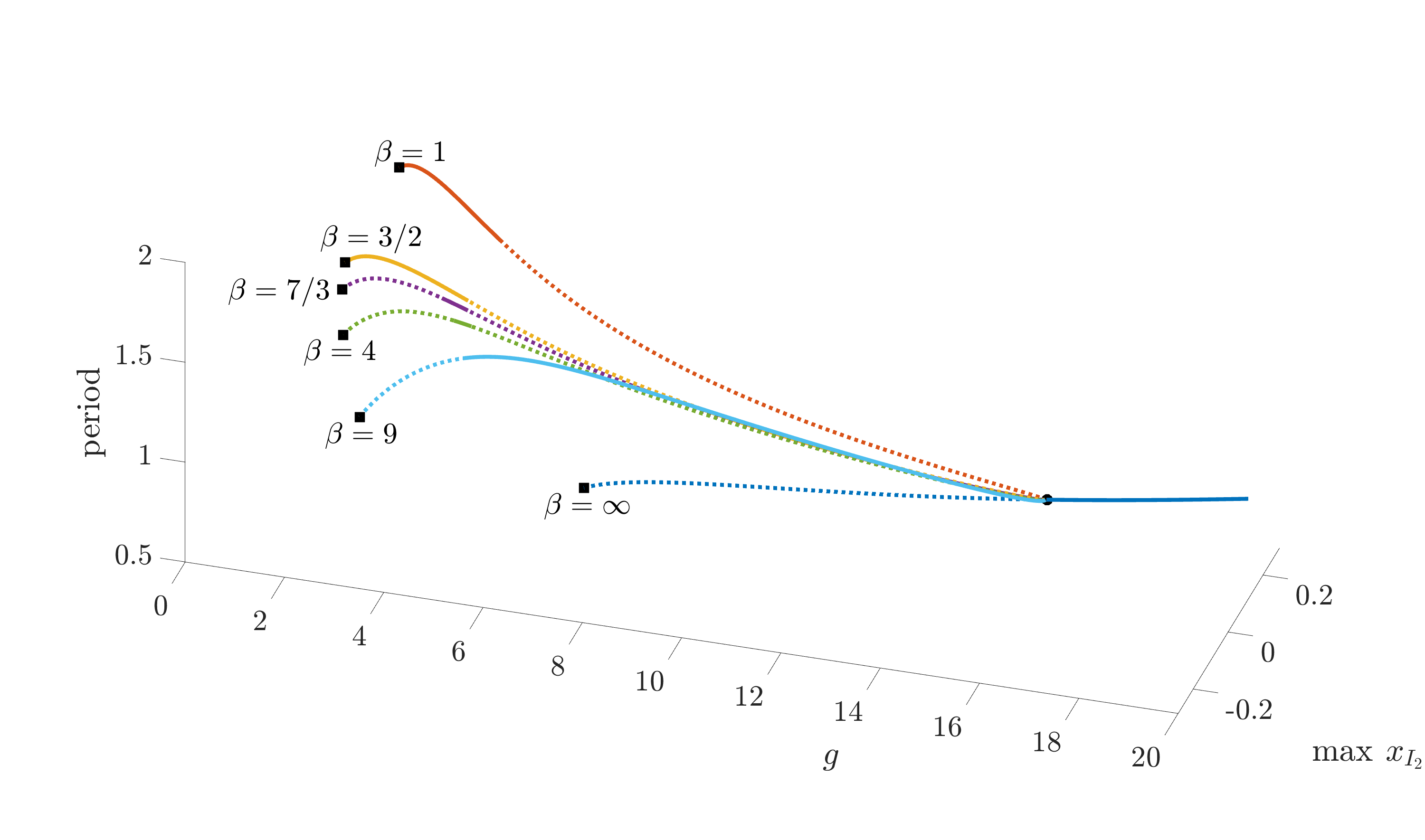}
    \caption{Period of limit cycle and $\max \: x_{I_2}$ versus $g$ for periodic solutions arising from Hopf bifurcations. Stable limit cycles are indicated with solid lines. The symmetric pitchfork of limit cycles is indicated with a filled circle. Hopf bifurcations are indicated with filled squares. Parameters are: $N = 50$,  $\alpha = 4$, $\mu_{EE} = 0.7$.}
    \label{fig:periodvsg50}
\end{figure}

This is a vector space of dimension 2, and it is fixed by the subgroup $\Sigma = S_1 \times S_{n_{I_1}} \times S_{n_{I_2}}$ of $\Gamma$ (see \cite[Section 3.3]{Barreiro2017}). Since $\dim \Fix_V(\Sigma) = \dim V = 2$, it follows from the Equivariant Hopf theorem \cite[Theorem 4.1]{GSS88Vol2} that there is a branch of small-amplitude, periodic solutions emanating from this Hopf bifurcation point for which the isotropy subgroup is $\Sigma$, i.e. the inhibitory cells are split into two clusters of sizes $n_{I_1}$ and $n_{I_2}$. This is the exact same symmetry as the $I_1/I_2$ branch from which these limit cycles bifurcate. For that reason, we can classify these periodic orbits in terms of the ratio $\beta = n_{I_1}/n_{I_2}$. Results from numerical parameter continuation (\cref{fig:periodvsg} and \cref{fig:periodvsg50}) indicate that this Hopf bifurcation is supercritical, and the limit cycles exist for $g > g_H(\beta)$. 

A plot of the period of these limit cycles with increasing $g$ is shown in \cref{fig:periodvsg} for $N=20$ (see also \cite[Fig. 2]{Barreiro2017}) and \cref{fig:periodvsg50} for $N=50$. There is a critical value $g = g^*$ where all of the limit cycle branches meet (see dark band in bottom panel of \cref{fig:periodvsg}). For $g > g^*$, the only remaining limit cycle is the $\beta = \infty$ limit cycle, which has become stable. The point $g = g^*$ is a symmetric pitchfork bifurcation of limit cycles, which we can see by examining the Floquet multipliers of the linearization about the $\beta = \infty$ limit cycle branch (see right panel of \cref{fig:periodvsg}). These Floquet multipliers are computed using AUTO, and are all real. In addition to a single Floquet multiplier at 1 which is always present, there is a Floquet multiplier $\rho_E$ with multiplicity $n_E - 1$, a Floquet multiplier $\rho_I$ with multiplicity $n_I - 1$, and Floquet multiplier $\rho_1$ with multiplicity 1. At $g = g^*$, the Floquet multiplier $\rho_I$ with multiplicity $n_I - 1$ passes through 1. As $g$ decreases though $g^*$, the $\beta = \infty$ limit cycle loses stability and gives rise to limit cycles with symmetry corresponding to each $I_1/I_2$ branch. This is analogous to the pitchfork bifurcation of the fixed point $\xvec = 0$ at $g = g_0$, which loses stability when the eigenvalue $\lambda_I$ with multiplicity $n_I - 1$ passes through the origin.

\subsection{Behavior of the \texorpdfstring{$I_1/I_2$}{I1/I2} branch for large \texorpdfstring{$g$}{g}}\label{sec:stab_largeg}

We have characterized the three-cluster fixed point solutions on the $I_1/I_2$ branches near the symmetric pitchfork bifurcation point at $g = g_0$. Next, we will show that these branches are unstable for sufficiently large $g$. Fix $\beta = n_{I_1}/n_{I_2}$, and let $\xvec = (x_E, x_{I_1}, x_{I_2})$ be a solution to \cref{eq:reducedsystemI1I2} for $g > g_0$; this solution depends on $g$. Recall from \cref{sec:solI1I2} that $\xvec$ is bounded for all $g$. Let $\xvec^*$ be the corresponding fixed point of \cref{eq:reducedmatrixform}. To determine the stability of $\xvec^*$, we will look at the eigenvalues of $D\tilde{F}(\xvec^*)$ for large $g$. The sum of these eigenvalues is 
\[
\text{Trace }D\tilde{F}(\xvec^*) = \frac{(n_E-1) \mu_{EE}}{\sqrt{N}} g \sech^2( g x_E ) - (n_I+1).
\]
We will show that for sufficiently large $g$, $\text{Trace }D\tilde{F}(\xvec^*) > 0$, and thus at least one eigenvalue has positive real part. To do this, we analyze the behavior of $x_E$ and $\sech(gx_E)$ as $g \rightarrow \infty$. First, we consider the case when $x_E \rightarrow 0$. There are three possibilities for the behavior of $\sech(gx_E)$, only two of which can occur.
\begin{enumerate}[(i)]
    \item $x_E \rightarrow 0$, and $g x_E \rightarrow 0$ (e.g. $x \sim g^{-\beta}$, for $\beta > 1$): then $\sech^2(gx_E) \rightarrow 1$, and so $g \sech^2(gx_E) \rightarrow \infty$.
    \item $x_E \rightarrow 0$, but $g x_E \rightarrow C$, for $C>0$ (e.g. $x \sim g^{-1}$): then $g\sech^2(gx_E) \sim g\sech^2(C) \rightarrow \infty$.
    \item $x_E \rightarrow 0$, but $g x_E \rightarrow \infty$ (e.g. $x \sim g^{-\beta}$, for $0 < \beta < 1$). If $x_E \sim g^{-\beta}$, then $g \sech^2(gx_E) \rightarrow 0$; this would seem to result in a negative trace as $g \rightarrow \infty$. However, we will show this cannot happen. Because $\sech^2(gx_E) \rightarrow 0$, it follows that $\tanh^2( g x_E) \rightarrow 1$, which implies $\tanh(g x_E) \rightarrow 1$. We use the first line from \cref{eq:reducedsystemI1I2} to obtain a lower bound for $x_E$ as follows. 
    Since $x_{I_1}$ and $x_{I_2}$ have opposite signs for $g > g_0$ (see the end of \cref{app:I1I2sol}), we can state that $\tanh(gx_{I_1})\leq 1$ and $\tanh(gx_{I_2})\leq 0$, and therefore that
    \begin{align*}
    x_E & = \frac{\mu_{EE}}{\sqrt{N}} \left[ (\alpha n_1-1) \tanh(gx_E) - \alpha \frac{\beta}{\beta+1} n_I \tanh(gx_{I_1}) - \alpha \frac{1}{\beta+1} n_I \tanh(gx_{I_2})\right]\\
    & \geq \frac{\mu_{EE}}{\sqrt{N}} \left[ (\alpha n_1-1)  - \alpha \frac{\beta}{\beta+1} n_I  \right] = \frac{\mu_{EE}}{\sqrt{N}}\left[ \alpha(n_I - n_{I_1}) - 1 \right] \geq \frac{\mu_{EE}}{\sqrt{N}}\left( \alpha - 1 \right),
    \end{align*}
    since $n_{I_1} \leq n_I - 1$. As long as we take $\alpha > 1$ (which is typically the case), $x_E$ is bounded away from 0 for all $g > g_0$, thus contradicting our original assumption that $x_E \rightarrow 0$.
\end{enumerate} 
We have shown that if $x_E \rightarrow 0$, $\text{Trace }D\tilde{F}(\xvec^*) > 0$ for sufficiently large $g$, which implies that $D\tilde{F}(\xvec^*)$ always has an eigenvalue with positive real part. 

The remaining possibility is that $x_E \rightarrow \hat{x}_E \neq 0$. In \cref{app:stab_largeg}, we show that this cannot occur. Therefore, all $(x_E,x_{I_1},x_{I_2})$ satisfy $x_E \rightarrow 0$ as $g\rightarrow \infty$, and so all equilibria on the $I_1/I_2$ branches are unstable for sufficiently large $g$. We note that this does not say anything about the stability of equilibria on any branches which may bifurcate from the $I_1/I_2$ branches. However, the results of extensive numerical timestepping simulations suggest that there are no stable equilibria for sufficiently large $g$.

\section{Excitatory clusters, weight parameters balanced}\label{sec:Eclusters}

We now allow the excitatory cells to be grouped into $n_C$ clusters of size $p$, where $p = \lfloor N f/n_C \rfloor$. We will take $p > 1$ to ensure that each excitatory cluster contains more than one cell, and we will also assume $n_C \geq \alpha$ (e.g. $n_C \geq 4$ for the standard value of $\alpha = 4$). Since we are interested in the behavior of the system for large $N$ and for a large number of clusters (e.g. $n_C$ scales with $\sqrt{N}$), this is not a significant restriction. Cells will be connected within, but not between, clusters. 
For simplicity, and relying on \cref{prop:tildeHeig}, we will focus only on the reduced system \cref{eq:reducedmatrixform}.
The right-hand side of \cref{eq:reducedmatrixform} is now $\Gamma$-equivariant for $\Gamma = S_{n_C} \times \, S_{n_I}$, where $n_C > 1$. That is, we can permute the labels of the excitatory clusters and the labels of the inhibitory cells without changing the equation.
We choose the weights so that the network is balanced. 
\begin{align*}
\mu_{EE} &= n_C \mu && \mu_{IE} = \mu \\
\mu_{EI} &= -\alpha \mu && \mu_{II} = -\alpha.
\end{align*}
The expression for $\mu_{EE}$ compensates for the fact that each excitatory cell has fewer excitatory connections. The eigenvalues of $\tilde{H}$ (right panel of \cref{fig:Heigpattern}) are:
\begin{itemize}
\item $\lambda_I := \alpha \mu > 0$ with multiplicity $n_I - 1$.
\item $\lambda_C := (p-1) n_C \mu > 0$, with multiplicity $n_C - 1$.
\item A complex conjugate pair of eigenvalues $\lambda_0 \pm i \omega_0$, with 
\begin{equation*}
    \lambda_0 := \frac{1}{2}\mu(\alpha - n_C), \quad 
    \omega_0 := \frac{1}{2}\mu \sqrt{ \alpha + n_C} \sqrt{ n_C(4 p - 1) - \alpha },
\end{equation*}
where we used the fact that $\alpha n_I = n_E = p n_C$.
\end{itemize}
Since $\lambda_E < 0$ and $\lambda_0 \leq 0$ (as a consequence of taking $n_C \geq \alpha$), the corresponding eigenvalues of $DF(0)$ will always be negative, and thus will not affect the stability of the fixed point at 0. The eigenvalues which determine stability of the origin are $\lambda_I$ and $\lambda_C$. We note that since $p > 1$, $0 < \lambda_I < \lambda_C$.

As in \cref{sec:E1I1}, we will determine the bifurcations which occur as $g$ is increased, together with the structures which emerge at these bifurcation points. First, the origin loses stability in a symmetric pitchfork bifurcation, after which point there is a branch of equilibria for every possible division of the excitatory clusters into two groups. This is similar to what occurs in the unclustered case, except the bifurcation involves the excitatory clusters instead of the inhibitory cells. As before, we derive leading order formulas for these branches, and show which of them are initially stable. As $g$ is further increased, instead of a Hopf bifurcation, there is another symmetric pitchfork bifurcation on each of these branches, in which the inhibitory cells split into two groups. For large $g$, there is a collection of stable fixed points, which we can locate using the limiting behavior of the system.

\subsection{Bifurcations of the origin}\label{sec:Eclusterbiforigin}

As the bifurcation parameter $g$ increases from 0, the first bifurcation occurs when 
the set of $n_C - 1$ eigenvalues $\lambda_C^*(g)$ of $D\tilde{F}(0)$ corresponding to $\lambda_C$ crosses the imaginary axis at
\begin{equation}
    g = g_C := \frac{\sqrt{N}}{(p-1) n_C \mu}.
\end{equation}
The corresponding eigenspace is the set of all zero-sum vectors with support in the excitatory clusters only, i.e. 
\[ 
V \equiv \ker(D\tilde{F})_{\Zerovec,g_C} = {\rm span} \, \left\{ \left[ 
\vvec_C \; \underbrace{\begin{matrix}0 & \cdots & 0\end{matrix}}_{n_I} \right] \right\}, \qquad \vvec_C \perp \Onevec_{n_{C}},
\]
which has dimension $n_C - 1$.
We can check that $\Gamma$ acts irreducibly on $V$, similarly to \cref{sec:symmpitch}. We then find subgroups $\Sigma$ of $\Gamma$ which satisfy the hypothesis of the Equivariant Branching Lemma by breaking the excitatory clusters up into two clusters $C_1$ and $C_2$ of sizes $n_{C_1}$ and $n_{C_2}$, where $n_{C_1} + n_{C_2} = n_C$. For each such decomposition, this describes a subgroup 
\begin{equation}
\Sigma_C = S_{n_{C_1}} \times S_{n_{C_2}} \times S_{n_I}
\end{equation}
of $\Gamma$. The subgroup $\Sigma_C$ has the fixed-point subspace 
\begin{eqnarray}
\Fix_V(\Sigma_C) & = & {\rm span} \, \left\{ \left[
\underbrace{\begin{matrix}1 & \cdots & 1\end{matrix}}_{n_{C_1}} \;
\underbrace{\begin{matrix}-\frac{n_{C_1}}{n_{C_2}} & \cdots & -\frac{n_{C_1}}{n_{C_2}} \end{matrix}}_{n_{C_2}}  \;
\underbrace{\begin{matrix}0 & \cdots & 0 \end{matrix}}_{n_I} 
 \right] \right\},
\end{eqnarray}
which has dimension 1.
It follows from the Equivariant Branching Lemma that there is a branch of equilibria emerging at the symmetric pitchfork bifurcation point $g=g_C$ for 
all such subgroups $\Sigma_C$, i.e. for every possible division of the excitatory clusters into exactly two groups of size $n_{C_1}$ and $n_{C_2}$. All cells are synchronized within each excitatory cluster. Each such branch may be characterized by the number 
\begin{equation}\label{eq:betac}
\beta_C = \frac{n_{C_1}}{n_{C_2}},
\end{equation}
which gives the ratio of the sizes of the two groups of excitatory clusters. Without loss of generality, we may take $n_{C_1} \geq n_{C_2}$, so that $\beta_C \geq 1$. At the start of each $C_1/C_2$ branch, the inhibitory cells are synchronized. This is the case since, near $g = g_C$, no other eigenvalues have crossed through the origin, thus no bifurcations involving the inhibitory cells have occurred. The solution on each $C_1/C_2$ branch is then given as $(x_{E_1}, x_{E_2}, x_{I})$. Due to the odd symmetry of \cref{eqn:sys_Basic}, there is a corresponding $C_1/C_2$ branch for each $\beta_C$ with solution $(-x_{E_1}, -x_{E_2}, -x_{I})$, which we will ignore for simplicity. Similar to what is discussed in \cref{sec:symmpitch}, a division of the excitatory clusters into more than two groups will lead to a fixed-point subspace of dimension 2 or greater, thus a branch with this symmetry is not guaranteed to exist by the Equivariant Branching Lemma. Such branches may occur, but as with the unclustered case, numerical evidence strongly suggests that all of them are unstable.

As $g$ is further increased, the eigenvalue $\lambda_I^*(g)$ with multiplicity $n_I-1$ crosses the imaginary axis at $g = g_0$, where $g_0$ is defined by \cref{eq:pitchlocation}. 
A second symmetric pitchfork bifurcation occurs at this point, this time involving the inhibitory cells. This is almost identical to what occurs in the unclustered case (\cref{sec:symmpitch}). Briefly, the corresponding eigenspace is the set of all zero-sum vectors with support in the inhibitory cells only, i.e. 
\[ 
V \equiv \ker(D\tilde{F})_{\Zerovec,g^*} = {\rm span} \, \left\{ \left[ 
\underbrace{\begin{matrix}0 & \cdots & 0\end{matrix}}_{n_C} \; \vvec_I \right] \right\}, \qquad \vvec_I \perp \Onevec_{n_{I}},
\]
which has dimension $n_I-1$.
We then break the inhibitory cells up into two groups $I_1$ and $I_2$ of sizes $n_{I_1}$ and $n_{I_2}$, where $n_{I_1} + n_{I_2} = n_I$, which describes a subgroup $\Sigma_I = S_{n_C} \times S_{n_{I_1}} \times S_{n_{I_2}}$ of $\Gamma$. The fixed-point subspace of $\Sigma_I$ is then given by
\begin{eqnarray}
\Fix_V(\Sigma_I) & = & {\rm span} \, \left\{ \left[ \underbrace{\begin{matrix}0 & \cdots & 0\end{matrix}}_{n_C} \;
\underbrace{\begin{matrix}1 & \cdots & 1\end{matrix}}_{n_{I_1}} \;
\underbrace{\begin{matrix}-\frac{n_{I_1}}{n_{I_2}} & \cdots & -\frac{n_{I_1}}{n_{I_2}} \end{matrix}}_{n_{I_2}} \right] \right\},
\end{eqnarray}
which has dimension 1. As in \cref{eq:pitchlocation}, it follows from Equivariant Branching Lemma that there is an $I_1/I_2$ branch of equilibria emerging at the symmetric pitchfork bifurcation point $g=g_0$ for every possible division of the inhibitory cells into exactly two groups of size $n_{I_1}$ and $n_{I_2}$.
An important distinction from the previous section is that there will be no Hopf bifurcation of the origin, since the complex conjugate pair of eigenvalues cannot cross the imaginary axis.

\subsection{Solutions on \texorpdfstring{$C_1/C_2$}{C1/C2} branch}

First, we derive leading order expressions for the solutions along the $C_1/C_2$ branches for $g$ close to $g_C$. The simplest case occurs when $n_C$ is even and $\beta_C = 1$, in which case $n_{C_1}=n_{C_2}$. On this branch, $x_{E_2} = -x_{E_1}$, i.e. there are two equally sized groups of excitatory clusters with equal and opposite activity, and all the inhibitory cells have synchronized activity $x_I = 0$. Taking $x_{E_1} = x_E$, $x_{E_2} = -x_E$, and $x_I = 0$ in \cref{eq:reducedsystem} and simplifying, we obtain the single equation $\tanh(g x_E) = g_C x_E$. As in \cref{sec:E1I1}, $x_E$ is given, to leading order, by
\begin{align}\label{eq:xEapprox}
x_E &= \sqrt{ \frac{3(g - g_C) }{g^3}} && g \geq g_C,
\end{align}
for $g$ close to $g_C$. For $\beta_C > 1$, we find the solution along each $C_1/C_2$ branch by reducing \cref{eqn:sys_Basic} to the 3-dimensional system
\begin{equation}\label{eq:cluster3system}
 \begin{aligned}
 \begin{bmatrix} x_{E_1} \\ x_{E_2} \\ x_{I} \end{bmatrix} 
 &= \frac{\mu}{\sqrt{N}} 
 \begin{bmatrix} 
    (p-1)n_C & 0 & -p n_C  \\
    0  & (p-1)n_C & -p n_C \\
    p n_C \frac{\beta_C}{\beta_C+1} &
    p n_C \frac{1}{\beta_C+1} &
    -(p n_C - \alpha)
 \end{bmatrix}
 \begin{bmatrix} \tanh(g x_{E_1}) \\\tanh ( g x_{E_2} ) \\\tanh(g x_{I})\end{bmatrix},
 \end{aligned}
\end{equation}
where we used $\alpha n_I = n_E = p n_C$. The variables $x_{E_1}$ and $x_{E_2}$ are the activities of the two groups of excitatory clusters, and $x_I$ is the activity of the inhibitory cells, which are synchronized since $g$ is close to $g_C$. 
The system \cref{eq:cluster3system} is the restriction of \cref{eq:reducedmatrixform} to the fixed-point subspace corresponding to the subgroup $S_{n_{C_1}} \times \, S_{n_{C_2}} \times \, S_{n_I}$ of $\Gamma$.
Following the same procedure as in \cref{sec:E1I1}, we obtain the following approximations for $x_{E_1}$, $x_{E_2}$, and $x_I$
\begin{align}\label{eq:XE1}
x_{E_1} &= \pm \sqrt{ \frac{ 3(g - g_C) }{ (1 - \beta_C + \beta_C^2 )g^3}} + \mathcal{O}\left( \frac{1}{N^2}\right), \quad
x_{E_2} = -\beta_C x_{E_1} + \mathcal{O}\left( \frac{1}{N^2} \right), \quad
x_I = \mathcal{O}\left( \frac{1}{N^2} \right)
&& g \geq g_C,
\end{align}
for $g$ close to $g_C$, which reduces to \cref{eq:xEapprox} when $\beta = 1$.

\subsection{Stability and bifurcations along \texorpdfstring{$C_1/C_2$}{C1/C2} branch}\label{sec:C1C2stability}

\begin{figure}
    \centering
    \includegraphics[width=6cm]{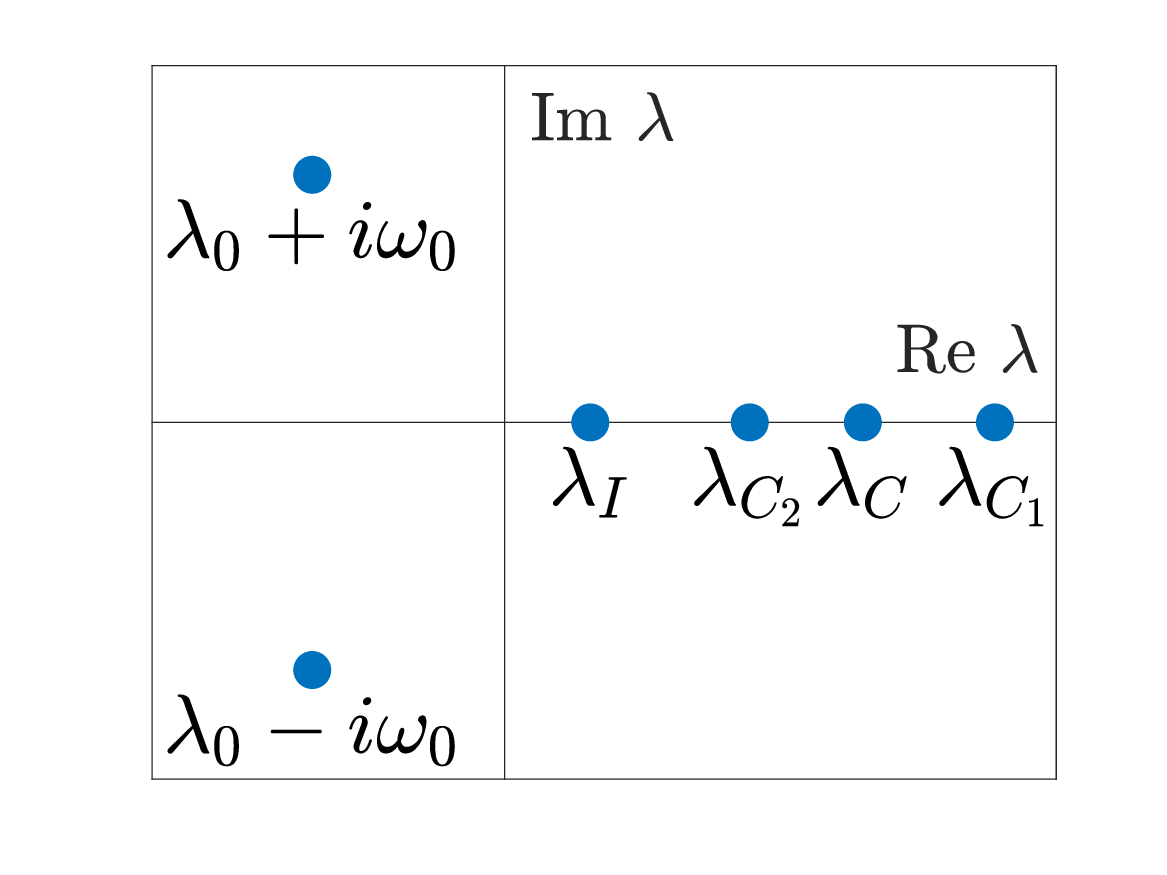}
    \caption{The eigenvalue pattern of the connectivity matrix $H(\xvec^*)$ for fixed points $\xvec^*$ on a $C_1/C_2$ branch with $\beta_C > 1$. The notation for the eigenvalues is explained below \cref{prop:H3Ceig}.}
    \label{fig:HstareigEcluster}
\end{figure}

We now analyze the stability of the $C_1/C_2$ branches for $g$ close to $g_C$. Choose any $\beta_C \geq 1$, so that $n_{C_1} = \frac{\beta_C}{\beta_C+1}n_C$ and $n_{C_2} = \frac{1}{\beta_C+1}n_C$. Let $\xvec = (x_{E_1}, x_{E_2}, x_{I})$ be a solution to \cref{eq:cluster3system}. We look at the linearization $D\tilde{F}(\xvec^*)$, where $\xvec^* = (x_{E_1}, \dots, x_{E_1}, x_{E_2}, \dots, x_{E_2}, x_{I}, \dots, x_{I})^T$, where $x_{E_1}$ and $x_{E_2}$ are repeated $n_{C_1}$ and $n_{C_2}$ times, respectively, and $x_I$ is repeated $n_I$ times. Stability will depend on the eigenvalues of $\tilde{H}(\xvec^*)$. A cartoon showing the location of these eigenvalues is given in \cref{fig:HstareigEcluster}. 

We follow the same procedure as in \cref{sec:I1I2stability}. First, we linearize the reduced system \cref{eq:cluster3system} about $(x_{E_1}, x_{E_2}, x_{I})$ to get the Jacobian
\begin{equation}\label{eq:J3C}
J_3(\xvec) = \frac{g}{\sqrt{N}} H_3(\xvec) - I_3,
\end{equation}
where 
\begin{equation}\label{eq:H3C}
H_3(\xvec) = \mu
\begin{bmatrix} 
    (p-1)n_C \sech^2(g x_{E_1}) & 0 & -p n_C \sech^2(g x_{I}) \\
    0  & (p-1)n_C \sech^2(g x_{E_2}) & -p n_C \sech^2(g x_{I}) \\
    p n_C \frac{\beta_C}{\beta_C+1} \sech^2(g x_{E_1}) &
    p n_C \frac{1}{\beta_C+1} \sech^2(g x_{E_2}) &
    -(p n_C - \alpha) \sech^2(g x_{I})
 \end{bmatrix}
\end{equation}
and $I_3$ is the $3 \times 3$ identity matrix. We have the following proposition concerning the eigenvalues of $H_3(\xvec)$ and $\tilde{H}(\xvec^*)$. The proof is omitted since it is similar to that of \cref{prop:H3eig}.

\pagebreak

\begin{proposition}\label{prop:H3Ceig}
Let $\xvec = (x_{E_1}, x_{E_2}, x_{I})$ be a solution to \cref{eq:cluster3system} and $\xvec^*$ the corresponding fixed point of \cref{eq:reducedmatrixform}, and let $H_3(\xvec)$ and $\tilde{H}(\xvec^*)$ be defined by \cref{eq:H3C} and \cref{eq:tildeHxstar}. Then
\begin{compactenum}[(i)]
    \item Every eigenvalue of $H_3(\xvec)$ is an eigenvalue of $\tilde{H}(\xvec^*)$.
    \item $\tilde{H}(\xvec^*)$ has the following additional eigenvalues:
    \begin{itemize}
        \item $\lambda_{C_1} := (p-1) n_C \mu \sech^2(g x_{E_1})$ with multiplicity $n_{C_1} - 1$.
        \item $\lambda_{C_2} := (p-1) n_C \mu \sech^2(g x_{E_2})$ with multiplicity $n_{C_2} - 1$.
        \item $\lambda_{I} := \alpha \mu \sech^2(g x_{I})$ with multiplicity $n_{I} - 1$.
    \end{itemize}
\end{compactenum}
\end{proposition}

\begin{figure}
    \centering
    \includegraphics[width=7.8cm]{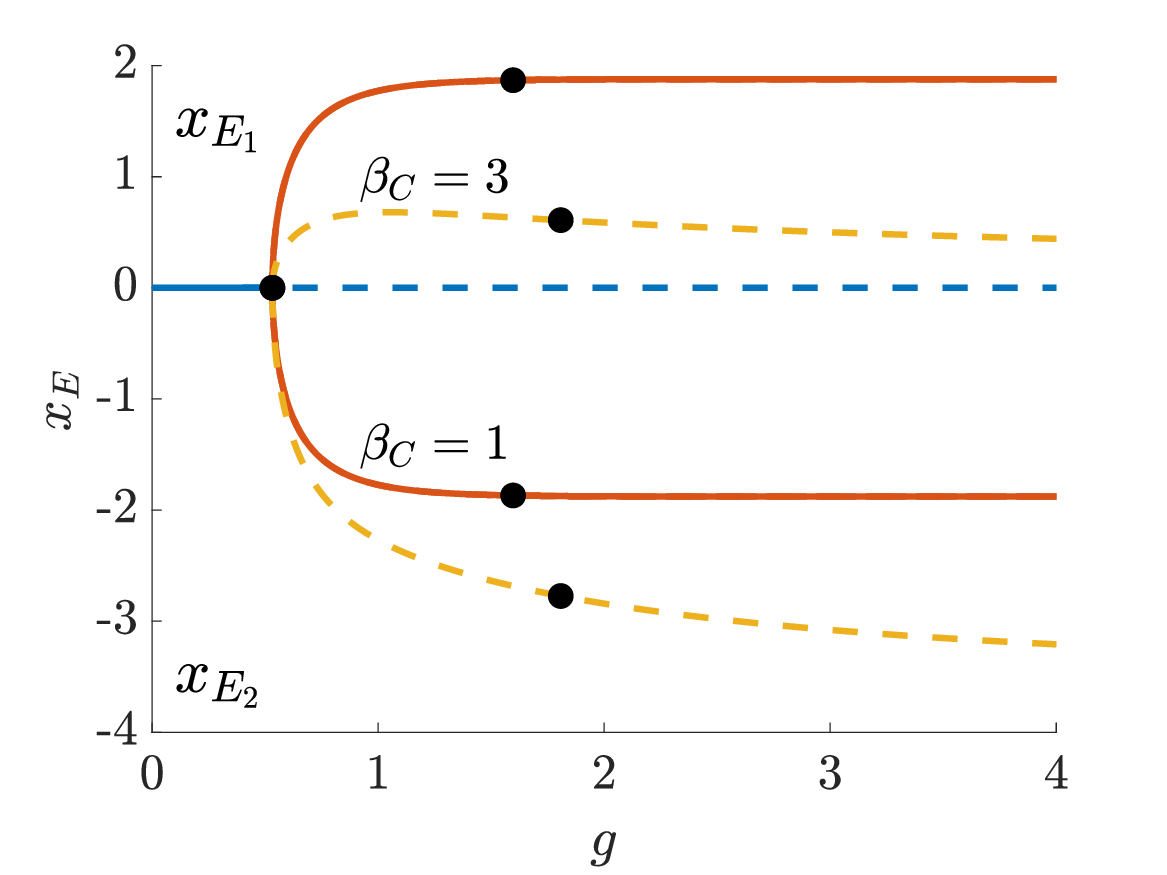}\hspace{-0.5cm}
    \vspace{-0.5cm}
    \includegraphics[width=9cm]{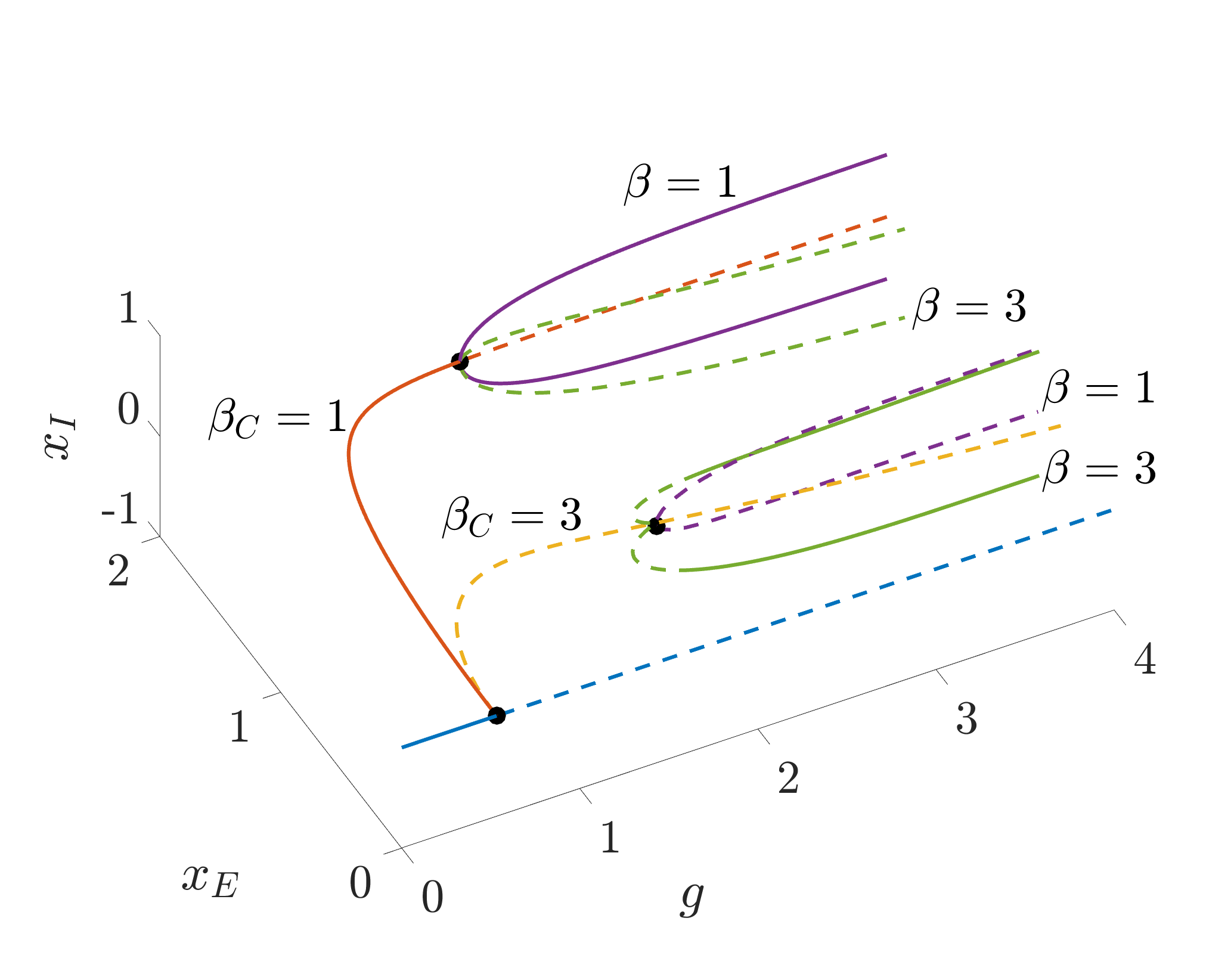} 
    \caption{When the excitatory cells are clustered ($n_C>1$), the first nontrivial fixed points are those for which the excitatory cells, rather than inhibitory cells, separate into two groups. Left: excitatory cell activity $x_{E_1}$ and $x_{E_2}$ on $C_1/C_2$ branches of equilibria of \cref{eqn:sys_Basic} with excitatory clustering for all possible values of $\beta_C$. The symmetric pitchfork bifurcations at $g = g_C$ and along the $C_1/C_2$ branches are indicated with filled circles. (To avoid clutter, the $I_1/I_2$ branches after the symmetric pitchfork bifurcation on the $C_1/C_2$ branch are not shown). Right: $I_1/I_2$ branches bifurcate from the $C_1/C_2$ branches ( to avoid clutter only $x_{E_1}$ is shown). Stable fixed points are indicated with solid lines. Parameters are: $N = 20$, $n_C = 4$, $p = 4$, $n_I = 4$, $\alpha = 4$, $\mu_{EE} = 0.7$.}
    \label{fig:clusterBD1}
\end{figure}

\begin{figure}
    \centering
    \includegraphics[width=8.25cm]{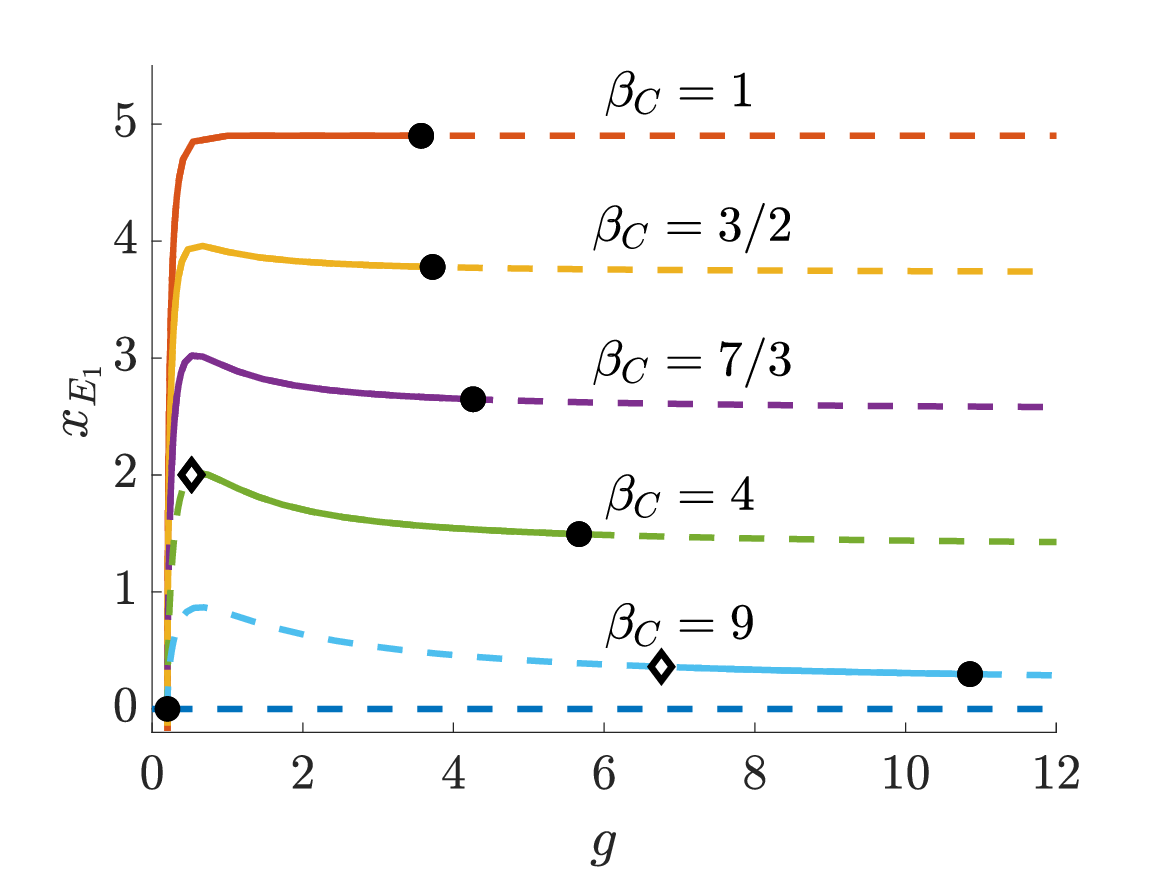}\hspace{-0.5cm}
    \includegraphics[width=8.25cm]{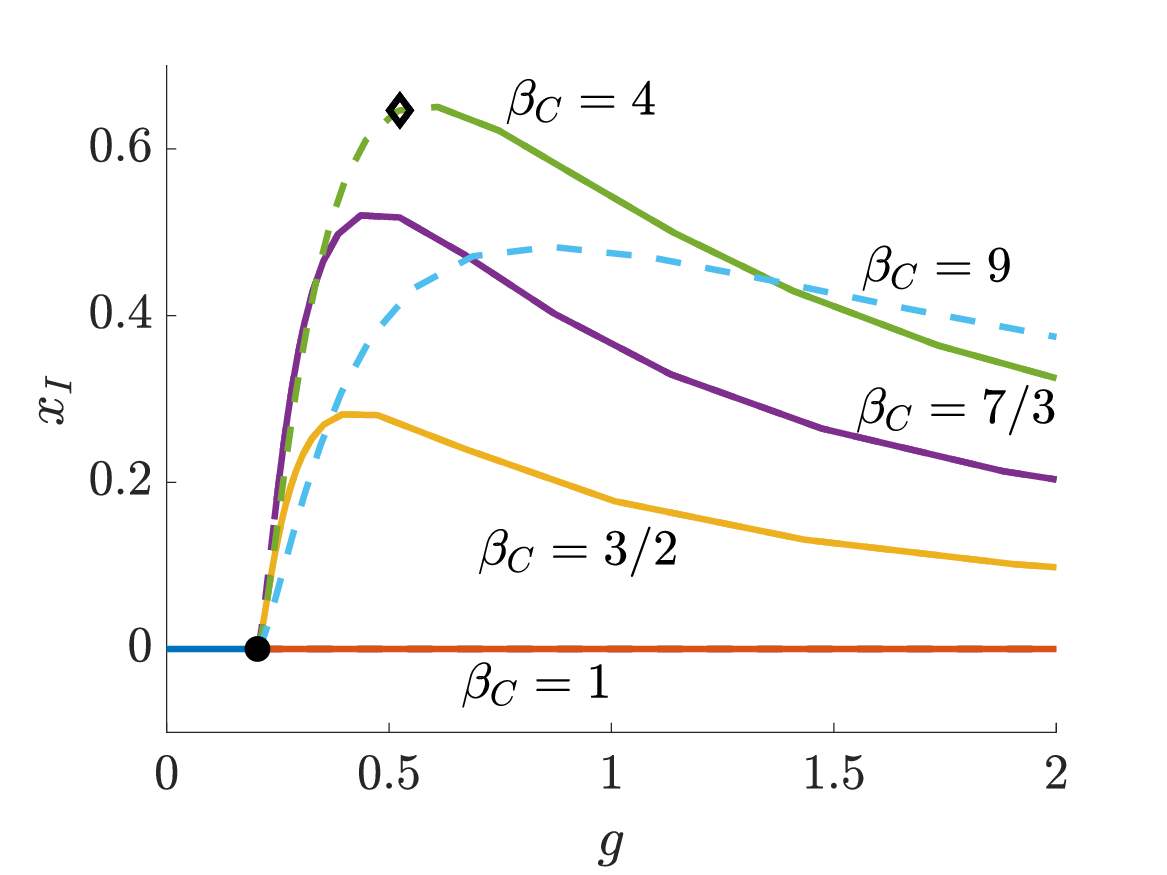} \\
    \vspace{-1cm}
    \hspace{-1cm} 
    \includegraphics[width=17.25cm]{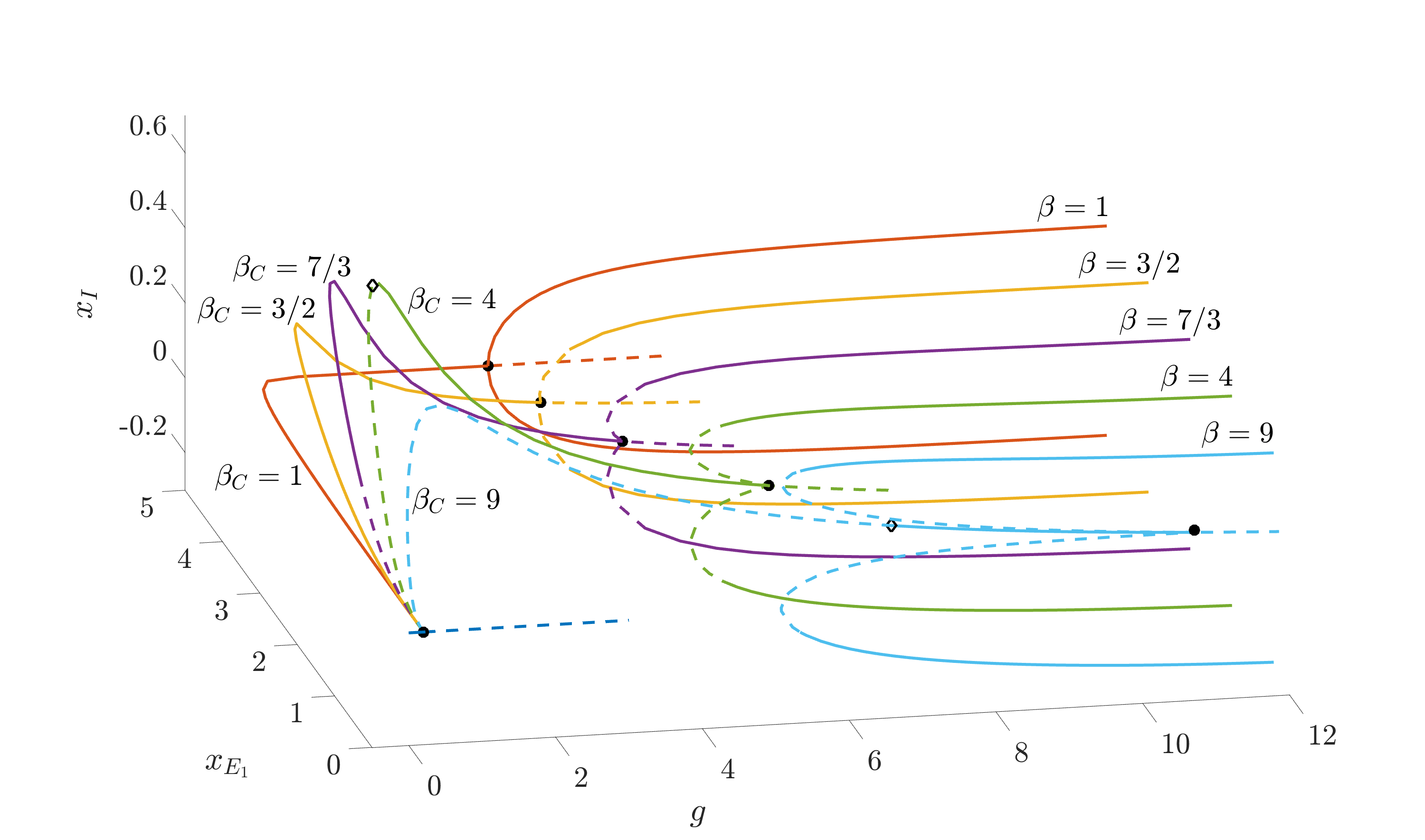}
    \caption{Bifurcation diagram of all possible $C_1/C_2$ branches, and selected $C_1/C_2/I_1/I_2$ branches, for a moderate value of $N$. Top: $C_1/C_2$ branches of equilibria of \cref{eqn:sys_Basic} with excitatory clustering for all possible values of $\beta_C$. Top left: $x_{E_1}$ vs $g$. Top right: $x_I$ vs $g$, zoomed into a narrower range of $g$ to show stability of $C_1/C_2$ branches near $g = g_C$. Symmetric pitchfork bifurcations at $g = g_C$ and along the $C_1/C_2$ branches are indicated with filled circle. To avoid clutter, the $C_1/C_2/I_1/I_2$ branches are not shown. Bottom: $x_I$ and $x_{E_1}$ vs. $g$, for $C_1/C_2/I_1/I_2$ branches bifurcating from the $C_1/C_2$ branches. The only $I_1/I_2$ branches shown here are the ones which are eventually stable, which in this case are those with $\beta = \beta_C$ (see \cref{table:validbeta}). Stable fixed points are indicated with solid lines, unstable fixed points with dashed line. Unstable $C_1/C_2$ branches for $\beta_C = 4$ and $\beta_C = 9$ become stable at the points indicated with the diamond. Parameters are: $N = 100$, $n_C = 10$, $p = 8$, $n_I = 20$, $\alpha = 4$, $\mu_{EE} = 0.7$.}
    \label{fig:clusterBD2}
\end{figure}

We note that the eigenvalues $\lambda_{C_1}$ and $\lambda_{C_2}$ split off from $\lambda_C$ at the pitchfork bifurcation point $g = g_C$; if $\xvec^* = 0$, then $\lambda_{C_1} = \lambda_{C_2} = \lambda_C$. To determine the stability of $\xvec^*$ for $g$ close to $g_C$, we first compute the eigenvalues of $D\tilde{F}(\xvec^*)$ corresponding to $\lambda_{C_1}$, $\lambda_{C_2}$, and $\lambda_I$. We find (see \cref{app:C1C2stability}) that the cluster-associated eigenvalue $\lambda_{C_1}^*(g)$ is negative for $1 \leq \beta_C < 2$ and positive for $\beta_C > 2$; $\lambda_{C_2}^*(g)$ is negative for $\beta_C  > 1/2$; and $\lambda_{I}^*(g)$ is negative for all $\beta_C$ for $N$ sufficiently large. The behavior of $\lambda_{C_1}^*(g)$ implies that the $C_1/C_2$ branches are initially unstable for $\beta_C > 2$ (see \cref{fig:clusterBD1} and \cref{fig:clusterBD2}). 

The remaining eigenvalues of $D\tilde{F}(\xvec^*)$ are the eigenvalues of $J_3(\xvec)$. Following the same procedure as in \cref{sec:I1I2stability} (see \cref{app:C1C2stability} for details), we find that, since we are taking $n_C \geq \alpha$, the eigenvalues of $J_3(\xvec)$ all have negative real part for $g$ close to $g_C$. Thus the $C_1/C_2$ branches are initially stable for $1 \leq \beta_C \leq 2$ (see the top panel of \cref{fig:clusterBD1} as well as \cref{fig:clusterBD2}).

As $g$ is further increased from $g_C$, there is a second symmetric pitchfork bifurcation on each $C_1/C_2$ branch as the eigenvalue $\lambda_I^*(g)$ of $D\tilde{F}(\xvec^*)$ with multiplicity $n_I-1$ crosses through the origin (see bifurcation diagram in \cref{fig:clusterBD1} and \cref{fig:clusterBD2}). 

The behavior at this bifurcation is exactly the same as for the second symmetric pitchfork bifurcation at the origin. The corresponding eigenspace $V$ is the set of all zero-sum vectors with support in the inhibitory cells only, which has dimension $n_I-1$. As above, we break the inhibitory cells up into two groups $I_1$ and $I_2$ of sizes $n_{I_1}$ and $n_{I_2}$, where $n_{I_1} + n_{I_2} = n_I$.
This describes a subgroup $\Sigma_I = S_{n_{C_1}} \times S_{n_{C_2}} \times S_{n_{I_1}} \times S_{n_{I_2}}$ of $\Gamma$, where we recall that $n_{C_1}$ and $n_{C_2}$ are fixed on this $C_1/C_2$ branch. The fixed-point subspace of $\Sigma_I$ is then given by
\begin{eqnarray}
\Fix_V(\Sigma_I) & = & {\rm span} \, \left\{ \left[ 
\underbrace{\begin{matrix}0 & \cdots & 0\end{matrix}}_{n_{C_1}} \;
\underbrace{\begin{matrix}0 & \cdots & 0\end{matrix}}_{n_{C_2}} \;
\underbrace{\begin{matrix}1 & \cdots & 1\end{matrix}}_{n_{I_1}} \;
\underbrace{\begin{matrix}-\frac{n_{I_1}}{n_{I_2}} & \cdots & -\frac{n_{I_1}}{n_{I_2}} \end{matrix}}_{n_{I_2}} \right] \right\},
\end{eqnarray}
which has dimension 1. It follows from the Equivariant Branching Lemma that, on every $C_1/C_2$ branch, there is an $I_1/I_2$ branch of solutions for every possible division of the inhibitory cells into exactly two clusters.

We can characterize these branches using the parameter $\beta = n_{I_1}/n_{I_2}$, as we did in the previous section. When $\beta_C = 1$, $x_I = 0$, and this bifurcation takes place at 
\begin{equation}\label{eq:gpitchinhbeta1}
g_I = \frac{\sqrt{N}}{\alpha \mu}.
\end{equation}
For $\beta_C > 1$, this bifurcation takes place at $g$ much greater than $g_C$, thus the approximation \cref{eq:XE1} no longer holds. To locate these bifurcations, we will examine the behavior of the system as $g$ becomes large. We note here that evidence from numerical parameter continuation suggests that there are no Hopf bifurcations along the $C_1/C_2$ branches; furthermore, numerical timestepping experiments suggest that there are no stable periodic orbits for any value of $g$. In addition, numerical experiments strongly suggest that there are no stable equilibria on any secondary branches.

\subsection{\texorpdfstring{$C_1/C_2$}{C1/C2} branches for large \texorpdfstring{$g$}{g}} 

\begin{figure}
    \centering
    \includegraphics[width=8.25cm]{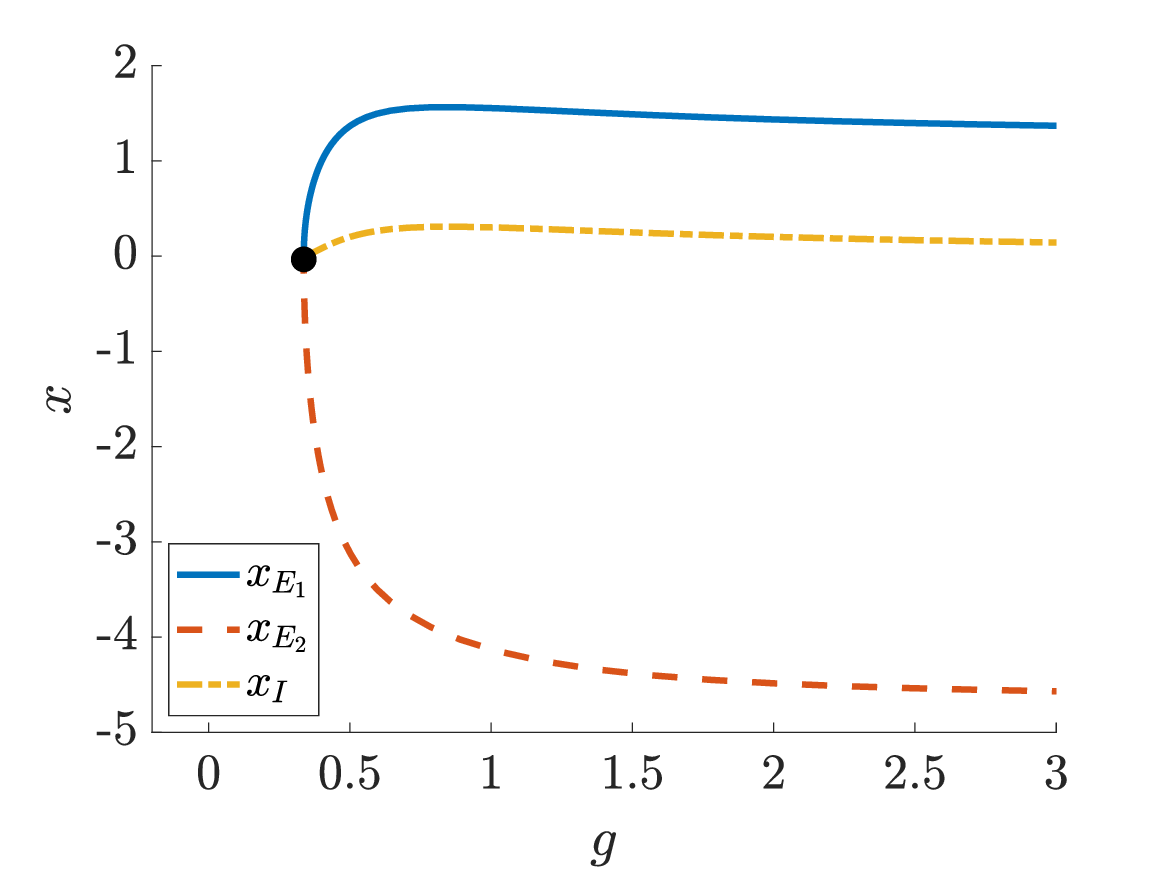}\hspace{-0.5cm}
    \includegraphics[width=8.25cm]{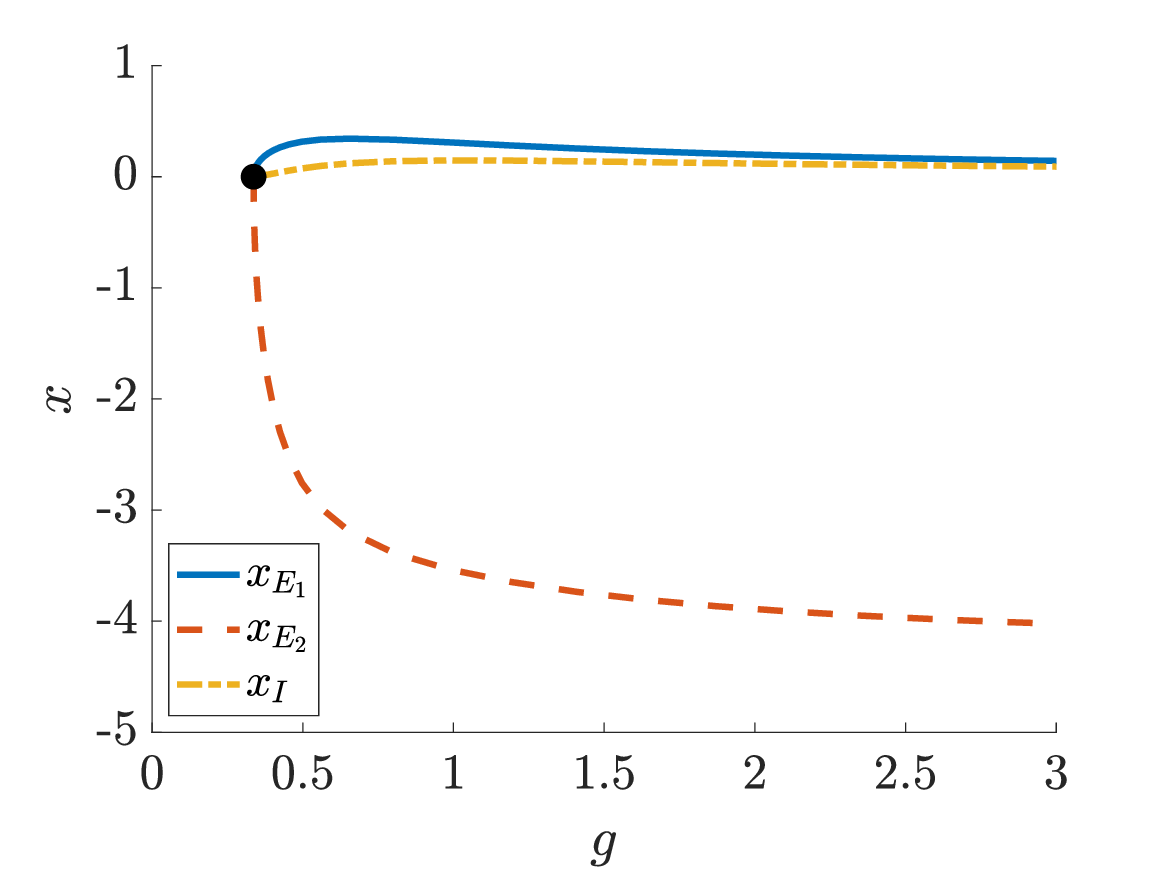}
    \caption{The saturation ($g \gg 1$) behavior of fixed points on a $C_1/C_2$ branch depends on the clustering parameter $\beta_C$. Left: $x_{E_1}$, $x_{E_2}$, and $x_I$ vs $g$ on $C_1/C_2$ branches for $1 < \beta_C < \beta_C^*$. Right: $\beta_C > \beta_C^*$. Parameters are: $N = 50$, $n_C = 10$, $p = 4$, $n_I = 10$, $\alpha =4$, $\mu = 0.7$. Given these parameters $\beta_C^* = 5.15385$: here we illustrate $\beta_C = 7/3$ (left) and $\beta_C = 9$ (right).}
    \label{fig:betacstar}
\end{figure}

We look at the behavior of solutions on the $C_1/C_2$ branches as $g$ becomes large. This will depend on the ratio $\beta_C = n_{C_1}/n_{C_2}$. When $\beta_C = 1$, $x_{E_2} = -x_{E_1} := x_E$, and $x_I = 0$ for all $g \geq g_C$. Numerical parameter continuation suggests that $x_{E} \rightarrow \hat{x}_{E} > 0$ as $g \rightarrow \infty$, which implies that $\tanh(g x_{E}) \rightarrow 1$. It follows from the first row of \cref{eq:cluster3system} that 
\begin{equation}\label{eq:xEhat}
\hat{x}_{E} = \frac{\mu}{\sqrt{N}}(p-1)n_C.
\end{equation}
For $\beta_C > 1$, numerical parameter continuation suggests $x_I \rightarrow 0$ as $g \rightarrow \infty$, but $\tanh(g x_I) \rightarrow \hat{y}_I \neq 0$. There are two patterns for the limiting behavior on the $C_1/C_2$ branches, which depend on whether $\beta_C < \beta_C^*$ or $\beta_C > \beta_C^*$, for a critical value
\begin{equation}\label{eq:betaCstar}
    \beta_C^* = \frac{(n_C p - \alpha )(2 p - 1) + \alpha p}{n_C p + \alpha(p-1)}.
\end{equation}
(See \cref{app:C1C2_largeg} for a derivation of $\beta_C^*$). These are illustrated in \cref{fig:betacstar}.
\begin{itemize}
    \item Case 1: ($1 < \beta_C < \beta_C^*$) $x_{E_1} \rightarrow \hat{x}_{E_1} > 0$ and $x_{E_2} \rightarrow \hat{x}_{E_2} < 0$. 
    \item Case 2: ($\beta_C > \beta_C^*$) $x_{E_1} \rightarrow 0$ with $\tanh(g x_{E_1}) \rightarrow \hat{y}_{E_1} \neq 0$, and $x_{E_2} \rightarrow \hat{x}_{E_2} < 0$.\\
\end{itemize}
As $N \rightarrow \infty$, $n_C p = f N \rightarrow \infty$ as well. If both $p$ and $n_C$ scale as $\sqrt{N}$, then the only significant terms in the numerator and denominator of \cref{eq:betaCstar} are of order $N$ or larger, in which case $\beta_C^* \rightarrow 2 p-1$ as $N \rightarrow \infty$. 

\begin{figure}
    \centering
    \includegraphics[width=8.25cm]{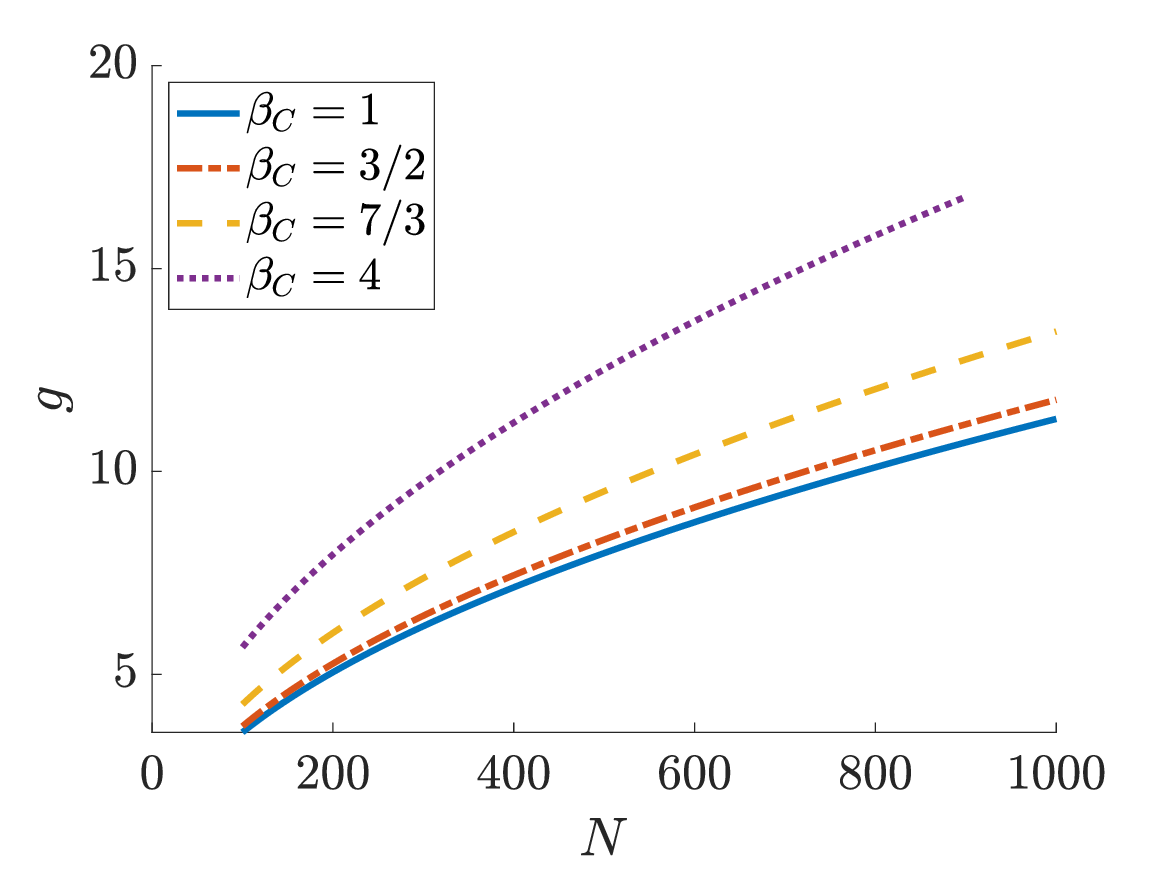}\hspace{-0.5cm}
    \includegraphics[width=8.25cm]{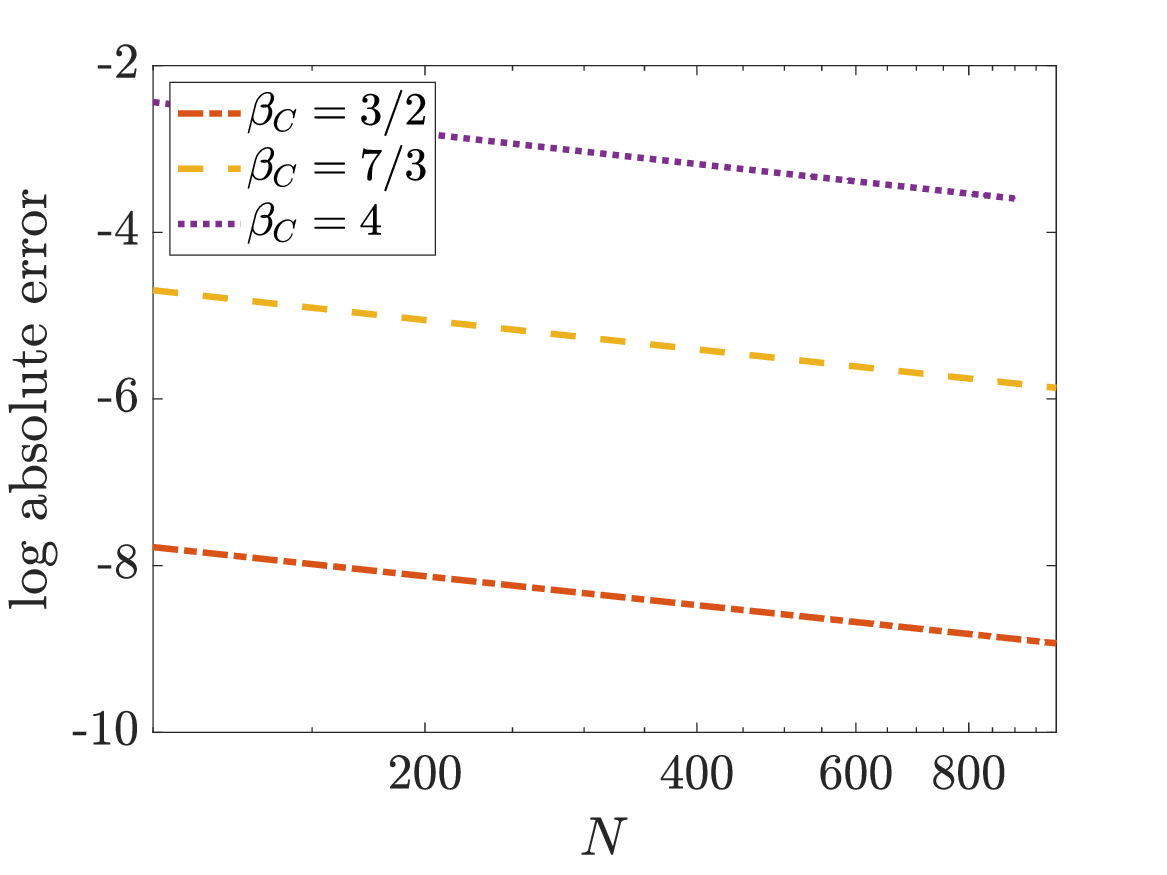}
    \caption{Location of the symmetric pitchfork bifurcation points on $C_1/C_2$ branches. Left:  $g_I(\beta_C)$ vs. $N$ for various $\beta_C$. Right: semi-log plot of the absolute error of approximation \cref{eq:gbetaC} vs $N$ for various $\beta_C$. The slope of each line is approximately -0.5 (validating the error term $\mathcal{O}(N^{-1/2})$). Parameters are: $n_C = 10$, $\alpha = 4$, $\mu = 0.7$.}
    \label{fig:pitcherror}
\end{figure}

In \cref{app:C1C2_largeg}, we derive formulas for $x_{E_1}$, $x_{E_2}$, and $x_{I}$ for both of these cases. We then use these formulas to find the location of the symmetric pitchfork bifurcation points on the $C_1/C_2$ branches when $\beta_C>1$ and $N$ is large. If we take both $p$ and $n_C$ to scale as $\sqrt{N}$, we can assume $\beta_C < \beta_C^*$, as discussed above. At this bifurcation, the eigenvalue $\lambda_I^*(g)$ of $D\tilde{F}(\xvec^*)$ with multiplicity $n_I-1$ crosses through 0. Using the identity $\sech^2(g x_{I}) = 1 - \tanh^2(g x_{I}) \rightarrow 1 - \hat{y}_I^2$ as $g \rightarrow \infty$ together with \cref{eq:yihat}, the symmetric pitchfork bifurcation on the $C_1/C_2$ branch is located, to leading order, at 
\begin{align}\label{eq:gbetaC}
    g_I(\beta_C) &= \frac{\sqrt{N}}{4 \alpha \mu} \frac{(1 + \beta_C)^2}{\beta_C} 
\end{align}
for $N$ large. When $\beta_C = 1$, this reduces to \cref{eq:gpitchinhbeta1}. See \cref{fig:clusterBD2} and the left panel of \cref{fig:pitcherror} for the location of the symmetric pitchfork bifurcations on the $C_1/C_2$ branches. Numerical simulation validates this formula, and suggests that the error term in \cref{eq:gbetaC} has order $\mathcal{O}(N^{-1/2})$ (\cref{fig:pitcherror}, right panel). We note that for $N$ large, $g_I(\beta_C)$ is quadratic in $\beta_C$, has a local minimum at $\beta_C = 1$, and is increasing for $\beta_C > 1$. We can see in \cref{fig:clusterBD2} and \cref{fig:pitcherror} that the location of the symmetric bifurcation points $g_I({\beta_C})$ increases with $\beta_C$.

For sufficiently large $N$, each $C_1/C_2$ branch will be stable immediately preceding the pitchfork bifurcation at $g_I(\beta_C)$. To see this, we evaluate the remaining eigenvalues of $D\tilde{F}(\xvec^*)$ when $g = g_I(\beta_C)$. As $N \rightarrow \infty$, $g_I(\beta_C) \rightarrow \infty$, thus $\sech(g_I(\beta_C) x_{E_j}) \rightarrow 0$ for $j = 1, 2$. It follows that for $g = g_I(\beta_C)$, $\lambda_{C_j} \rightarrow 0$, thus $\lambda_{C_j}^*(g) \rightarrow -1$ for $j = 1, 2$. By the same argument, taking $N \rightarrow \infty$ will zero out the first two columns of \cref{eq:H3C} when $g = g_I(\beta_C)$. Thus, in the limit $N \rightarrow \infty$, $H_3(\xvec^*)$ will have a pair of eigenvalues at 0 and an additional eigenvalue at $-(p n_C - \alpha) \sech^2 g x_I \leq 0$. The corresponding eigenvalues of $D\tilde{F}(\xvec^*)$ will be negative. 

As a example, consider the $N=100$ system shown in \cref{fig:clusterBD2}. The $C_1/C_2$ branches for $\beta_C = 7/3$, 4, and 9 start unstable, but regain stability before the symmetric pitchfork bifurcation points. This does not necessarily occur for small values of $N$ (see \cref{fig:clusterBD1} for $N=20$, where this does not happen).

\subsection{Stability of \texorpdfstring{$C_1/C_2/I_1/I_2$}{C1/C2/I1/I2} solutions for large \texorpdfstring{$g$}{g}}\label{sec:C1C2I1I2_largeg}

After the symmetric pitchfork bifurcation point on the $C_1/C_2$ branches, both the excitatory clusters and inhibitory cells have split into two populations. We are interested in stable fixed points when $g$ is large. In particular, we seek fixed point branches in which the excitatory clusters are split into two populations with ratio $\beta_C = n_{C_1}/n_{C_2}$, and the inhibitory cells are also split into two populations with ratio $\beta = n_{I_1}/n_{I_2}$. This reduces \cref{eqn:sys_Basic} to the system of equations
\begin{equation}\label{eq:cluster4system}
    \begin{aligned}
    \begin{bmatrix} x_{E_1} \\ x_{E_2} \\ x_{I_1} \\ x_{I_2} \end{bmatrix} 
    &= \frac{\mu}{\sqrt{N}} 
    \begin{bmatrix} 
       (p-1)n_C & 0 & -\alpha \frac{\beta}{\beta+1}n_I &  -\alpha \frac{1}{\beta+1}n_I \\
       0  & (p-1)n_C & -\alpha \frac{\beta}{\beta+1}n_I &  -\alpha \frac{1}{\beta+1}n_I \\
       p n_C \frac{\beta_C}{\beta_C+1} &
       p n_C \frac{1}{\beta_C+1} &
       -\alpha \left(\frac{\beta}{\beta+1}n_I-1\right) &  -\alpha \frac{1}{\beta+1}n_I \\
       p n_C \frac{\beta_C}{\beta_C+1} &
       p n_C \frac{1}{\beta_C+1} &
       -\alpha \frac{\beta}{\beta+1}n_I & -\alpha \left(\frac{1}{\beta+1}n_I - 1 \right)
    \end{bmatrix}
    \begin{bmatrix} \tanh(g x_{E_1}) \\ \tanh ( g x_{E_2} ) \\\tanh(g x_{I_1}) \\\tanh(g x_{I_2})  \end{bmatrix},
    \end{aligned}
\end{equation}
which is the restriction of \cref{eq:reducedmatrixform} to the fixed-point subspace corresponding to the subgroup $S_{n_{C_1}} \times \, S_{n_{C_2}} \times \, S_{n_{I_1}} \times \, S_{n_{I_2}}$ of $\Gamma$.
Parameter continuation suggests that as $g \rightarrow \infty$, $(x_{E_1}, x_{E_2}, x_{I_1}, x_{I_1}) \rightarrow (\hat{x}_{E_1}, \hat{x}_{E_2}, \hat{x}_{I_1}, \hat{x}_{I_2})$, where $\hat{x}_{E_1}, \hat{x}_{I_1} > 0$ and $\hat{x}_{E_2}, \hat{x}_{I_2} < 0$. Such solutions exist for 
\begin{equation}\label{eq:bccondition}
    \frac{2 \beta n_I - \beta - 1}{2 n_I + \beta + 1} < \beta_C < \frac{2 \beta n_I + \beta + 1}{2 n_I - \beta - 1}.
\end{equation}
for all valid $\beta$ satisfying $1 \leq \beta < 2p-1$. See \cref{app:C1C2I1I2_largeg} for detailed calculations.

For some small values of $N$, a list of all valid pairs of $(\beta, \beta_C)$ which satisfy \cref{eq:bccondition} is given in \cref{table:validbeta}. (A value of $\beta$ or $\beta_C$ is valid for a particular $N$ only if the ratio of inhibitory cells or excitatory clusters is possible for that value of $N$). In the specific case where $n_C = n_I$, it follows from \cref{eq:bccondition} that 
\[
n_{I_1} - \frac{1}{2} < n_{C_1} < n_{I_1} + \frac{1}{2}.
\]
Since $n_{C_1}$ must be an integer, $n_{C_1} = n_{I_1}$, which implies $\beta_C = \beta$.

The fixed point $\xvec^*$ corresponding to each of these $(\beta, \beta_C)$ is eventually stable for sufficiently large $g$, since as $g\rightarrow \infty$, $H(\xvec^*)$ approaches the $\Zerovec$ matrix, thus the Jacobian $DF(\xvec^*)$ approaches $-I$, which has a single eigenvalue of $-1$ with multiplicity $N$. The solutions corresponding to the top row of \cref{table:validbeta} are shown in the bottom panel of \cref{fig:clusterBD1}, and the solutions corresponding to the bottom row are shown in the bottom panel of \cref{fig:clusterBD2}; we can see from the figures that the corresponding fixed points are all stable for sufficiently large $g$. Numerical experiments strongly suggest that there are no stable equilibria for large $g$ other than these.

\begin{table}
\centering
    \begin{tabular}{lllll}
        \toprule
        $N$ & $n_I$ & $n_C$ & $p$ & $(\beta, \beta_C)$ \\
        \midrule
        20 & 4 & 4 & 4 & (1, 1), (3, 3) \\
        25 & 5 & 5 & 4 & (3/2, 3/2), (4, 4) \\
        25 & 5 & 4 & 5 & (4, 3) \\
        35 & 7 & 7 & 4 & (4/3, 4/3), (5/2, 5/2), (6, 6) \\
        35 & 7 & 4 & 7 & (5/2, 3) \\
        50 & 10 & 10 & 4 & (1, 1), (3/2, 3/2), (7/3, 7/3), (4, 4) \\
        100 & 20 & 10 & 8 & (1, 1), (3/2, 3/2), (7/3, 7/3), (4, 4), (9, 9) \\
        \bottomrule
    \end{tabular}
    \vspace{0.25cm}
    \caption{Valid pairs $(\beta, \beta_C)$ which satisfy \cref{eq:bccondition}, for selected values of $N$, $n_C$ and $n_I$. (Note that $\alpha = 4$ in all cases, which determines $n_I$ and $p n_C$).}
    \label{table:validbeta}
\end{table}

\subsection{Excitatory clusters with weight parameters unchanged}

We briefly consider a system with excitatory clusters, but in which we have not adjusted the excitatory weight strengths, i.e. $\mu_{EI} = -\alpha \mu_{EE}$, $\mu_{II} = -\alpha \mu_{EE}$, and $\mu_{IE} = \mu_{EE}$. In this case, the two eigenvalues of $\tilde{H}$ with positive real part are $\lambda_I = \alpha \mu_{EE}$ and $\lambda_C = (p-1)\mu_{EE}$. If $\lambda_C > \lambda_I$, which occurs when $n_C < \frac{f N}{\alpha+1}$, the behavior is qualitatively the same as for the case balanced weight parameters discussed above. If  $\lambda_C < \lambda_I$, which occurs when $n_C > \frac{f N}{\alpha+1}$, the order of the two symmetric pitchfork bifurcations is reversed. As $g$ is increased, the inhibitory cells bifurcate from the origin first, followed by the excitatory clusters.

\subsection{Restored self-coupling} \label{sec:restore_selfCoup_Eclusters}
We can restore self-coupling of neurons with each excitatory cluster by replacing the matrix $(p-1) \mu_{EE} I_{n_C}$ in the upper left block of \cref{eq:tildeH} with $p \mu_{EE} I_{n_C}$. The eigenvalues of $\tilde{H}$ are then given by:
\begin{itemize}
\item $\lambda_I := \alpha \mu > 0$ with multiplicity $n_I - 1$
\item $\lambda_C := p n_C \mu > 0$, with multiplicity $n_C - 1$.
\item A complex conjugate pair of eigenvalues $\lambda_0 \pm i \omega_0$, with 
\begin{equation*}
    \lambda_0 := \frac{1}{2}\mu \alpha, \quad 
    \omega_0 := \frac{1}{2}\mu \sqrt{ \alpha( 4 n_C p - \alpha ) },
\end{equation*}
\end{itemize}
The eigenvalue pattern is similar to that in the right panel of \cref{fig:Heigpattern}, except the complex conjugate pair $\lambda_0 \pm i \omega_0$ has positive real part. As a consequence, there will be a Hopf bifurcation at the origin at $g_H = \sqrt{N}/\alpha \mu$. Parameter continuation with AUTO indicates that the resulting limit cycle has all excitatory clusters synchronized and all inhibitory cells synchronized, and is unstable for $g > g_H$. In addition, timestepping simulations suggest that there are no stable limit cycles for any value of $g$. The pattern of symmetric pitchfork bifurcations, first at the origin and then on each $C_1/C_2$ branch, is the same as for the case with no self-coupling.

\section{Inhibitory clusters}\label{sec:inhibitoryclusters}

We will briefly consider the case where the inhibitory cells are clustered, while the excitatory cells remain unclustered. Suppose the inhibitory cells are grouped into $n_{C_I}$ inhibitory clusters of size $p_I$, so that $n_I = n_{C_I} p_I$. We perform the same reduction as in \cref{sec:simplermodel} to obtain the matrix $\tilde{H}$.
Since there is a single cluster of excitatory cells, they will always be synchronized. For the choice of weights $\mu_{EI} = -\alpha \mu_{EE}$, $\mu_{II} = -\alpha \mu_{EE}$, and $\mu_{IE} = \mu_{EE}$, the eigenvalues of $\tilde{H}$ are:
\begin{itemize}
\item $\lambda_I := \alpha \mu_{EE} > 0$ with multiplicity $(p_I-1) \times n_{C_I} = n_I - n_{C_I}$.
\item $\lambda_{C_I} := -(p_I-1)\alpha \mu_{EE} < 0$, with multiplicity $n_{C_I}-1$.
\item A complex conjugate pair of eigenvalues $\lambda_0 \pm i \omega_0$, with 
\begin{align*}
    \lambda_0 &:= \frac{1}{2}\mu_{EE} \left[ \alpha( 1 + p_I(n_{C_I}-1)) -1 \right]
      \\
    \omega_0 &:= \sqrt{a^2 \left(\left(-3 n_{C_I}^2+2 n_{C_I}+1\right) p_I^2-2 (n_{C_I}+1) p_I+1\right)-2 a
    (n_{C_I}p_I +p_I -1)+1},
\end{align*}
where we used the fact that $n_E = \alpha n_{C_I} p_I$.
\end{itemize}

\begin{figure}
    \centering
    \begin{tabular}{c}
    \includegraphics[width=6cm]{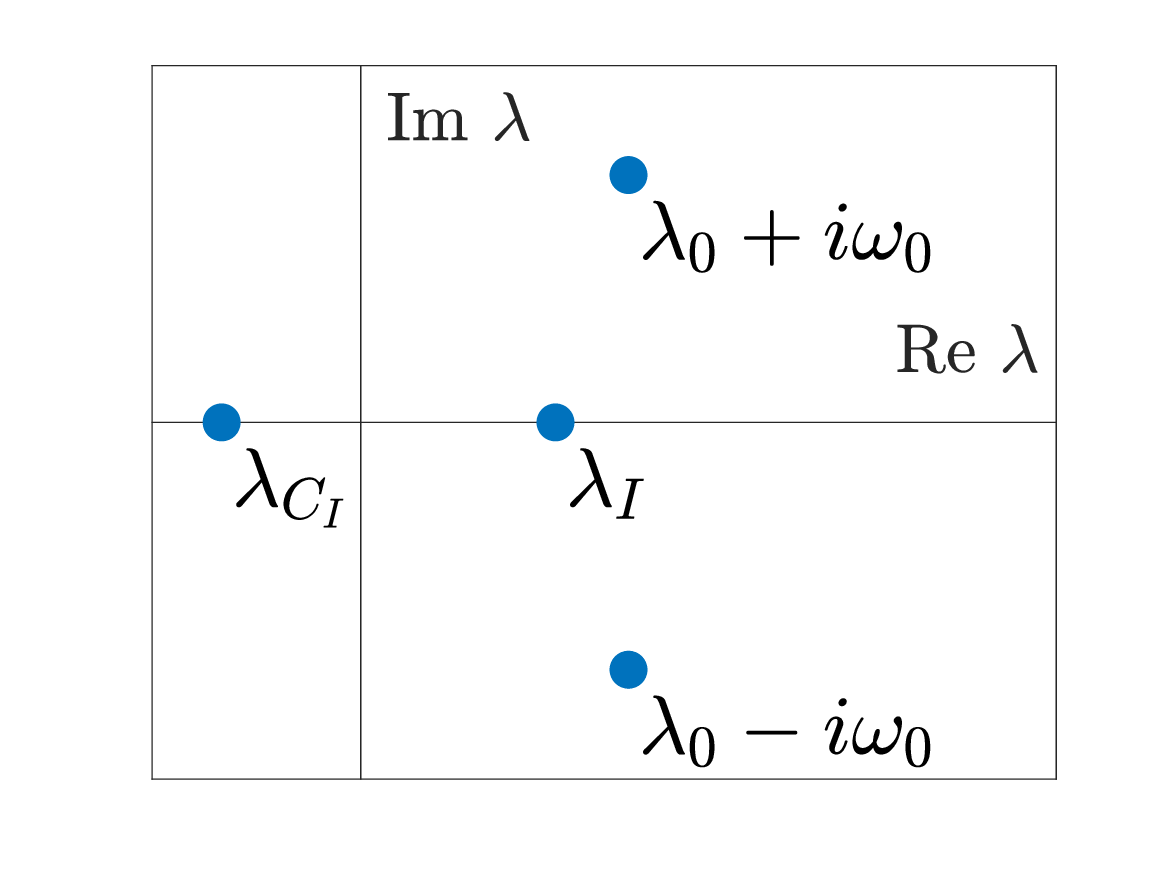}
    \end{tabular}
    \caption{Eigenvalue pattern of the matrix $\tilde{H}$ for a single excitatory cluster and multiple inhibitory clusters.}
    \label{fig:HeigpatternIcluster}
\end{figure}

\noindent This eigenvalue pattern is shown in \cref{fig:HeigpatternIcluster}.
The two eigenvalues with positive real part are $\lambda_I$ and $\lambda_0 + i \omega_0$, so these are the only eigenvalues which will cause bifurcations as $g$ is varied. We note that $\lambda_0 > \lambda_I$, thus the first bifurcation which will occur at the origin is a Hopf bifurcation at 
\[
g_H = \frac{2 \sqrt{N}}{\mu_{EE} \left[ \alpha( 1 + p_I(n_{C_I}-1)) -1 \right] }
\]
when the complex pair $\lambda_0 + i \omega_0$ crosses the real axis.
The behavior at this bifurcation is identical to that at the Hopf bifurcation at the origin in the unclustered case (\cref{sec:periodic}). Briefly, the corresponding eigenspace to $\lambda_0 + i \omega_0$ is
\[ 
V \equiv  \ker(DF)_{\Zerovec,g_H}  = {\rm span} \, 
\left\{ \left[ 1 \; \underbrace{\begin{matrix}0 & \cdots & 0\end{matrix}}_{n_I} \right],
\left[ 0 \; \underbrace{\begin{matrix}1 & \cdots & 1\end{matrix}}_{n_I} \right]
 \right\},
\]
which is fixed by $\Gamma = S_1 \times S_{n_I}$. Since $\dim \Fix_V(\Gamma) = \dim V = 2$, it follows from the Equivariant Hopf theorem \cite[Theorem 4.1]{GSS88Vol2} that there is a branch of limit cycles emanating from this Hopf bifurcation point for which the isotropy subgroup is $\Gamma$, which implies that the inhibitory neurons are all synchronized (we recall that the excitatory neurons are always synchronized).

\begin{figure}
    \centering
    \includegraphics[width=8.25cm]{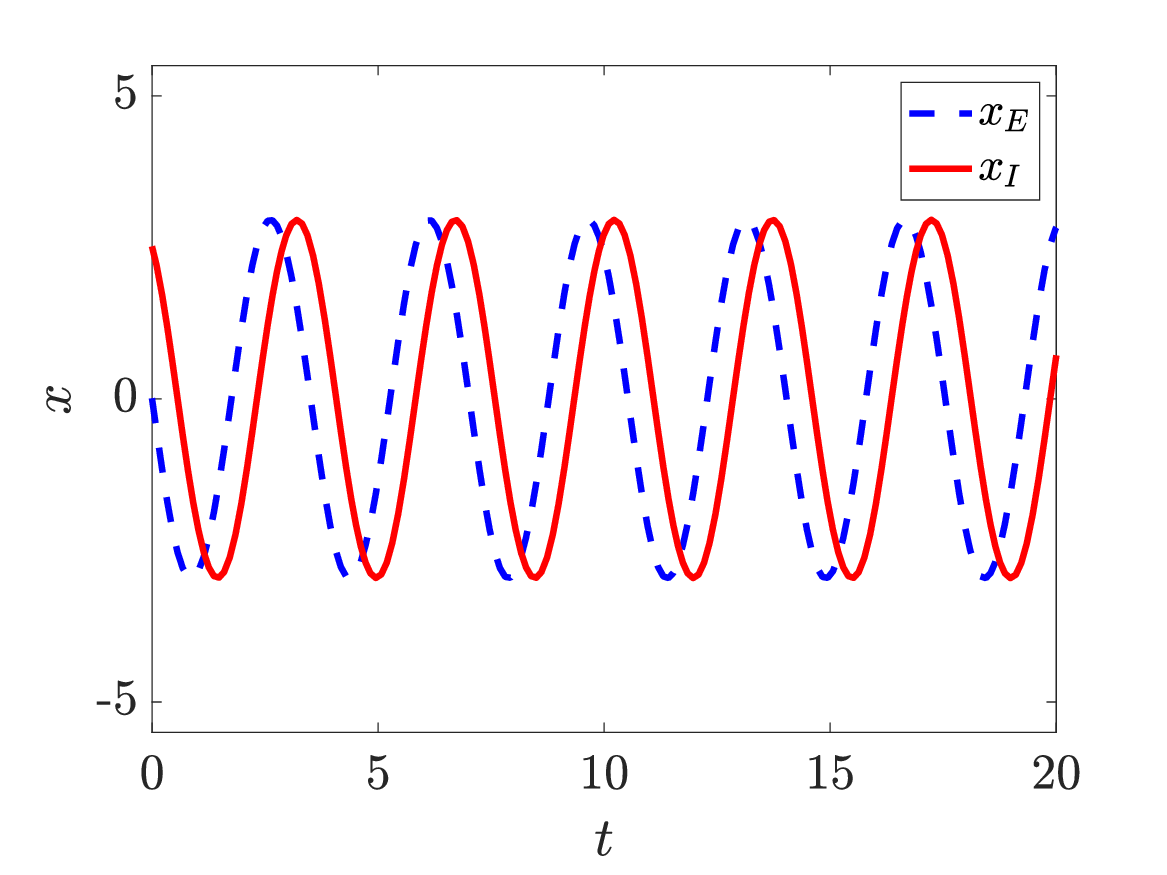}\hspace{-0.5cm}
    \includegraphics[width=8.25cm]{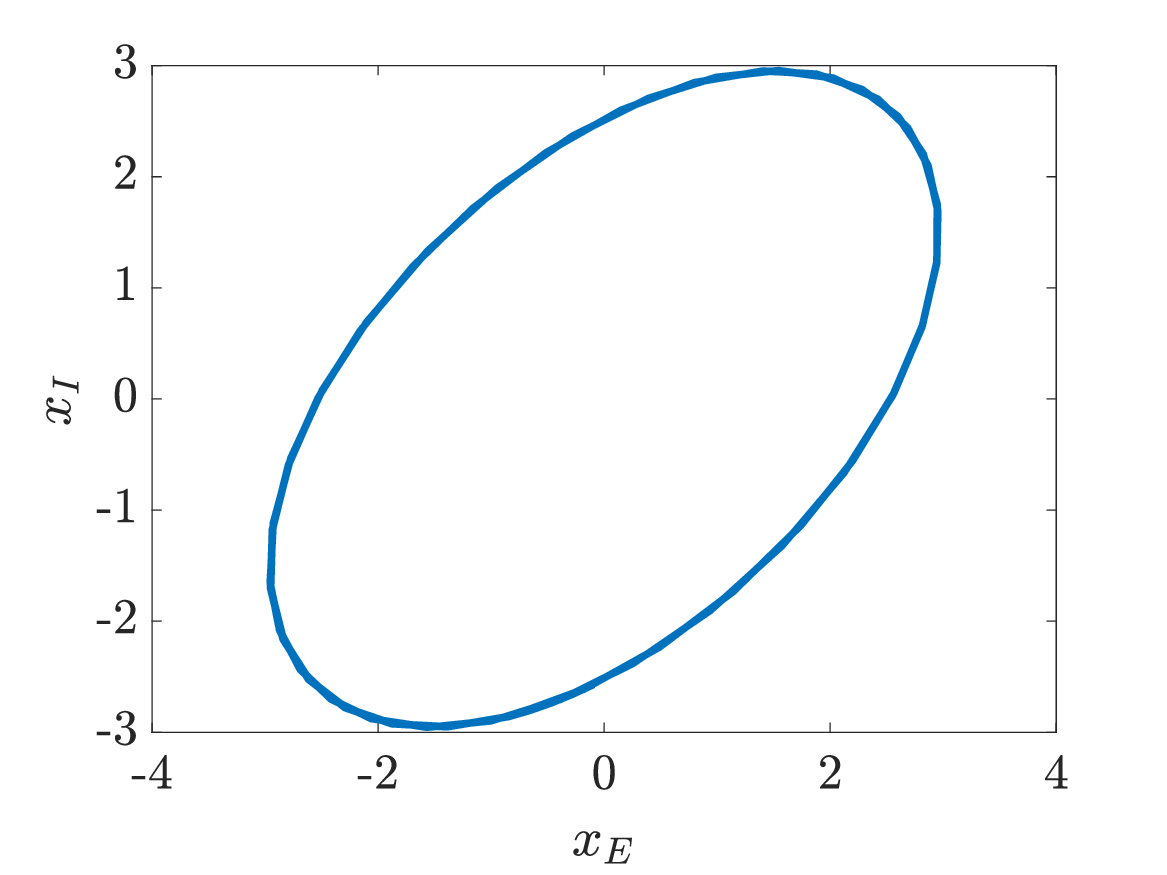}
    \caption{Limit cycle arising from the Hopf bifurcation at origin for \cref{eqn:sys_Basic} with inhibitory cell clustering and $\mu_{II} = -\alpha \mu_{EE}$. Both the excitatory cell activity $x_E$ and inhibitory cell activity $x_I$ are synchronized. Notably, the period does not increase with $N$. Left: $x_E$ and $x_I$ vs $t$. Right: $x_I$ vs $x_E$. Parameters are: $N=1600$, $N_{C_I} = 20$, $p_I = 20$, $\alpha = 4$, $\mu_{EE}= 0.7$, $g = 1.02 g_H$. The period of the limit cycle is 1.792.} 
    \label{fig:limitcycleIC}
\end{figure}

We are interested in what occurs for large $N$ and large $n_{C_I}$. As an example, let $n_{C_I}$ scale with $\sqrt{N}$ by taking $n_{C_I} = p_I = \sqrt{n_I} = \sqrt{(1-f)N}$. For this scaling, as $N$ increases, the Hopf bifurcation takes place at $g_H \approx \frac{2}{f \mu_{EE} \sqrt{N}}$, and we also have $\omega_0 \approx \frac{\sqrt{3}}{2}f N \mu_{EE}$. This implies that at $g = g_H$, $DF(0)$ has a complex conjugate pair of eigenvalues with real part of 0 and imaginary part of approximately $\sqrt{3}$. See \cref{fig:limitcycleIC} for an illustration of this limit cycle when $N=1600$, $n_{C_I}=20$, and $g$ is slightly larger than $g_H$. The frequency of the limit cycle is 1.792, which is less than $5\%$ away from $\sqrt{3}$. Thus, for large $N$, the frequency of the limit cycle emerging at the Hopf bifurcation of the origin is asymptotically constant as $N$ increases. This contrasts to the case where the inhibitory and excitatory cells are unclustered, where the frequency of the limit cycle scales as $\sqrt{N}$. Numerical timestepping experiments suggest that this limit cycle is stable for $g > g_H$.

\section{Discussion}

\begin{figure}
    \centering
    \begin{tabular}{cc}
    \includegraphics[width=10cm]{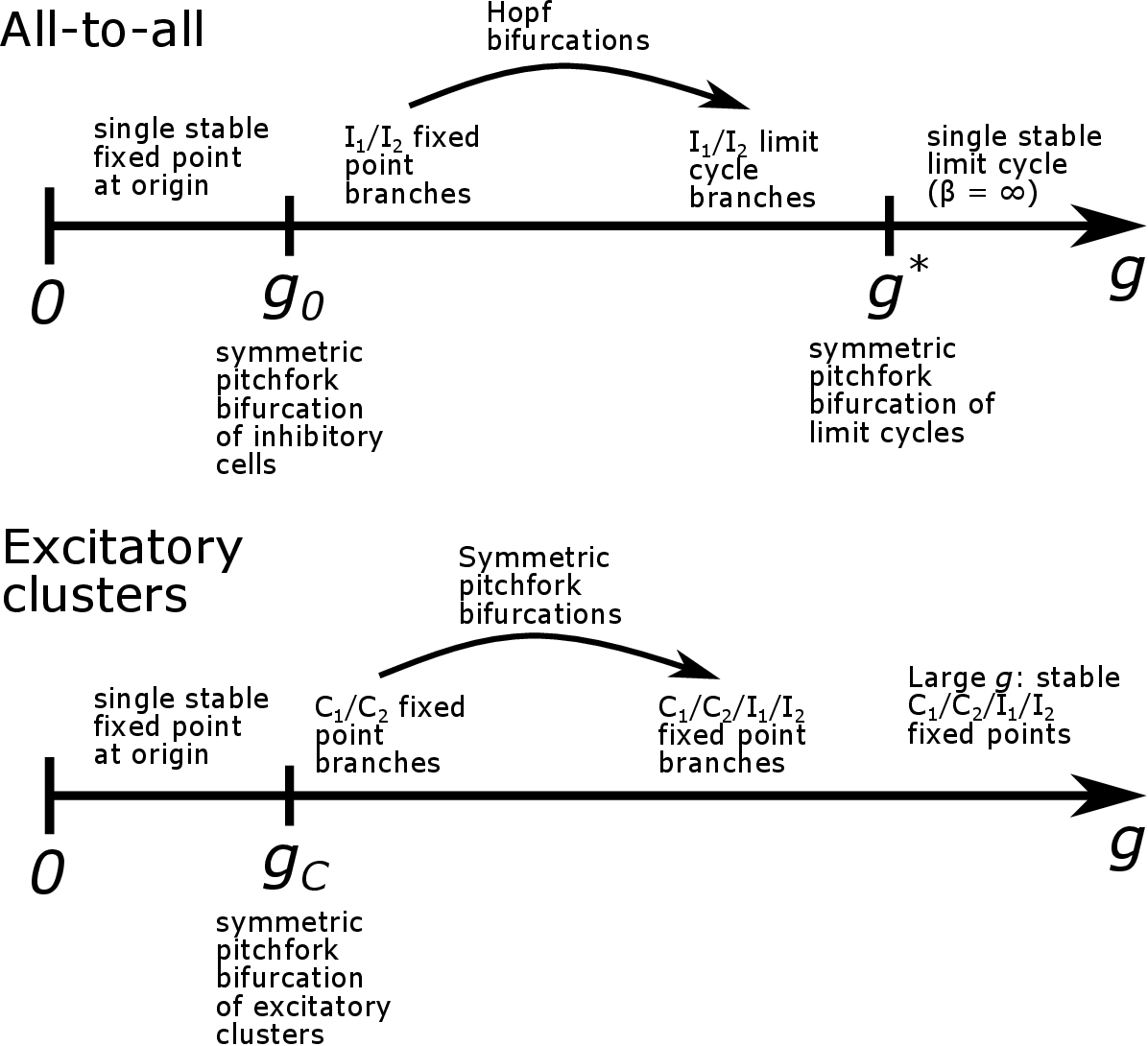}
    \end{tabular}
    \caption{Cartoon summary of fixed points, limit cycles, and bifurcations as $g$ increases from 0 for all-to-all connected network (top) and network with excitatory clustering (bottom).} 
    \label{fig:bdsummary}
\end{figure}

In this paper, we analyze a family of clustered excitatory-inhibitory neural networks, and, in particular, the underlying bifurcation structures that arise because of permutation symmetries in the network. For the simplest case, an all-to-all connected network which excludes self-connections, we extend the results in \cite{Barreiro2017} to provide a more complete picture of the bifurcations in the system, as well as estimates for the locations of the bifurcation points and the corresponding branches of equilibria which emanate from these bifurcations. For $g$ close to 0, the origin is a stable equilibrium. As $g$ is increased, the origin becomes unstable in a symmetric pitchfork bifurcation at $g=g_0$, at which a new branch of equilibria emerges for each possible division of the inhibitory cells into two synchronized clusters of sizes $n_{I_1}$ and $n_{I_2}$ (the $I_1/I_2$ branches). We characterize each $I_1/I_2$ branch by the ratio $\beta = n_{I_1}/n_{I_2}$. We then derive a leading order estimate for the equilibria on each $I_1/I_2$ branch for $g$ close to $g_0$ and show that, for large $N$, these branches are stable for $1\leq \beta < 2$, but unstable otherwise ($\beta \ge 2$). Furthermore, we show that the equilibria on the $I_1/I_2$ branches are all unstable for sufficiently large $g$. Along each $I_1/I_2$ branch, a Hopf bifurcation creates a branch of periodic orbits, wherein the inhibitory cells maintain their division into the same two synchronized clusters; the frequency of these limit cycles increases with $N$. We use our estimates for the $I_1/I_2$ branches to locate these Hopf bifurcations, to leading order, and show that they approach $g_0$ for large $N$. All these periodic orbit branches merge at a symmetric pitchfork bifurcation of limit cycles, at some large value of the bifurcation parameter $g=g^*$; for $g>g^*$, there is a single stable limit cycle for which the excitatory population and inhibitory population are each synchronized. See the top figure in \cref{fig:bdsummary} for a cartoon summary.

We next consider the case where the excitatory cells are broken into clusters of equal size. The connection weights between excitation cells in the same cluster are normalized so that the network is still approximately balanced. In this case, as $g$ is increased from 0, the origin becomes unstable in a symmetric pitchfork bifurcation point at $g=g_C$. In contrast with the previous case, this bifurcation involves the excitatory clusters instead of the inhibitory cells. For $g>g_C$, there is a branch of solutions corresponding to each possible division of the excitatory clusters into synchronized groups of sizes of sizes $n_{C_1}$ and $n_{C_2}$ (the $C_1/C_2$ branches). We characterize each branch by the ratio $\beta_C = n_{C_1}/n_{C_2}$. Near $g = g_C$, each solution branch is stable for $1 \le \beta_C < 2$, and is otherwise unstable. Along each $C_1/C_2$ branch, there is a further symmetric pitchfork bifurcation, in which the inhibitory cells split into two clusters of sizes $n_{I_1}$ and $n_{I_2}$ (with ratio $\beta = n_{I_1}/n_{I_2}$), yielding equilibria in which both the excitatory clusters and the inhibitory cells are split into two groups (the $C_1/C_2/I_1/I_2$ branches). Unlike the previous case, there are no Hopf bifurcations along these branches. For large $g$, the only branches that remain stable are those for which $\beta_C$ is close to $\beta$, in the precise sense we describe in \cref{sec:C1C2I1I2_largeg}; in other words, the excitatory clusters and the inhibitory cells must break up in a similar way. See the bottom figure in \cref{fig:bdsummary} for a cartoon summary. Finally, we briefly consider a network in which the inhibitory cells are clustered, rather than the excitatory cells. Here we find that, as in the case with all-to-all coupling, the origin loses stability in a Hopf bifurcation; however, in contrast to the all-to-all case, the frequency of the resulting limit cycle does not increase with $N$.

\subsection{Relationship to other work}
The population-clustered systems we consider in \cref{sec:E1I1} are similar to a simple version of the Wilson–Cowan equations (reviewed in \cite{et10,Chow_Y_JNeurophys_2020}), which can likewise be interpreted in terms of coupled neural populations. Other authors have derived and analyzed similar systems for balanced networks as a mean-field limit from large networks: however, recent examples differ from the current work because of the scaling of the deterministic part of the connectivity matrix. We retain "strong“ coupling as a function of system size ($1/\sqrt{N}$) as in \cite{RA06}, vs. “weak” scaling ($1/N$) \cite{hermann_etal_2012,kadmon_HS_2015,Schuessler_etal_PRR_2020}. In \cite{hermann_etal_2012}, for example, connectivity matrices are chosen with entries $J_{ij} \sim N(J/N, \sigma^2/N)$.  As $N \rightarrow \infty$, the mean connectivity ($1/N$) goes to zero faster than the typical random deviation from the mean ($1/\sqrt{N})$; thus  outgoing synaptic weights will no longer be single-signed, in violation of Dale’s Law. One consequence of weak scaling seems to be that oscillations are observed at the population but not necessary the cell level \cite{Ginzburg_HS_1994,Brunel_Hakim_1999}; in contrast, the limit cycles we describe in \cref{sec:periodic} are observed at both the cell and the population level.
	
In works that do use strong scaling, the coherent fluctuations that are observed require a perfect orthogonality condition \cite{delMolino_etal_PRE_2013,Landau_HS_PCB_2018} or an external forcing \cite{Landau_HS_PRR_2021} to balance. Furthermore, the nonrandom part of the connectivity matrix is low rank; this is not the case in the current work, in which some examples are low rank but most are not. Contrasting the excitatory clustering with and without self-coupling (\cref{sec:Eclusters} and \cref{sec:restore_selfCoup_Eclusters}), for example, we observe the same pattern of stable fixed points although one is low rank and the other is not.

Other recent studies of balanced neural network models do not include a deterministic mean connectivity matrix, but instead regulate correlations through the probability of small network motifs, such as reciprocal connections \cite{marti2018correlations} or common input/diverging motifs \cite{hu2014local,recanatesi2019dimensionality,dahmen2021strong}. The frequency of motifs can be shown to regulate cross-correlations \cite{hu2014local}, time scales \cite{marti2018correlations}, and dimensionality of the network response \cite{recanatesi2019dimensionality,dahmen2021strong}.  This last observation is particularly interesting in light of the many experimental studies documenting low-dimensional neural activity (reviewed in the Introduction). A natural next question is to investigate networks which are partly structured (having highly connected clusters as in the current studies) but partially random.  A promising avenue is to apply Hu et al.’s resumming theory to multi-population networks, to investigate whether the low-dimensional structures we find in the current work manifest in a network of coupled Gaussian processes, and as the network connections themselves becomes more random \cite{hu2014local}. Another related work \cite{stern2014dynamics} studies the clustered architecture we considered in \cref{sec:Eclusters}, but without structured inhibition, and studies the persistence of fixed points as randomness is added to the connection matrix.

\subsection{Future directions}

Future directions include better characterizing the periodic orbits which arise from the Hopf bifurcations in the network with all-to-all coupling case. It may be possible to determine their stability pattern, as well as to locate the bifurcation point at $g=g^*$. Some assumptions about our network can be relaxed; for example the use of the $\tanh$ function is not essential to any calculations that do not explicitly invoke odd symmetry, and could be replaced by another saturating nonlinearity. Another direction includes exploring other network topologies, such as unequal cluster sizes, spatial connectivity, or hierarchical clustering \cite{rosenbaum17,Ebsch_RR_PlosCB_2018}. 

Finally, the ultimate goal of these investigations must be to apply these insights to real networks, which will not be perfectly symmetric and which may be modeled by allowing a random perturbation to the connection matrix (i.e. $H \rightarrow H + \epsilon A$). 
The right-hand side of \cref{eqn:sys_Basic} is locally Lipschitz continuous in $\R^N$; therefore hyperbolic fixed points and periodic orbits will remain when the connectivity matrix is perturbed by a random matrix, i.e. $G=H + \epsilon A$ for small $\epsilon$. However, the range of $\epsilon$ for which a hyperbolic structure persists is not known \textit{a priori}. 
We conjecture that the perturbed system will continue to exhibit fixed points and periodic orbits that are found in the unperturbed system, even when the perturbations are large enough that the spectrum of the connectivity matrix ``masks" the underlying symmetry. In our previous study, we found that stable trajectories in the unperturbed all-to-all clustered system accurately predicted which solutions would be observed in the perturbed system \cite{Barreiro2017}. This highlights the importance of determining not only existence but \emph{stability} in the unperturbed system. We look forward to exploring this question in future work.

\pagebreak

\appendix

\section{Solutions along \texorpdfstring{$I_1/I_2$}{I1/I2} branches: detailed calculations}\label{app:I1I2sol}

Here we derive leading order expressions for the equilibria along the $I_1/I_2$ branches for $g$ close to $g_0$. We begin with the simplest case, which is when $n_I$ is even and $\beta = 1$. 
Taking $x_{I_1} = x_I$, $x_{I_2} = -x_I$, and $x_E = 0$ in \cref{eq:reducedsystemI1I2} and simplifying, we obtain the single equation (\cite[{Eq. 16}]{Barreiro2017})
\[
-x_I + \frac{\alpha \mu_{EE} }{\sqrt{N}} \tanh(g x_I) = 0, 
\]
which simplifies to 
\begin{equation}\label{eq:xIbrancheq}
\tanh(g x_I) - g_0 x_I = 0.
\end{equation}
Defining $f(x_I) := \tanh(g x_I) - g_0 x_I$, we note that $f(0) = 0$, $f'(0) = g - g_0$, and $f(x_I) \rightarrow -\infty$ as $x_I \rightarrow \infty$. When $g > g_0$, $f(x_I)$ is initially increasing, thus it follows from the continuity of $f$ and the intermediate value theorem that \cref{eq:xIbrancheq} has a solution with $x_I > 0$ for all $g > g_0$. Furthermore, $x_I \rightarrow 1/g_0$ as $g \rightarrow \infty$.

To obtain an approximation of this solution for $g$ close to $g_0$, we expand the LHS of \cref{eq:xIbrancheq} in Taylor series about $x_I = 0$ and $g = g_0$ and simplify to get
\begin{equation}\label{eq:tanhTaylor}
(g-g_0) x_I - \frac{(g x_I)^3}{3} + \frac{2(g x_I)^5}{15} + \mathcal{O}\left( x_I^7 \right) = 0.
\end{equation}
We note that the remainder term in $(g-g_0)$ is transcendentally small in the sense of \cite{Holmes2012}. Keeping up to cubic terms in $x_I$, equation \cref{eq:tanhTaylor} simplifies to
\[
x_I \left( (g - g_0) - \frac{g^3}{3} x_I^2 \right) = 0.
\]
Solving the non-zero solution for $x_I$ results in the expression \cref{eq:xIapprox}.
We can obtain a higher-order approximation by keeping up to fifth-order terms in \cref{eq:tanhTaylor} to get
\begin{equation*}
x_I \left( (g - g_0) - \frac{g^3}{3} x_I^2 + \frac{2 g^5}{15} x_I^4 \right) = 0,
\end{equation*}
which is $x_I$ multiplied by a quadratic in $x_I^2$. To find the nonzero solution for $x_I$, we solve this quadratic for $x_I^2$ and take square roots, yielding \cref{eq:xIapprox5}.

For $\beta > 1$, as $N \rightarrow \infty$, numerical continuation with the parameter continuation software package AUTO \cite{AUTO} suggests that \cref{eq:reducedsystemI1I2} has a solution of the form
\begin{equation}\label{eq:I1I2asymp}
    x_{I_2} = -\beta x_{I_1} + \mathcal{O}\left( \frac{1}{N^2} \right), \quad 
    x_{I_1} = \mathcal{O}\left( \frac{1}{N} \right), \quad
     x_E = \mathcal{O}\left( \frac{1}{N^2} \right)
\end{equation}
for $g$ close to $g_0$. Subtracting the second and third equations in \cref{eq:reducedsystemI1I2}, we get
\[
 x_{I_1} - x_{I_2} = \frac{\alpha}{\mu_{EE}\sqrt{N}}\left( \tanh(g x_{I_1}) - \tanh(g x_{I_2}) \right).
 \]
Substituting \cref{eq:I1I2asymp} as an ansatz, expanding the $\tanh$ terms in a Taylor series about $x_{I_1} = 0$ to cubic order, and simplifying, we obtain the formula given in \cref{eq:XI1}.

Finally, we show that $x_{I_1}$ and $x_{I_2}$ have opposite signs for all $g > g_0$. Since \cref{eq:reducedsystemI1I2} is smooth in $(x_E, x_{I_1}, x_{I_2})$ and $g$, the solutions $x_{I_1}$ and $x_{I_2}$ are smooth in $g$. For $g$ close to $g_0$, $x_{I_1}$ and $x_{I_2}$ have opposite signs; if this is not the case for some $g > g_0$, either $x_{I_1}$ or $x_{I_2}$ must pass through 0. We will show that this cannot happen. Suppose $x_{I_1} = 0$ for some $g^* > g_0$. Substituting this into \cref{eq:reducedsystemI1I2} and subtracting the second row from the first, we have $x_E = -\frac{\mu_{EE}}{\sqrt{N}} \tanh(g^* x_E)$, which is impossible unless $x_E = 0$. If $x_E = 0$, then $x_{I_2} = -\frac{\mu_{EE}}{\sqrt{N}} \alpha(n_{I_2} - 1) \tanh(g^* x_{I_2})$, which is again impossible unless $x_{I_2} = 0$. Thus $x_{I_1}=0$ implies $(x_E, x_{I_1}, x_{I_2}) = 0$. This would mean that the $I_1/I_2$ branch would intersect the zero solution in another bifurcation point at $g^* > g_0$, which we know does not occur, since we have found all bifurcation points of the origin. The case where $x_{I_2} = 0$ for some $g^* > g_0$ is similar.

\section{Stability and bifurcations along \texorpdfstring{$I_1/I_2$}{I1/I2} branches: detailed calculations}\label{app:I1I2stab}

To determine the stability of $\xvec^*$ for $g$ close to $g_0$, we start by computing the eigenvalues of $D\tilde{F}(\xvec^*)$ corresponding to $\lambda_{I_1}$ and $\lambda_{I_2}$. Substituting \cref{eq:XI1} for $x_{I_1}$, using the Taylor series expansion $\sech^2 x = 1 - x^2 + \mathcal{O}(x^4)$, and simplifying, the eigenvalue $\lambda_{I_1}^*(g)$ of $D\tilde{F}(\xvec^*)$ is located at
\begin{align*}
    \lambda_{I_1}^*(g) &= \frac{g-g_0}{g} \left( 1 - \frac{3}{1-\beta+\beta^2 }\right) + \mathcal{O}\left(\frac{1}{N^2} \right) && g > g_0,
\end{align*}
which is negative for $1 \leq \beta < 2$ and positive for $\beta > 2$. Similarly, the eigenvalue $\lambda_{I_2}^*(g)$ of $D\tilde{F}(\xvec^*)$ corresponding to $\lambda_{I_2}$ is located at 
\begin{align*}
    \lambda_{I_2}^*(g) &= \frac{g-g_0}{g} \left( 1 - \frac{3 \beta^2}{1-\beta+\beta^2 }\right) + \mathcal{O}\left(\frac{1}{N^2} \right) && g > g_0,
\end{align*}
which is negative for $\beta > 1/2$ and thus does not affect stability. 

It remains to find leading order expressions for the eigenvalues of $H_3(\xvec)$. When $\xvec = 0$, the matrix $H_3(0)$ has a single eigenvalue at $\lambda_I$ and a complex conjugate pair of eigenvalues $\lambda_0 \pm \omega_0$, where these are defined at the beginning of \cref{sec:E1I1}. These do not depend on $\beta$. For $\xvec$ small but nonzero, we use a perturbation method to approximate the eigenvalues of $H_3(\xvec)$. We substitute the expressions \cref{eq:I1I2asymp} into characteristic polynomial for $H_3(\xvec)$, keeping only terms of up to order $1/N$, so that the leading order expression only involves $x_{I_1}$. We then use the Taylor expansion $\sech^2(g x_{I_1}) = 1 - (g x_{I_1})^2 + \mathcal{O}( x_{I_1}^4 )$, keeping only terms up to quadratic order. For each eigenvalue $\lambda$ of $H_3(\xvec)$, we use a power series ansatz 
\begin{equation}\label{eq:lambdaansatz}
\lambda + \epsilon x_{I_1}^2 + \mathcal{O}(x_{I_1})^4.
\end{equation}
We substitute this ansatz into the characteristic polynomial for $H_3(\xvec)$ and solve for $\epsilon$ by matching the coefficients of $x_{I_1}^2$. (This computation, and the remaining computations in this section, were performed with the aid of Wolfram Mathematica). Using this method for $\lambda = \lambda_I$, $H_3(\xvec^*)$ has a real eigenvalue located at
\[
\lambda_I = \alpha \mu_{EE} \left(1 - (1-\beta+\beta^2)g^2 x_{I_1}^2 \right) + \mathcal{O}\left(\frac{1}{N^2} \right).
\]
Substituting the estimate \cref{eq:XI1} for $x_{I_1}$ and simplifying, the eigenvalue $\lambda_I^*(g)$ of $J_3(\xvec)$ corresponding to $\lambda_I$ is located at
\begin{align*}
\lambda_I^*(g) = \frac{\alpha \mu_{EE} g}{\sqrt{N}} \left( 1 - \frac{3(g-g_0)}{g}\right) - 1 &= -2\left( \frac{g - g_0}{g_0} \right) + \mathcal{O}\left(\frac{1}{N^2} \right) && g \geq g_0,
\end{align*}
which is always negative, and thus does not affect stability.

Finally, we use this method to locate the eigenvalue of $H_3(\xvec)$ corresponding to $\lambda_0 \pm \omega_0$. In doing so, we will find a Hopf bifurcation on each $I_1/I_2$ branch. $H_3(\xvec)$ has a complex conjugate pair of eigenvalues, where the real part is given by
\begin{equation}
\lambda_0(g, \beta) = \frac{\mu_{EE}}{2}\left( \alpha - 1 + \alpha \beta g^2 (n_I-1) x_{I_1}^2 \right) + \mathcal{O}\left(\frac{1}{N} \right).
\end{equation}
We can get more accurate approximations for $\lambda(g, \beta)$ by taking higher powers of $x_{I_1}$ in our power series ansatz \cref{eq:lambdaansatz}. For example, when $\beta=1$, we can obtain the fourth-order approximation
\[
\lambda_0(g, 1) = \frac{\mu_{EE}}{2}\left( \alpha - 1 + \alpha g^2 (n_I-1) x_{I_1}^2 - \frac{2}{3}\alpha g^4 (n_I-1) x_{I_1}^4 \right) + \mathcal{O}\left(\frac{1}{N^2} \right).
\]
Similar fourth-order approximations can be obtained when $\beta>1$, but the resulting coefficient of $x_{I_1}^4$ is significantly more complicated.
Substituting \cref{eq:XI1} for $x_{I_1}$ and simplifying, $J_3(\xvec)$ has a complex conjugate pair of eigenvalues $\lambda_0^*(g) \pm i \omega_0^*(g)$, where
\begin{equation}\label{eq:lambdagbeta}
\lambda_0^*(g, \beta) = \frac{\mu_{EE} g}{2 \sqrt{N}}\left[ \alpha - 1 + \alpha \beta (n_I-1) \frac{ 3(g - g_0) }{ (1 - \beta + \beta^2 )g} \right] - 1 + \mathcal{O}\left(\frac{1}{N} \right).
\end{equation}
To locate the Hopf bifurcation on each $I_1/I_2$ branch, which occurs when the complex pair of eigenvalues crosses the imaginary axis, we solve $\lambda_0^*(g, \beta) = 0$ for $g$, substitute $g_0 = \sqrt{N}/\alpha \mu_{EE}$, and simplify to obtain the expression in \cref{eq:ghopfformula}.

\section{Proof of \texorpdfstring{\cref{prop:limitcycle}}{Proposition}} \label{sec:limitcycleproof}

First, we show that that \cref{eq:2dimsystem} has no fixed points other than the origin. To do this, we make the change of variables $(y_1, y_2) = (\tanh(g x_1),\tanh(g x_2))$, and note that it is equivalent to show that the system of equations
\begin{equation}\label{eq:2dimsystemy}
\begin{aligned}
g_1(y_1, y_2) := -\frac{1}{g}\tanh^{-1}(y_1) + \frac{\mu_{EE}}{\sqrt{N}}\left((n_E - 1) y_1 - \alpha n_I y_2 \right) = 0 \\
g_2(y_1, y_2) := -\frac{1}{g}\tanh^{-1}(y_2) + \frac{\mu_{EE}}{\sqrt{N}}\left( n_E y_1 - \alpha (n_I - 1) y_2 \right) = 0
\end{aligned}
\end{equation}
has no solution other than $(y_1, y_2) = (0,0)$. The first equation $g_1(y_1, y_2) = 0$ is satisfied when
\begin{equation}\label{eq:y2sol}
y_2 = y_2^*(y_1) := \frac{ g \mu_{EE} (n_E-1) y_1 - \sqrt{N} \tanh^{-1}(y_1)}{\alpha g \mu_{EE} n_I}.
\end{equation}
To show that \cref{eq:2dimsystemy} has no solutions other than the origin, we substitute \cref{eq:y2sol} into $g_2(y_1, y_2)$ to get 
\begin{equation}
\begin{aligned}
g_2(y_1, y_2^*(y_1)) &= \frac{\mu_{EE}( n_E + n_I - 1) y_1 }{n_I \sqrt{N}} + \frac{n_I - 1}{g n_I} \tanh^{-1}(y_1) \\
&\qquad+ \frac{1}{g} \tanh^{-1} \left( \frac{ \sqrt{N} \tanh^{-1}(y_1) - g \mu_{EE}(n_E -1) y_1}{ \alpha g \mu_{EE} n_I } \right).
\end{aligned}
\end{equation}    
We will show that $g_2(y_1, y_2^*(y_1)) > 0$ for $y_1 > 0$. Since $g_2(y_1, y_2^*(y_1))$ is an odd function in $y_1$, this will imply that $g_2(y_1, y_2^*(y_1)) < 0$ for $y_1 < 0$, from which the desired result will follow. Since $\tanh^{-1}y_1 \geq y_1$ for $y_1 \geq 0$, it suffices to show that
\begin{equation}\label{eq:h}
\begin{aligned}
h(y_1) &:= \frac{\mu_{EE}( n_E + n_I - 1) y_1 }{n_I \sqrt{N}} + \frac{n_I - 1}{g n_I} \tanh^{-1}(y_1) \\
&\qquad+ \frac{1}{g} \tanh^{-1} \left( \frac{ \sqrt{N} \tanh^{-1}(y_1) - g \mu_{EE}(n_E -1) y_1}{ \alpha g \mu_{EE} n_I } \right) > 0
\end{aligned}
\end{equation}
for $y_1 > 0$. Since $h(0) = 0$, we will show that $h'(0) > 0$ for $y_1 > 0$. Computing the derivative with the assistance of Mathematica,
\begin{equation*}\label{eq:hprime}
\begin{aligned}
h'(y_1) &= 
\frac{(N-1)\mu_{EE}}{(1-f) N^{3/2} } + \frac{1}{g(1 - y_1^2)} 
- \frac{
 f \mu_{EE} N ( g \mu_{EE} ( f N - 1) - \sqrt{N}}{f^2 g^2 \mu_{EE}^2 N^2 - (\sqrt{N} + 
 g \mu_{EE}(1 - f N))^2 y_1^2} \\
 &\geq \frac{(N-1)\mu_{EE}}{(1-f) N^{3/2} } + \frac{1}{g(1 - y_1^2)} 
- \frac{1}{g \left( 1 - \left( \frac{ f N - 1}{f N} \right)^2 y_1^2 \right)} \\
&\geq \frac{(N-1)\mu_{EE}}{(1-f) N^{3/2} } > 0.
\end{aligned}
\end{equation*}
We have therefore shown that \cref{eq:2dimsystem} has no fixed points other than the origin.

\begin{figure}
    \centering
    \begin{tabular}{c}
    \includegraphics[width=8cm]{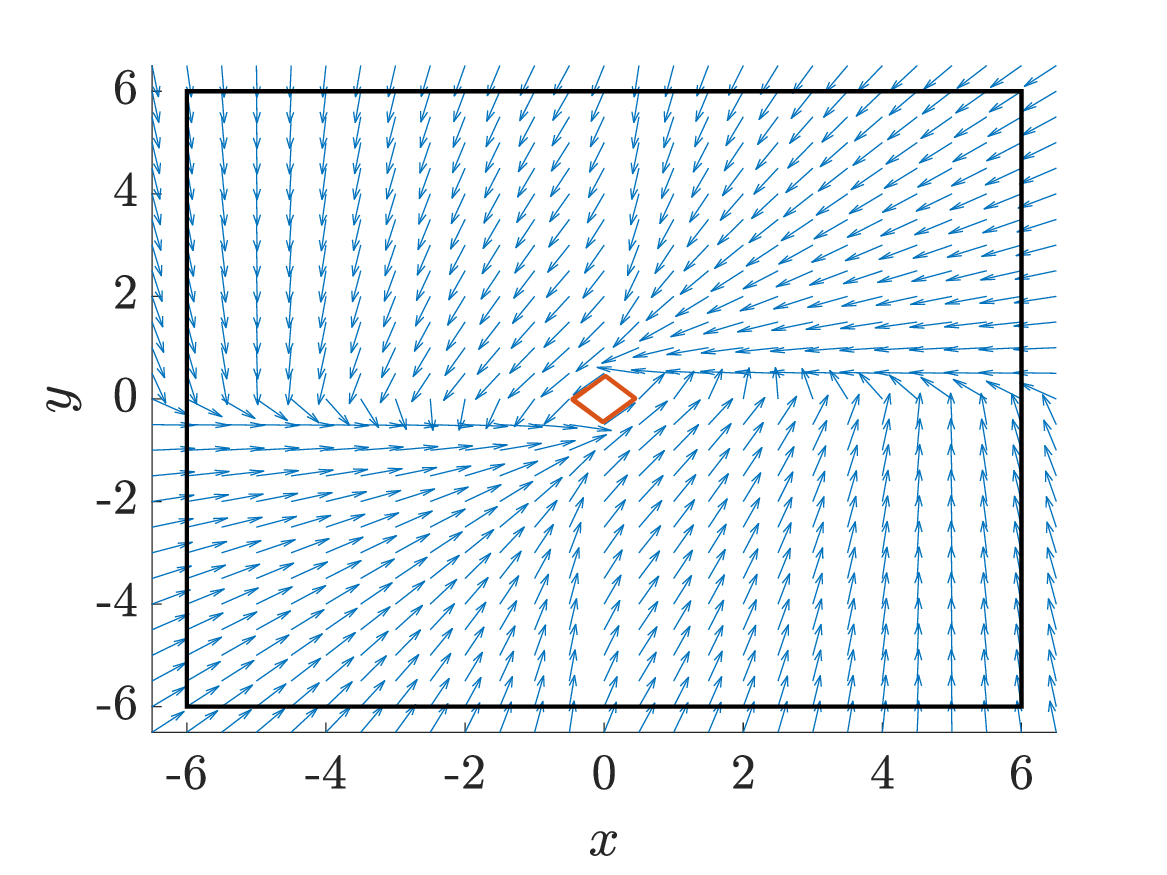}
    \end{tabular}
    \caption{Slope fields for \cref{eq:2dimsystem}, with a small limit cycle visible in center. Slope field points inward on black box, which is the trapping region for the Poincar{\'e}-Bendixson theorem. Parameters: $N = 20$, $g = 5$, $\alpha = 4$, and $\mu_{EE} = 0.7$.}
    \label{fig:nullclines}
\end{figure}

The Linearization of \cref{eq:2dimsystem} about the origin is the $2 \times 2$ matrix
\[
J = \frac{g \mu_{EE}}{\sqrt{N}}
\begin{bmatrix} 
n_E - 1 & -\alpha n_I \\
n_E & -\alpha(n_I - 1)
\end{bmatrix} - I_2,
\]
which has a complex conjugate pair of eigenvalues $\frac{g}{\sqrt{N}}(\lambda_0 \pm i \omega_0) - 1$, where $\lambda_0$ and $\omega_0$ are defined in \cref{sec:E1I1}. This pair crosses through the imaginary axis at $g = g_H$, where $g_H$ is defined by \cref{eq:0hopflocation}, leading to a Hopf bifurcation in the reduced system \cref{eq:2dimsystem}, and the origin is repelling for $g > g_H$. To show there is a limit cycle for all $g > g_H$, we use the Poincar{\'e}-Bendixson theorem \cite[Chapter 16]{Coddington1955}. For a trapping region, we draw a square around the origin with corners $(-a, -a)$ and $(a, a)$. On the line $x = a$, for $a$ large,
\[
\dot{x} \leq -a + \frac{2 n_E}{\sqrt{N}} = -a + 2 f \sqrt{N},
\]
which can be made negative by taking $a$ sufficiently large. Similarly, we can take $a$ sufficiently large so that the vector field defined by \cref{eq:2dimsystem} points inward at all points on the square (\cref{fig:nullclines}). Since the origin is repelling for $g > g_H$ and is the only fixed point of the system, it follows from the Poincar{\'e}-Bendixson theorem that there is a limit cycle surrounding the origin for $g > g_H$. We note that although the limit cycle from \cref{prop:limitcycle} is stable in the two-dimensional system \cref{eq:2dimsystem}, the theorem says nothing about its stability in the full system \cref{eqn:sys_Basic}.

\section{Stability of the \texorpdfstring{$I_1/I_2$}{I1/I2} branch for large \texorpdfstring{$g$}{g}: detailed calculations }\label{app:stab_largeg} 

Here we prove our assertion, made in \cref{sec:stab_largeg}, that $x_E \rightarrow 0$ as $g\rightarrow \infty$ along any $I_1/I_2$ solution branch.
Suppose, instead, that $x_E \rightarrow \hat{x}_E \neq 0$ as $g \rightarrow \infty$. Without loss of generality, we can take $\hat{x}_E > 0$, since by odd symmetry of \cref{eqn:sys_Basic}, there will be a corresponding solution with $\hat{x}_E < 0$. This implies that $\tanh x_{E} \rightarrow 1$. There are four cases to consider:\\
\begin{itemize}
    \item Case 1: $x_{I_1} \rightarrow \hat{x}_{I_1} \neq 0$ and $x_{I_2} \rightarrow \hat{x}_{I_2} \neq 0$.
    \item Case 2: $x_{I_1} \rightarrow \hat{x}_{I_1} \neq 0$ and $x_{I_2} \rightarrow 0$.
    \item Case 3: $x_{I_1} \rightarrow 0$ and $x_{I_2} \rightarrow \hat{x}_{I_2} \neq 0$.
    \item Case 4: $x_{I_1} \rightarrow 0$ and $x_{I_2} \rightarrow 0$.\\
\end{itemize}

The computations to follow were done with the assistance of Wolfram Mathematica. For Case 1, $\tanh(g x_{I_1}) \rightarrow \pm 1$ and $\tanh(g x_{I_2}) \rightarrow \pm 1$. We can then use \cref{eq:reducedsystemI1I2} to solve for $(\hat{x}_E, \hat{x}_{I_1}, \hat{x}_{I_2})$. The signs of these solutions are all inconsistent, as we can see in \cref{table:signtable}. 

\begin{table}
\centering
    \begin{tabular}{llll}
        \toprule
        sgn $(\hat{x}_E, \hat{x}_{I_1}, \hat{x}_{I_2})$ & $\frac{\sqrt{N}}{\mu_{EE}}\hat{x}_E$ & $\frac{\sqrt{N}}{\mu_{EE}}\hat{x}_{I_1}$ & $\frac{\sqrt{N}}{\mu_{EE}}\hat{x}_{I_2}$ \\
        \midrule
        $(1,1,1)$   & $-1 < 0$ & $\alpha$ & $\alpha$ \\
        $(1,-1,-1)$ & $2 \alpha n_I - 1$ & $\alpha(2 n_I - 1) > 0 $ & $\alpha(2 n_I - 1) > 0$ \\
        $(1,1,-1)$  & $2 \alpha n_{I_2} - 1$ & $\alpha(2 n_{I_2} + 1)$ & $\alpha(2 n_{I_2} - 1) > 0$  \\
        $(1,-1,1)$  & $2 \alpha n_{I_1} - 1$ & $\alpha(2 n_{I_1} - 1) > 0$ & $\alpha(2 n_{I_1} + 1) > 0$  \\
        \bottomrule
    \end{tabular}
    \label{table:signtable}
    \vspace{0.25cm}
    \caption{Sign table showing that all solutions for nonzero $(\hat{x}_E, \hat{x}_{I_1}, \hat{x}_{I_2})$ are inconsistent.}
\end{table}

For Case 2, if $\tanh( g \hat{x}_{I_2} ) \rightarrow 0$, the solution $(\hat{x}_E, \hat{x}_{I_1}, \hat{x}_{I_2})$ from \cref{eq:reducedsystemI1I2} is inconsistent using the same argument as in Case 1. The only remaining possibility is $\tanh( g \hat{x}_{I_2} ) \rightarrow \hat{y}_{I_2}$, where $0 < |\hat{y}_{I_2}| < 1$. In the limit $g \rightarrow \infty$, \cref{eq:reducedsystemI1I2} becomes
\begin{equation*}
 \begin{aligned}
 \begin{bmatrix} \hat{x}_E \\ \hat{x}_{I_1} \\ 0 \end{bmatrix} 
 &= \frac{\mu_{EE}}{\sqrt{N}} 
 \begin{bmatrix} (\alpha n_I - 1) & -\alpha \frac{\beta}{\beta+1} n_I & - \alpha \frac{1}{\beta+1} n_I  \\
    \alpha n_I  & -\alpha \left(\frac{\beta}{\beta+1} n_I-1\right) & - \alpha \frac{1}{\beta+1} n_I  \\
    \alpha n_I & -\alpha \frac{\beta}{\beta+1} n_I & -\alpha \left(\frac{1}{\beta+1} n_I-1\right)
 \end{bmatrix}
 \begin{bmatrix} 1 \\ \pm 1 \\ \hat{y}_{I_2} \end{bmatrix}.
 \end{aligned}
 \end{equation*}
The consistency condition (from the third row) can only be satisfied if $\hat{y}_{I_2} = \frac{n_I}{n_I - (\beta + 1)} > 1$ (for $\hat{x}_{I_1} > 0$) or $\hat{y}_{I_2} = \frac{n_I(1 + 2 \beta)}{n_I - (\beta + 1)} > 1$ (for $\hat{x}_{I_1} < 0$), both of which are impossible. Case 3 is similar.

For Case 4, if $\tanh( g \hat{x}_{I_1} ) \rightarrow 0$ or $\tanh( g \hat{x}_{I_2} ) \rightarrow 0$, the solution $(\hat{x}_E, \hat{x}_{I_1}, \hat{x}_{I_2})$ from \cref{eq:reducedsystemI1I2} is inconsistent using the same argument as in Case 1. The remaining possibility is $\tanh( g \hat{x}_{I_1} ) \rightarrow \hat{y}_{I_1}$  and $\tanh( g \hat{x}_{I_2} ) \rightarrow \hat{y}_{I_2}$, where $0 < |\hat{y}_{I_1}| , |\hat{y}_{I_2}| < 1$. In the limit $g \rightarrow \infty$, \cref{eq:reducedsystemI1I2} becomes
\begin{equation*}
 \begin{aligned}
 \begin{bmatrix} \hat{x}_E \\ 0 \\ 0 \end{bmatrix} 
 &= \frac{\mu_{EE}}{\sqrt{N}} 
 \begin{bmatrix} (\alpha n_I - 1) & -\alpha \frac{\beta}{\beta+1} n_I & - \alpha \frac{1}{\beta+1} n_I  \\
    \alpha n_I  & -\alpha \left(\frac{\beta}{\beta+1} n_I-1\right) & - \alpha \frac{1}{\beta+1} n_I  \\
    \alpha n_I & -\alpha \frac{\beta}{\beta+1} n_I & -\alpha \left(\frac{1}{\beta+1} n_I-1\right)
 \end{bmatrix}
 \begin{bmatrix} 1 \\ \hat{y}_{I_1} \\ \hat{y}_{I_2} \end{bmatrix}.
 \end{aligned}
 \end{equation*}
 The consistency conditions (from the second and third rows) can only be satisfied if $\hat{y}_{I_1} = \hat{y}_{I_2} = \frac{n_I}{n_I - 1} > 1$, which is impossible.

\section{Stability and bifurcations along \texorpdfstring{$C_1/C_2$}{C1/C2} branch: detailed calculations}\label{app:C1C2stability}

Following the procedure in \cref{sec:I1I2stability} and \cref{app:I1I2stab}, we substitute the expressions from \cref{eq:XE1} into the the formulas for the eigenvalues in \cref{prop:H3Ceig} and simplify to obtain leading order expressions for the corresponding eigenvalues of $D\tilde{F}(\xvec^*)$
\begin{equation}\label{eq:DFclustereigs}
\begin{aligned}
\lambda_{C_1}^*(g) &= \frac{g-g_C}{g} \left( 1 - \frac{3}{1-\beta_C+\beta_C^2 }\right) \\
\lambda_{C_2}^*(g) &= \frac{g-g_C}{g} \left( 1 - \frac{3 \beta_C^3}{1-\beta_C+\beta_C^2 }\right) \\
\lambda_{I}^*(g) &= \frac{\alpha \mu g}{\sqrt{N}} - 1.
\end{aligned}
\end{equation}
The eigenvalue $\lambda_{C_1}^*(g)$ is negative for $1 \leq \beta_C < 2$ and positive for $\beta_C > 2$; $\lambda_{C_2}^*(g)$ is negative for $\beta_C  > 1/2$; and $\lambda_{I}^*(g)$ is negative for all $\beta_C$ for $N$ sufficiently large.

It remains to find leading order expressions for the eigenvalues of $H_3(\xvec)$, given by \cref{eq:H3C}. When $\xvec = 0$, the matrix $H_3(0)$ has a single eigenvalue at $\lambda_C$ and a complex conjugate pair of eigenvalues $\lambda_0 \pm i \omega_0$, where these are defined at the beginning of \cref{sec:Eclusters}. Using the same asymptotic procedure as in \cref{sec:I1I2stability} and \cref{app:I1I2stab}, $H_3(\xvec^*)$ has a real eigenvalue corresponding to $\lambda_C$ located at
\[
\lambda_C(\xvec^*) = (p-1)n_C \mu  \left(1 - (1 - \beta_C+\beta_C^2)g^2 x_{E_1}^2 \right) + \mathcal{O}\left(\frac{1}{N^2} \right).
\]
Substituting the estimate \cref{eq:XE1} for $x_{E_1}$ and simplifying, the eigenvalue $\lambda_C^*(g)$ of $J_3(\xvec^*)$ corresponding to $\lambda_C$ is located, to leading order, at 
\begin{align*}
    \lambda_C^*(g) &= -2\left( \frac{g - g_C}{g_C} \right),
\end{align*}
for $g$ close to $g_C$. Since this eigenvalue is always negative, it will not affect stability. Similarly, $H_3(\xvec)$ has a complex conjugate pair of eigenvalues $\lambda_0 + i \omega_0$, where the real part is given by
\begin{equation*}
\lambda_0(g, \beta_C) = \frac{\mu}{2}\left[ (\alpha - n_C) - \beta_C g^2 n_C (p - 1) x_{E_1}^2 \right]
\end{equation*}
to leading order, for $g$ close to $g_C$. Since we are taking $n_C \geq \alpha$, this is always negative for $g$ close to $g_C$. 

\section{\texorpdfstring{$C_1/C_2$}{C1/C2} branches for large \texorpdfstring{$g$}{g}: detailed calculations}\label{app:C1C2_largeg} 
Here we provide details of the behavior of solutions on the $C_1/C_2$ branch as $g$ becomes large.  We claim there are two patterns for the limiting behavior on the $C_1/C_2$ branches, which depend on whether $\beta_C < \beta_C^*$ or $\beta_C > \beta_C^*$, for some critical value $\beta_C^*$, which we will determine below. These were illustrated in \cref{fig:betacstar}.\\
\begin{itemize}
    \item Case 1: ($1 < \beta_C < \beta_C^*$) $x_{E_1} \rightarrow \hat{x}_{E_1} > 0$ and $x_{E_2} \rightarrow \hat{x}_{E_2} < 0$. 
    \item Case 2: ($\beta_C > \beta_C^*$) $x_{E_1} \rightarrow 0$ with $\tanh(g x_{E_1}) \rightarrow \hat{y}_{E_1} \neq 0$, and $x_{E_2} \rightarrow \hat{x}_{E_2} < 0$.
\end{itemize}

\noindent For Case 1, since $\tanh(g x_{E_1}) \rightarrow 1$ and $\tanh(g x_{E_2}) \rightarrow -1$, we can solve for $\hat{y}_I$ using row 3 of \cref{eq:cluster3system} to get
\begin{equation}\label{eq:yihat}
    \hat{y}_I = \frac{\beta_C - 1}{\beta_C+1} \frac{p n_C}{p n_C - \alpha},
\end{equation} 
from which it follows that
\begin{equation}\label{eq:xilimiteq}
    x_I \rightarrow \frac{1}{g}\tanh^{-1}\left( \frac{\beta_C - 1}{\beta_C+1} \frac{p n_C}{p n_C - \alpha} \right) \text{ as } g \rightarrow \infty.
\end{equation}
Using \cref{eq:yihat} with rows 1 and 2 of \cref{eq:cluster3system},
\begin{equation}\label{eq:xEhat12}
\begin{aligned}
    \hat{x}_{E_1} &= \frac{\mu}{\sqrt{N}}\left( (p-1)n_C - \frac{\beta_C - 1}{\beta_C+1} \frac{p^2 n_C^2}{p n_C - \alpha} \right) \\
    \hat{x}_{E_2} &= \frac{\mu}{\sqrt{N}}\left( -(p-1)n_C - \frac{\beta_C - 1}{\beta_C+1} \frac{p^2 n_C^2}{p n_C - \alpha} \right),
\end{aligned}
\end{equation}
which reduce to \cref{eq:xEhat} when $\beta_C = 1$. Since $n_C p = f N \rightarrow \infty$ as $N \rightarrow \infty$, this simplifies to
\begin{equation}\label{eq:xEhat12IlargeN}
    \begin{aligned}
        x_{E_1} &\rightarrow\frac{\mu}{\sqrt{N}}\left( (p-1)n_C - \frac{\beta_C - 1}{\beta_C+1} p n_C \right) \\
        x_{E_2} &\rightarrow \frac{\mu}{\sqrt{N}}\left( -(p-1)n_C - \frac{\beta_C - 1}{\beta_C+1} p n_C \right) \\
        x_I &\rightarrow \frac{1}{g}\tanh^{-1}\left( \frac{\beta_C - 1}{\beta_C+1} \right)
    \end{aligned}
\end{equation}
as $g, N \rightarrow \infty$. For \cref{eq:xEhat12} to be valid, the consistency conditions $\hat{x}_{E_1} > 0$ and $\hat{x}_{E_2} < 0$ must be satisfied. Since $\hat{x}_{E_2} < 0$ always holds, \cref{eq:xEhat12} is consistent as long as
\begin{equation}\label{eq:betaCineq}
(p-1)n_C - \frac{\beta_C - 1}{\beta_C+1}\frac{p^2 n_C^2}{p n_C - \alpha}  > 0.
\end{equation}
Solving for $\beta_C$, this results in the condition $\beta_C < \beta_C^*$, where $\beta_C^*$ is defined in \ref{eq:betaCstar}.

For Case 2, we can solve for $\hat{y}_{E_1}$ and $\hat{y}_I$ using rows 2 and 3 of \cref{eq:cluster3system} to get
\begin{equation}\label{eq:ye1hatyihat}
    \begin{aligned}
        \hat{y}_{E_1} &= \frac{n_C p^2 }{ \alpha(1+\beta_C)(p-1) + n_C p(1 + \beta_C - p)} \\
        \hat{y}_{I} &= \frac{n_C p(p-1) }{ \alpha(1+\beta_C)(p-1) + n_C p(1 + \beta_C - p)},
    \end{aligned}
\end{equation}
from which it follows that
\begin{equation}
    \begin{aligned}
        x_{E_1} &\rightarrow \frac{1}{g} \tanh^{-1} \left( \frac{n_C p^2 }{ \alpha(1+\beta_C)(p-1) + n_C p(1 + \beta_C - p)}  \right) \\
        x_{E_2} &\rightarrow \frac{\mu}{\sqrt{N}}\left( -(p-1)n_C - \frac{n_C^2 p^2(p-1) }{ \alpha(1+\beta_C)(p-1) + n_C p(1 + \beta_C - p)}\right) \\
        x_{I} &\rightarrow \frac{1}{g} \tanh^{-1} \left(\frac{n_C p(p-1) }{ \alpha(1+\beta_C)(p-1) + n_C p(1 + \beta_C - p)} \right).
    \end{aligned}
\end{equation}
as $g \rightarrow \infty$. We note that for $\beta_C > \beta_C^*$, we cannot take $N \rightarrow \infty$ with $n_C$ held fixed, since for sufficiently large $N$, we will always have $\beta_C < \beta_C^*$. 

\section{Stable excitatory clusters for large \texorpdfstring{$g$}{g}: detailed calculations} 
\label{app:C1C2I1I2_largeg}
We begin with the ansatz (suggested by numerical continuation) that as $g \rightarrow \infty$, $(x_{E_1}, x_{E_2}, x_{I_1}, x_{I_1}) \rightarrow (\hat{x}_{E_1}, \hat{x}_{E_2}, \hat{x}_{I_1}, \hat{x}_{I_2})$, where $\hat{x}_{E_1}, \hat{x}_{I_1} > 0$ and $\hat{x}_{E_2}, \hat{x}_{I_2} < 0$. 
With these assumptions, equation \cref{eq:cluster4system} reduces to 
\begin{equation}\label{eq:xhat}
    \begin{aligned}
        \hat{x}_{E_1} &= \frac{\mu}{\sqrt{N}}\left[ (p-1)n_C - \alpha \frac{\beta-1}{\beta+1}n_I \right] \\
        \hat{x}_{E_2} &= \frac{\mu}{\sqrt{N}}\left[ -(p-1)n_C - \alpha \frac{\beta-1}{\beta+1}n_I \right] \\
        \hat{x}_{I_1} &= \frac{\mu}{\sqrt{N}}\left[  \frac{\beta_C-1}{\beta_C+1} p n_C - \alpha \left( \frac{\beta-1}{\beta+1}n_I - 1 \right) \right] \\
        \hat{x}_{I_2} &= \frac{\mu}{\sqrt{N}}\left[  \frac{\beta_C-1}{\beta_C+1} p n_C - \alpha \left( \frac{\beta-1}{\beta+1}n_I + 1 \right) \right],
    \end{aligned}
\end{equation}
since $\tanh(g x_{E_1}), \tanh(g x_{I_1}) \rightarrow 1$ and $\tanh(g x_{E_2}), \tanh(g x_{I_2}) \rightarrow -1$ as $g \rightarrow \infty$. Equation \cref{eq:xhat} gives the limiting solutions $(\hat{x}_{E_1}, \hat{x}_{E_2}, \hat{x}_{I_1}, \hat{x}_{I_1})$ as long as the consistency conditions $\hat{x}_{E_1}, \hat{x}_{I_1} > 0$ and $\hat{x}_{E_2}, \hat{x}_{I_2} < 0$ are satisfied. Since $\mu > 0$, the consistency conditions reduce to
\begin{equation}\label{eq:consistency}
    \begin{aligned}
        &(p-1)n_C - \alpha \frac{\beta-1}{\beta+1}n_I > 0 \\
        -&(p-1)n_C - \alpha \frac{\beta-1}{\beta+1}n_I < 0 \\
        &\frac{\beta_C-1}{\beta_C+1} p n_C - \alpha \left( \frac{\beta-1}{\beta+1}n_I - 1 \right) > 0 \\
        &\frac{\beta_C-1}{\beta_C+1} p n_C - \alpha \left( \frac{\beta-1}{\beta+1}n_I + 1 \right) < 0.
    \end{aligned}
\end{equation}
The first pair of inequalities in \cref{eq:consistency} is satisfied if and only if
\[
    \left| \frac{\beta-1}{\beta+1} \right| < \frac{(p-1)n_C}{\alpha n_I} = 1 - \frac{1}{p},
\]
where we used the fact that $n_C p = n_E = \alpha n_I$. Since we are taking $\beta \geq 1$, this simplifies to $1 \leq \beta < 2p$. 
Similarly, the second pair of inequalities in \cref{eq:consistency} is satisfied if and only if 
\[
    \alpha \left( \frac{\beta-1}{\beta+1}n_I - 1 \right) < \frac{\beta_C-1}{\beta_C+1} p n_C < \alpha \left( \frac{\beta-1}{\beta+1}n_I + 1 \right),
\]
which simplifies to \cref{eq:bccondition}.

\bibliographystyle{siamplain}
\bibliography{main.bib}

\end{document}